\documentclass[%
 reprint,
superscriptaddress,
nofootinbib,
 amsmath,amssymb,
 aps,
 prd,
]{revtex4-2}

\usepackage[usenames,dvipsnames]{xcolor}
\clearpage{}%

\newcommand{\rs}{r_{*}}
\newcommand{\rp}{r_{+}}
\newcommand{\rmi}{r_{-}}
\newcommand{\diff}{d}
\newcommand{\bigO}{\mathcal{O}}

\usepackage{graphicx}%
\graphicspath{{./}{figures/}}
\usepackage[caption=false,labelformat=simple]{subfig}

\usepackage{dcolumn}%
\usepackage{booktabs}
\usepackage{bm}%
\usepackage{hyperref}%
\hypersetup{
	colorlinks=true,
	citecolor=NavyBlue,
	linkcolor=NavyBlue,
	urlcolor=NavyBlue,
}
\usepackage{cancel}
\usepackage{slashed}
\allowdisplaybreaks[4]
\usepackage[normalem]{ulem}
\usepackage{braket}
\usepackage{nicefrac}
\usepackage[acronym]{glossaries}
\newacronym{GSN}{GSN}{generalized Sasaki-Nakamura}
\newacronym{SN}{SN}{Sasaki-Nakamura}
\newacronym{MST}{MST}{Mano-Suzuki-Takasugi}
\newacronym{NP}{NP}{Newman-Penrose}
\newacronym{BH}{BH}{black hole}
\newacronym{GW}{GW}{gravitational-wave}
\newacronym{ODE}{ODE}{ordinary differential equation}
\newacronym{PDE}{PDE}{partial differential equation}
\newacronym{GDT}{GDT}{Generalized Darboux Transformation}
\newacronym{EMRI}{EMRI}{extreme mass-ratio inspiral}
\newacronym{LIGO}{LIGO}{Laser Interferometer Gravitational-Wave Observatory}
\newacronym{LISA}{LISA}{Laser Interferometer Space Antenna}
\newacronym{DECIGO}{DECIGO}{Deci-hertz Interferometer Gravitational wave Observatory}
\newacronym{AD}{AD}{automatic differentiation}
\newacronym{IBP}{IBP}{integration by parts}
\newacronym{SWSH}{SWSH}{spin-weighted spheroidal harmonic}
\newacronym{BL}{BL}{Boyer-Lindquist}
\newacronym{FT}{FT}{Fourier transform}
\newacronym{ZFL}{ZFL}{zero frequency limit}
\newacronym{QNM}{QNM}{quasi-normal mode}
\usepackage{amsfonts}
\usepackage{mathrsfs}

\makeatletter
\newcommand*{\glsplainhyperlink}[2]{%
  \colorlet{currenttext}{.}%
  \colorlet{currentlink}{\@linkcolor}%
  \hypersetup{linkcolor=currenttext}%
  \hyperlink{#1}{#2}%
  \hypersetup{linkcolor=currentlink}%
}
\let\@glslink\glsplainhyperlink
\makeatother

\begin{document}

\title{Gravitational radiation from Kerr black holes using the Sasaki-Nakamura formalism: Waveforms and fluxes at infinity}
\author{Yucheng Yin}
\email{yucheng.yin@nbi.ku.dk}
\affiliation{%
Center of Gravity, Niels Bohr Institute, Blegdamsvej 17, 2100 Copenhagen, Denmark}%
\affiliation{Department of Astronomy, School of Physics, Peking University, 100871 Beijing, China}%
\affiliation{Kavli Institute for Astronomy and Astrophysics at Peking University, 100871 Beijing, China}%
\author{Rico K.~L. Lo}
\email{kalok.lo@nbi.ku.dk}
\affiliation{%
Center of Gravity, Niels Bohr Institute, Blegdamsvej 17, 2100 Copenhagen, Denmark}%
\author{Xian Chen}
\affiliation{Department of Astronomy, School of Physics, Peking University, 100871 Beijing, China}%
\affiliation{Kavli Institute for Astronomy and Astrophysics at Peking University, 100871 Beijing, China}%

\date{\today}%

\begin{abstract}
In linear perturbation theory for Kerr black holes, there are two equivalent formalisms, namely the Teukolsky and the Sasaki-Nakamura (SN) formalism.
Typically, one defaults to the Teukolsky formalism, especially when calculating extreme mass ratio inspiral waveforms, and uses the SN formalism when dealing with extended sources, as it offers superior convergence when employing the Green's function method for calculating the inhomogeneous solution.
In this work, we present a new scheme for solving the inhomogeneous SN equation, based on integration by parts, that eliminates the extra radial integration step required in the standard formulation to construct the source term for convolution with the SN variable.
We derive also a SN source term that is valid for point particles on arbitrary motions around Kerr black holes.
Our approach enables efficient computations of gravitational waveforms within the SN formalism in \emph{all} cases, from compact to extended sources.
We validate our scheme and code implementation against the literature and find excellent agreement, achieving comparable performance without employing any special optimization techniques.
\end{abstract}

\maketitle

\section{Introduction \label{sec:intro}}
It has now been a decade since the first direct detection of a \gls{GW} signal by the \gls{LIGO} \cite{LIGOScientific:2016aoc} that marks the beginning of \gls{GW} astronomy.
Over the past ten years, we are able to understand much more about astrophysics, such as the population properties of stellar-mass compact binary system \cite{LIGOScientific:2025pvj}, fundamental physics, such as \gls{BH} mechanics \cite{LIGOScientific:2025rid}, and more, by analyzing the gravitational waveforms observed by ground-based detectors such as \gls{LIGO} \cite{LIGOScientific:2014pky}, Virgo \cite{VIRGO:2014yos} and KAGRA \cite{Somiya:2011np}.

The next big leap in the field of \gls{GW} astronomy would be the commission of space-based \gls{GW} detectors such as the \gls{LISA} \cite{Colpi:2024xhw}.
These space-based detectors target a much lower frequency band---in the millihertz range---compared to those ground-based ones.
As a result, they are sensitive to different astrophysical sources of \glspl{GW}.
One such sources is the \glspl{EMRI} \cite{2018LRR....21....4A}, which are the gravitational radiation emitted by massive \glspl{BH} when perturbed by smaller bodies, such as stars and \glspl{BH} that are much lighter, orbiting around them.

In contrast to the gravitational waveforms coming from the merger of a stellar-mass compact binary system that we can observe at ground-based detectors effectively infinitely far away, the \gls{GW} signals coming from \glspl{EMRI} can remain detectable for months to years instead of just mere seconds \cite{Babak:2017tow, Berry:2019wgg}.
Therefore, it is crucial for us to be able to compute these \gls{EMRI} waveforms accurately, such that we can compare these theoretical predictions with observations and extract properties about the sources of those \glspl{EMRI}.

\subsection{Primer in solving for gravitational waveforms from Kerr black holes using Green's functions}
The gravitational waveforms that we observe at spatial infinity contain two modes, namely the plus polarization $h_{+}$ and the cross polarization $h_{\times}$, respectively.
They are encoded in the perturbed Weyl scalar $\psi_4$ as \cite{Chandrasekhar:1985kt}
\begin{equation}
\frac{1}{2}\dfrac{\partial^2}{\partial t^2} \left( h_{+} - ih_{\times} \right) = \psi_4(r \to \infty).
\end{equation}
In his seminal work, Teukolsky showed that the equation governing the linear perturbation to the scalar $\psi_{4}$, which is a \acrlong{PDE}, can be solved using separation of variables and a \gls{FT} \cite{Teukolsky:1973ha}.
Schematically, in the Kinnersley tetrad, this decomposition can be written as
\begin{equation}
\rho^{-4}\psi_4(t, r, \theta, \varphi) = \sum_{\ell m \omega} R_{\ell m \omega}(r) {}_{-2}S_{\ell m \omega}(\theta, \varphi) e^{- i\omega t},
\end{equation}
where $\rho=-(r-ia\cos\theta)^{-1}$, and $(t,r,\theta,\varphi)$ are the \gls{BL} coordinates.
Throughout this paper, we use geometric units where $c = G = M = 1$.

For the angular sector $(\theta, \varphi)$, the solutions are known as the \glspl{SWSH} ${}_{-2}S_{\ell m \omega}(\theta, \varphi) = {}_{-2}S_{\ell m \omega}(\theta)e^{i m \varphi}$.
We refer readers to the Appendix A of Ref.~\cite{Lo:2023fvv} for more details.\footnote{Furthermore, a prime denotes a derivative with respect to $r$, an overhead dot denotes a derivative with respect to $t$, while a bar over a variable denotes its complex conjugate.
For the normalization conventions of the \gls{FT} and \glspl{SWSH} adopted in this paper, refer to Appendix~\ref{subsec:normalization_conventions}.}
As for the radial sector, the solutions $R_{\ell m \omega}(r)$ are governed by an \gls{ODE}, aptly referred to as the radial Teukolsky equation in literature, given by
\begin{equation}\label{Eq.TeukolskyRadialEquation}
    \left[\Delta^2\frac{\diff}{\diff  r}\left(\frac{1}{\Delta}\frac{\diff }{\diff  r}\right)-V_{\rm T}(r)\right]R_{\ell m\omega}(r)= - \mathcal{T}_{\ell m\omega}(r),
\end{equation}
with a potential $V_{\rm T}$ given by
\begin{equation}\label{Eq.TeukolskyEffectivePotential}
    V_{\rm T}(r)= -\dfrac{K^2 + 4i(r-1)K}{\Delta} + 8i\omega r + \lambda_{\ell m \omega},
\end{equation}
where $\Delta = (r - r_+)(r - r_-)$, $r_{\pm} = 1 \pm \sqrt{1 - a^2}$, $K = (r^2 + a^2)\omega - ma$, and
$\lambda_{\ell m\omega}$ is the separation constant from the angular sector. For the sake of simplicity, we will drop the $\ell m \omega$ subscript when there is no risk of confusion hereinafter.

Conceptually, the inhomogeneous radial Teukolsky equation in Eq.~\eqref{Eq.TeukolskyRadialEquation} can be solved using the Green's function method.
We start with two linearly independent homogeneous solutions that satisfy one of the two boundary conditions for the inhomogeneous solution $R^{\rm inhomo}$ that we want, respectively.
In this case, we impose the boundary conditions that the solution is purely ingoing at the horizon and purely outgoing at spatial infinity, and the corresponding homogeneous solutions are denoted by $R^{\rm in}(r)$ and $R^{\rm up}(r)$, respectively. Specifically, these solutions have the following asymptotic forms
\begin{align}
\label{eq:Rin}
	R^{\mathrm{in}}(r) & = \begin{cases}
		B^{\mathrm{trans}}_{\mathrm{T}}  \Delta^{2} e^{-i \kappa r_*}, & r \to r_+ \\
		B^{\mathrm{inc}}_{\mathrm{T}} \dfrac{e^{-i\omega r_*}}{r} + B^{\mathrm{ref}}_{\mathrm{T}} r^3 e^{i\omega r_*}, & r \to \infty \\
	\end{cases}, \\
\label{eq:Rup}
	R^{\mathrm{up}}(r) & = \begin{cases}
		C^{\mathrm{ref}}_{\mathrm{T}} \Delta^{2} e^{-i \kappa r_*} + C^{\mathrm{inc}}_{\mathrm{T}} e^{i \kappa r_*}, & r \to r_+ \\
		C^{\mathrm{trans}}_{\mathrm{T}} r^3 e^{i\omega r_*}, & r \to \infty
	\end{cases},
\end{align}
where $B^{\rm trans,inc, ref}_{\rm T}$ and $C^{\rm trans,inc, ref}_{\rm T}$ are the transmission, incidence, and reflection coefficients for the $R^{\rm in}$ and $R^{\rm up}$ solutions, respectively, and $\kappa = \omega-ma/(2\rp)$.

With these two solutions, we can construct a Green's function $G_{\rm T}(r, \tilde{r})$ as
\begin{equation}
G_{\rm T}(r, \tilde{r}) = \begin{cases}
\dfrac{1}{W_R} R^{\rm up}(r)R^{\rm in}(\tilde{r}), & r > \tilde{r} \\ 
\dfrac{1}{W_R} R^{\rm in}(r)R^{\rm up}(\tilde{r}), & r < \tilde{r} 
\end{cases}, 
\end{equation} 
where $W_R$ is the scaled Wronskian for the two Teukolsky solutions given by
\begin{equation}
	W_R = \dfrac{1}{\Delta} \left( R^{\rm in} \dfrac{\diff R^{\rm up}}{\diff r} - R^{\rm up} \dfrac{\diff R^{\rm in}}{\diff r} \right).
\end{equation}
The inhomogeneous solution $R^{\rm inhomo}(r)$ is then given by
\begin{equation}
\label{eq:inhomo_R_conv_integral_full}
\begin{aligned}
	R^{\rm inhomo}(r) = & \frac{R^{\rm up}(r)}{W_R} \int_{r_+}^{r} d\tilde{r}\, \dfrac{R^{\rm in}(\tilde{r}) \mathcal{T}(\tilde{r})}{\Delta^{2}(\tilde{r})} \\
	& + \frac{R^{\rm in}(r)}{W_R} \int_{r}^{\infty} d\tilde{r} \, \dfrac{R^{\rm up}(\tilde{r}) \mathcal{T}(\tilde{r})}{\Delta^{2}(\tilde{r})}
\end{aligned}.
\end{equation}
As $r \to \infty$, the solution becomes
\begin{equation}
\label{eq:inhomo_R_conv_integral}
    R^{\rm inhomo}_{\ell m \omega}(r \to \infty) = \underbrace{\frac{1}{2 i \omega B^{\mathrm{inc}}_{\mathrm{T}}} \int_{r_+}^{\infty} d\tilde{r}\, \dfrac{R^{\rm in}(\tilde{r}) \mathcal{T}(\tilde{r})}{\Delta^{2}(\tilde{r})}}_{ Z^{\infty}_{\ell m \omega} } r^3 e^{i\omega r_*},
\end{equation}
where we have substituted an expression for $W_{R}$, and we can see that $R^{\rm inhomo}_{\ell m \omega}$ indeed satisfies the purely outgoing boundary condition at spatial infinity.

Computationally, one will need to first solve the homogeneous radial Teukolsky equation to get $R^{\rm in, up}(r)$ and the Wronskian $W_{R}$ before performing the convolution integral over the source term.
This can be done using various methods, such as the \gls{MST} method \cite{Mano:1996vt, 10.1143/PTP.112.415, 10.1143/PTP.113.1165}, the \gls{SN} formalism \cite{SASAKI198268, 10.1143/PTP.67.1788, Lo:2023fvv}, and recently a method based on analytical series expansion \cite{Jiang:2025mna}.
When the source term $\mathcal{T}$ is compact (i.e., nonvanishing only at finite values of $r$), for instance, the bound motion of a test particle orbiting around a \gls{BH} for \gls{EMRI} waveform modeling, Eq.~\eqref{eq:inhomo_R_conv_integral_full} is perfectly fine for these kind of calculations (see, for example, Ref.~\cite{Drasco:2005kz}).

However, Eq.~\eqref{eq:inhomo_R_conv_integral_full} is no longer suitable for numerical computations when $\mathcal{T}$ is extended and does not decay fast enough. In these cases, the integral for $Z^{\infty}_{\ell m \omega}$ is divergent.\footnote{The Teukolsky formalism itself is still valid. It is just that Eq.~\eqref{eq:inhomo_R_conv_integral} is no longer a solution to the inhomogeneous radial Teukolsky equation. See Ref.~\cite{Poisson:1996ya} for a detailed explanation.}
A prototypical example of this scenario would be the radial infall of a test particle from infinity towards a \gls{BH}.
In fact, this was the very motivation that led to the development of the \gls{SN} formalism \cite{SASAKI198268, 10.1143/PTP.67.1788}, which gives a well-behaved source term and a convergent convolution integral even when the source is extended, on top of providing an efficient numerical scheme to solve the homogeneous Teukolsky equation in Eq.~\eqref{Eq.TeukolskyRadialEquation}.

\subsection{This work}%
In this paper, we revamp the \gls{SN} formalism for the driven case to take \emph{full} advantages of the formalism for computing gravitational radiation from Kerr \glspl{BH}. 
Specifically, we give a derivation of the source term for the inhomogeneous \gls{SN} equation that is valid for any equation of motion (i.e., not necessary a geodesic), and introduce a new scheme for solving gravitational waveforms that bypasses the additional integration that was required to obtain the appropriate source term, which is a common criticism of the formalism.

This paper is organized as follows.
In Sec.~\ref{subsec:ABC_of_SN}, we first review the basics of the \gls{SN} formalism.
Then in Sec.~\ref{subsec:IBP}, we describe our new scheme for solving the inhomogeneous \gls{SN} equation using integration by parts, followed by the recipes for computing gravitational waveforms and fluxes at infinity using the \gls{SN} formalism in Sec.~\ref{subsec:recipes}.
We present in Sec.~\ref{subsec:result_bound} and Sec.~\ref{subsec:result_unbound} our results for bound and unbound orbits, respectively.
Finally, we discuss some applications, limitations, and extensions of this work in Sec.~\ref{sec:discussions}.

\section{Sasaki-Nakamura formalism}
\label{Sec:SN_formalism}
Here we review the basics of the \gls{SN} formalism for the sake of completeness.
In essence, the formalism introduces a new variable $X_{\ell m \omega}$ in place of $R_{\ell m \omega}$ used in the Teukolsky formalism.
This new variable is constructed such that the \gls{ODE} it satisfies has a short-ranged potential and a source term that gives a convergent integral when using the Green's function method.
We refer readers to Refs.~\cite{SASAKI198268, 10.1143/PTP.67.1788} for the detailed construction of the variable.

\subsection{Basics of the Sasaki-Nakamura formalism\label{subsec:ABC_of_SN}}

The variable $X_{\ell m \omega}$ satisfies the \gls{SN} equation, which is given by
\begin{equation}\label{Eq.Inhomogeneous_SN}
    \left[\frac{\diff^2}{\diff {r_*}^2}- \mathcal{F}_{\ell m \omega}\frac{\diff}{\diff r_*}- \mathcal{U}_{\ell m \omega}\right]X_{\ell m\omega}(r_*)=\mathcal{S}_{\ell m\omega}(r),
\end{equation}
where
\begin{subequations}
    \begin{align}
        \mathcal{F}(r)&= \frac{\eta'}{\eta}\frac{\Delta}{r^2+a^2},\\
        \mathcal{U}(r)&=\frac{\Delta U_1}{(r^2+a^2)^2}+G^2+\frac{\Delta G'}{r^2+a^2}-\mathcal{F}G,\\
        G(r)&=-\frac{2(r-1)}{r^2+a^2}+\frac{r\Delta}{(r^2+a^2)^2},\\
        U_1(r)&=V_{\rm T} +\frac{\Delta^2}{\beta}\left[\left(2\alpha+\frac{\beta'}{\Delta}\right)'-\frac{\eta'}{\eta}\left(\alpha+\frac{\beta'}{\Delta}\right)\right],
    \end{align}
\end{subequations}
with
\begin{subequations}
    \begin{align}
        \alpha &=3i K'+\lambda+\frac{6\Delta}{r^2}-i\frac{K\beta}{\Delta^2},\\
        \beta &=\Delta\left(-2i K+\Delta'-\frac{4\Delta}{r}\right), \\
        \eta &=c_0+\frac{c_1}{r}+\frac{c_2}{r^2}+\frac{c_3}{r^3}+\frac{c_4}{r^4},
    \end{align}
\end{subequations}
and
\begin{subequations}
    \begin{align}
	c_0 & = -12i\omega + \lambda(2 + \lambda) -12 a\omega \left( a\omega - m \right), \\
	c_1 & = 8iam\lambda + 8ia^2 \omega( 3 - \lambda), \\
	c_2 & = -24 i a \left( a \omega - m\right) + 12 a^2 \left[ 1 - 2 \left( a \omega - m \right)^2 \right], \\
	c_3 & = 24 i a^3 \left( a \omega - m \right) - 24a^2, \\
	c_4 & = 12a^4.
    \end{align}
\end{subequations}
Moreover, the \gls{ODE} is written with respect to the tortoise coordinate $\rs$ given by
\begin{equation}\label{Eq.Tortoise}
    \begin{aligned}
        \rs(r)& = \int^{r} \frac{\tilde{r}^2+a^2}{\Delta}\diff \tilde{r},\\
        & = r+\frac{2\rp}{\rp-\rmi}\ln\frac{r-\rp}{2}-\frac{2\rmi}{\rp-\rmi}\ln\frac{r-\rmi}{2}.
    \end{aligned}
\end{equation}
The Teukolsky variable $R_{\ell m\omega}(r)$ in Eq.~\eqref{Eq.TeukolskyRadialEquation} can be constructed from the \gls{SN} variable $X_{\ell m\omega}\left(\rs(r)\right)$ in Eq.~\eqref{Eq.Inhomogeneous_SN} using
\begin{equation}\label{Eq.SNtransformationRtoX}
    R_{\ell m\omega}(r)=\Lambda^{-1}\left[X_{\ell m\omega}(\rs(r))\right]+\frac{\left(r^2+a^2\right)^{3/2}}{\eta}\mathcal{S}_{\ell m\omega},
\end{equation}
where $\Lambda^{-1}$ is the differential operator for the inverse \gls{SN} transformation defined as
\begin{equation}
    \Lambda^{-1}\left[X_{\ell m\omega}\right]=\frac{1}{\eta}\left[\frac{\alpha\Delta+\beta'}{\sqrt{r^2+a^2}}X_{\ell m\omega}-\frac{\beta}{\Delta}\left(\frac{\Delta X_{\ell m\omega}}{\sqrt{r^2+a^2}}\right)'\right].
\end{equation}
Comparing Eq.~\eqref{Eq.SNtransformationRtoX} with the homogeneous case, e.g., in Ref.~\cite{Lo:2023fvv}, we see that there is now an additional contribution coming from the source term $\mathcal{S}_{\ell m \omega}$. By construction, $\mathcal{S}_{\ell m \omega}$ decays fast enough as $r \to \infty$ such that it does not contribute to Eq.~\eqref{Eq.SNtransformationRtoX} when evaluated at spatial infinity.
Importantly, this implies that
\begin{equation}
\label{eq:inhomo_X_to_inhomo_R_asym_relation}
    R_{\ell m \omega}(r\to\infty) = \lim_{r \to \infty} \Lambda^{-1}\left[X_{\ell m\omega}(\rs(r))\right].
\end{equation}

The derivation of the expression relating the \gls{SN} source term $\mathcal{S}_{\ell m \omega}$ with the Teukolsky source term $\mathcal{T}_{\ell m \omega}$ can be found in Appendix \ref{App:Derivation_of_inhomogeneous_SN}.
Here, we just state the result, which is
\begin{equation}\label{Eq.Source_S}
    \mathcal{S}_{\ell m\omega}=\frac{\eta\Delta\mathcal{W}}{(r^2+a^2)^{3/2}r^2}\exp\left(-i\int^r\frac{K}{\Delta}\diff\tilde{r}\right),
\end{equation}
where we define an auxiliary function $\mathcal{W}(r)$ related to the Teukolsky source term $\mathcal{T}_{\ell m\omega}$ by
\begin{equation}\label{Eq.d2W}
    \frac{\diff^2 \mathcal{W}}{\diff r^2}=-\frac{r^2}{\Delta^2}\mathcal{T}_{\ell m\omega}(r)\exp\left(i\int^r\frac{K}{\Delta}\diff\tilde{r}\right).
\end{equation}
One in principle needs to integrate the \gls{ODE} in Eq.~\eqref{Eq.d2W} to obtain the \gls{SN} source term for Eq.~\eqref{Eq.Inhomogeneous_SN}.
The Teukolsky source term $\mathcal{T}_{\ell m \omega}$ itself for a point particle is given by \cite{2003LRR.....6....6S}
\begin{equation}\label{Eq.T_definition}
    \begin{aligned}
        \mathcal{T}_{\ell m\omega}(r)=\;&\mu\int_\gamma\diff \tau\ e^{i\omega t(\tau)-i m\varphi(\tau)} \\
        &\Delta^2 \left\{\left(A_{nn0}+A_{n\bar{m}0}+A_{\bar{m}\bar{m}0}\right)\delta(r-r(\tau))\right.\\
        &+\left[\left(A_{n\bar{m}1}+A_{\bar{m}\bar{m}1}\right)\delta(r-r(\tau))\right]'\\
        &\left.+\left[A_{\bar{m}\bar{m}2}\delta(r-r(\tau))\right]''\right\},
    \end{aligned}
\end{equation}
where $\mu$ is the mass of the particle and $\gamma$ denotes its trajectory.
The expressions of $A_{nn0}$, $A_{n\bar{m}0}$, $A_{\bar{m}\bar{m}0}$, $A_{n\bar{m}1}$, $A_{\bar{m}\bar{m}1}$, $A_{\bar{m}\bar{m}1}$ can be found in Appendix \ref{Appendix_A_and_W}.

We can solve the inhomogeneous \gls{SN} equation using the Green's function method.
Similarly, we need $X^{\rm in}$ and $X^{\rm up}$ that satisfy the purely ingoing boundary condition at the horizon and purely outgoing boundary condition at spatial infinity, respectively. Asymptotically, they are given by
\begin{equation}\label{Eq.Xin_Asymptotic}
    X^{\mathrm{in}}(\rs)=\begin{cases}
        B^{\mathrm{trans}}_{\mathrm{SN}}e^{-i \kappa \rs}& \rs \to-\infty\\
        B^{\mathrm{inc}}_{\mathrm{SN}}e^{-i\omega \rs}+B^{\mathrm{ref}}_{\mathrm{SN}}e^{i\omega \rs}& \rs \to\infty
    \end{cases},
\end{equation}
and
\begin{equation}\label{Eq.Xup_Asymptotic}
    X^{\mathrm{up}}(r_*)=\begin{cases}
        C^{\mathrm{ref}}_{\mathrm{SN}}e^{-i \kappa r_*}+C^{\mathrm{inc}}_{\mathrm{SN}}e^{i \kappa \rs }& \rs \to -\infty\\
        C^{\mathrm{trans}}_{\mathrm{SN}}e^{i \omega \rs} & \rs\to\infty
    \end{cases},
\end{equation}
where $B^{\rm trans,inc, ref}_{\rm SN}$ and $C^{\rm trans,inc, ref}_{\rm SN}$ are the transmission, incidence and reflection coefficients for the $X^{\rm in}$ and $X^{\rm up}$ solutions, respectively.
The inhomogeneous solution $X^{\rm inhomo}(\rs)$ is then given by
\begin{equation}
\label{eq:full_inhomo_X_using_green_func}
    \begin{aligned}
        X^{\rm inhomo}_{\ell m\omega}(\rs)=&\frac{X_{\ell m\omega}^{\mathrm{up}}(\rs)}{W_X}\int_{-\infty}^{\rs} X_{\ell m\omega}^{\mathrm{in}}(\tilde{r}_*)\frac{\mathcal{S}_{\ell m\omega}(\tilde{r}_*) }{\eta}\diff \tilde{r}_* \\
        &+\frac{X_{\ell m\omega}^{\mathrm{in}}(\rs)}{W_X}\int_{\rs}^\infty X_{\ell m\omega}^{\mathrm{up}}(\tilde{r}_*)\frac{\mathcal{S}_{\ell m\omega}(\tilde{r}_*)}{\eta}\diff\tilde{r}_*,
    \end{aligned}
\end{equation}
where $W_X$ is the scaled Wronskian\footnote{The scaled Wronskain defined for the \gls{SN} variable $X$ is in fact identical to that defined for the Teukolsky variable $R$. Refer to Appendix E of Ref.~\cite{Lo:2023fvv} for a proof.} defined by
\begin{equation}
    W_X = \dfrac{1}{\eta} \left[ X_{\ell m\omega}^{\mathrm{in}} \frac{\diff X_{\ell m\omega}^{\mathrm{up}}}{\diff \rs} - X_{\ell m\omega}^{\mathrm{up}} \frac{\diff X_{\ell m\omega}^{\mathrm{in}}}{\diff \rs} \right] =\frac{2i\omega}{c_0}B^{\mathrm{inc}}_{\mathrm{SN}}C^{\mathrm{trans}}_{\mathrm{SN}} .
\end{equation}

In particular, when $\rs \to \infty$, the inhomogeneous \gls{SN} solution becomes
\begin{multline}\label{Eq.X^Infty}
    X^{\rm inhomo}_{\ell m\omega}(\rs \to \infty) = \\ 
    \underbrace{\frac{c_0}{2i\omega B_{\mathrm{SN}}^{\mathrm{inc}}}\int_{-\infty}^\infty\frac{X_{\ell m\omega}^{\mathrm{in}}(\rs)\mathcal{S}_{\ell m\omega}(\rs)}{\eta}\diff \rs}_{X_{\ell m\omega}^{\infty}} e^{i\omega \rs}.
\end{multline}
Using Eq.~\eqref{eq:inhomo_X_to_inhomo_R_asym_relation}, one can relate the asymptotic amplitude at infinity $X^{\infty}_{\ell m \omega}$ with $Z^{\infty}_{\ell m \omega}$, which means
\begin{equation}
    R^{\rm inhomo}_{\ell m\omega}(r \to \infty) = -\frac{4\omega^2}{c_0}X_{\ell m\omega}^{\infty} r^3 e^{i \omega \rs}.
\end{equation}
The \gls{GW} polarizations $h_{+}$ and $h_{\times}$ can then be expressed as
\begin{multline}
\label{Eq.h}
    h_{+} - ih_{\times} = \\ -\frac{2}{r}\sum_{\ell m}\int_{-\infty}^\infty \frac{Z_{\ell m\omega}^{\infty}}{\omega^2}{_{-2}}S_{\ell m \omega}(\theta)e^{-i\omega(t-\rs) + im\varphi}\diff\omega,
\end{multline}
where
\begin{equation}\label{Eq.Z}
    Z_{\ell m\omega}^{\infty}=-\frac{4\omega^2}{c_0}X_{\ell m\omega}^{\infty}.
\end{equation}

\subsection{New scheme for solving the inhomogeneous Sasaki-Nakamura equation using integration by parts}
\label{subsec:IBP}
Conventionally, solving for gravitational waveforms $h_{+,\times}$ using the \gls{SN} formalism requires first integrating Eq.~\eqref{Eq.d2W} for $\mathcal{W}$ [and hence $\mathcal{S}$ through Eq.~\eqref{Eq.Source_S}], often numerically, and then integrating the convolution integral of some homogeneous solution $X$ with the source term $\mathcal{S}$ in Eq.~\eqref{Eq.X^Infty} for $X^{\infty}_{\ell m \omega}$.
Comparing with Eq.~\eqref{eq:inhomo_R_conv_integral}, where the source term $\mathcal{T}$ in the convolution integral often has an analytical expression, the \gls{SN} formalism seems to be at a disadvantage.
However, this does \emph{not} have to be the case. Here, we show that by using \gls{IBP} twice with the help of an auxiliary function, one can convert the convolution integral in the \gls{SN} formalism to use the Teukolsky source term $\mathcal{T}$.

We first define two auxiliary functions $Y_{\ell m\omega}^{\rm in/up}(r)$, respectively, where
\begin{equation}\label{Eq.ODEforY}
    Y_{\ell m\omega}^{\rm in/up\ \prime\prime}(r)\equiv \frac{X_{\ell m\omega}^{\rm in/up}(r)}{r^2\sqrt{r^2+a^2}}\exp\left(-i\int^r\frac{K}{\Delta}\diff r\right).
\end{equation}
Furthermore, we replace $X^{\rm in/up}_{\ell m \omega}$ with $Y^{\rm in/up}_{\ell m \omega}$ in Eq.~\eqref{eq:full_inhomo_X_using_green_func}, we can obtain
\begin{equation}
    \label{eq:full_inhomo_X_with_Y}
    \begin{aligned}
        X^{\rm inhomo}_{\ell m\omega}(\rs)=&\frac{X_{\ell m\omega}^{\mathrm{up}}(\rs)}{W_X}\int_{r_{+}}^{r(\rs)} Y^{\rm in\ \prime\prime}_{\ell m \omega} \mathcal{W}(r) \diff r \\
        &+\frac{X_{\ell m\omega}^{\mathrm{in}}(\rs)}{W_X} \int_{r(\rs)}^{\infty} Y^{\rm up\ \prime\prime}_{\ell m \omega} \mathcal{W}(r) \diff r.
    \end{aligned}
    \vspace{0.5em} %
\end{equation}
Now Eq.~\eqref{eq:full_inhomo_X_with_Y} is written in a suggestive form. We can apply \gls{IBP} twice to swap the differentiation (with respect to $r$) from $Y$ to $\mathcal{W}$, at the expense of picking up extra boundary terms where
\begin{multline}\label{eq:Xinhomo_IBP}
        X^{\rm inhomo}_{\ell m\omega}(\rs) = \frac{X_{\ell m\omega}^{\mathrm{up}}(\rs)}{W_X} \int_{r_{+}}^{r(\rs)} Y^{\rm in}_{\ell m \omega} \frac{\diff ^2\mathcal{W}(r)}{\diff r^2} \diff r \\
        + \frac{X_{\ell m\omega}^{\mathrm{up}}(\rs)}{W_X} \left[ Y^{\rm in \ \prime}_{\ell m\omega}(r)\mathcal{W}(r)-Y^{\rm in}_{\ell m\omega}(r)\mathcal{W}'(r) \right]_{\rp}^{r(\rs)} \\   
        + \frac{X_{\ell m\omega}^{\mathrm{in}}(\rs)}{W_X} \left[ Y^{\rm up \ \prime}_{\ell m\omega}(r)\mathcal{W}(r)-Y^{\rm up}_{\ell m\omega}(r)\mathcal{W}'(r) \right]^{\infty}_{r(\rs)} \\  
        +\frac{X_{\ell m\omega}^{\mathrm{in}}(\rs)}{W_X} \int_{r(\rs)}^{\infty} Y^{\rm up}_{\ell m \omega} \frac{\diff ^2\mathcal{W}(r)}{\diff r^2}  \diff r.
\end{multline}
Specifically, we are interested in the case when $\rs \to \infty$, i.e.,
\begin{equation}
\label{Eq.Xinf after IBP}
    \begin{aligned}
        X^{\infty}_{\ell m\omega} & = \frac{c_0}{2i\omega B^{\mathrm{inc}}_{\mathrm{SN}}}\left[Y^{\rm in \ \prime}_{\ell m\omega}(r)\mathcal{W}(r)-Y^{\rm in}_{\ell m\omega}(r)\mathcal{W}^{\prime}(r)\right]_{\rp}^\infty \\
        & \; +\frac{c_0}{2i\omega B^{\mathrm{inc}}_{\mathrm{SN}}}\int_{\rp}^\infty Y^{\rm in}_{\ell m\omega}(r)\frac{\diff ^2\mathcal{W}(r)}{\diff r^2}\diff r.
    \end{aligned}
\end{equation}
This is the key result of the paper---if we can discard the boundary terms (which later in the text we show that this is justified in some cases), then we can calculate $X^{\infty}_{\ell m\omega}$ without having to solve for $\mathcal{W}$.
Additionally, the new auxiliary function $Y^{\rm in}_{\ell m\omega}$ introduced here does \emph{not} depend on the source term and can be constructed easily from the homogeneous solution $X^{\rm in}_{\ell m\omega}$, which will be the subject of Sec.~\ref{subsubsec:Y(r)}.

By inserting Eq.~\eqref{Eq.d2W} into Eq.~\eqref{Eq.Xinf after IBP}, the convolution integral in our new scheme can be written as
\begin{widetext}
\begin{equation}\label{Eq.I}
    \begin{aligned}
        I&=\int_{\rp}^\infty Y^{\rm in}(r)\frac{\diff^2\mathcal{W}(r)}{\diff r^2}\diff r\\
        &=-\int_{\rp}^\infty Y^{\rm in}(r)\frac{r^2}{\Delta^2}\mathcal{T}_{\ell m\omega}(r)\exp\left(i\int^r\frac{K}{\Delta}\diff\tilde{r}\right)\diff r\\
        &=-\mu\int_{\rp}^\infty\int_{\gamma}  r^2Y^{\rm in}(r)\exp\left(i\int^r\frac{K}{\Delta}\diff\tilde{r}\right)\bigl[\left(A_{nn0}+A_{\bar{m}n0}+A_{\bar{m}\bar{m}0}\right)\delta(r-r(\tau))\\
        &\qquad+\left\{\left(A_{\bar{m}n1}+A_{\bar{m}\bar{m}1}\right)\delta(r-r(\tau))\right\}_{,r}+\left\{A_{\bar{m}\bar{m}2}\delta(r-r(\tau))\right\}_{,rr}\bigr]e^{i\omega t(\tau)-i m\varphi(\tau)}\diff\tau\diff r\\
        &=-\mu\int_\gamma\bigl[\mathcal{Y}(r)\left(A_{nn0}+A_{\bar{m}n0}+A_{\bar{m}\bar{m}0}\right)-\mathcal{Y}'(r)\left(A_{\bar{m}n1}+A_{\bar{m}\bar{m}1}\right)\\
        &\qquad\qquad\qquad +\mathcal{Y}''(r)A_{\bar{m}\bar{m}2}\bigr]_{r=r(\tau),\theta=\theta(\tau)}e^{i\omega t(\tau)-i m\varphi(\tau)}\diff\tau,
    \end{aligned}
\end{equation}
\end{widetext}
where we define for convenience\footnote{Near the completion of this work, we realized that the $\mathcal{Y}(r)$ functions constructed below have a close connection to the Teukolsky functions $R(r)$. 
By comparing Eq.~\eqref{Eq.I} with, for example, Eq.~(3.31) in Ref.~\cite{Drasco:2005kz}, we see that $\mathcal{Y}$ as defined in Eq.~\eqref{eq:mathcalY} is proportional to $R$, since both formulae are computing the same physical quantity (up to a known conversion factor; we opted not to write out the expression explicitly here).
Then, at least for $s = -2$, we can obtain an \gls{ODE} that allows us to solve for $Y$ \emph{directly} without knowing $X$ first as in Eq.~\eqref{Eq.ODEforY}, by writing $R^{{\rm in/up}}(r) \propto r^2 Y^{{\rm in/up}} \exp(i \int^{r} K/\Delta \, d\tilde{r})$ and substituting this into Eq.~\eqref{Eq.TeukolskyRadialEquation}. Therefore, the \gls{SN}-\gls{IBP} approach introduced here can be made even more efficient. We will pursue this idea in future publications.}
\begin{equation}
\label{eq:mathcalY}
    \mathcal{Y}(r)=r^2Y^{\rm in}(r)\exp\left(i \int^r\frac{K}{\Delta}\diff\tilde{r}\right).
\end{equation}
Notice that we have exchanged the order of the $\diff\tau$ integral and the $\diff r$ integral in the last equality of Eq.~\eqref{Eq.I}.
This allows us to eliminate the Dirac delta function and its derivative and evaluate the integral along the particle trajectory.

In fact, by simplifying the expression enclosed in the square brackets in the last equality of Eq.~\eqref{Eq.I}, we can obtain a very elegant expression for $I$ as
\begin{widetext}
\begin{equation}\label{Eq.I_SNIBP}
    I = -\mu\int_\gamma \left[\mathcal{N}^2(\tau)W_{nn}(\tau)+\mathcal{N}(\tau)\bar{\mathcal{M}}(\tau)W_{n\bar{m}}(\tau)+\bar{\mathcal{M}}^2(\tau)W_{\bar{m}\bar{m}}(\tau)\right]e^{i\omega t(\tau)-i m\varphi(\tau)}\diff\tau,
\end{equation}
\end{widetext}
where
\begin{subequations}\label{Eq.N_and_M}
    \begin{align}
        \mathcal{N} & =u^t-a\sin^2\theta u^\varphi+\frac{\Sigma}{\Delta}u^r,\\
        \bar{\mathcal{M}} & =i a\sin\theta u^t-i\left(r^2+a^2\right)\sin\theta u^\varphi+ \Sigma u^\theta,
    \end{align}
\end{subequations}
with $u$ denoting the four velocity of the particle and $\Sigma = r^2 + a^2 \cos^2 \theta$.
Note that Eq.~\eqref{Eq.I_SNIBP} holds also for nongeodesic motions.
The $W$ terms (not to be confused with $\mathcal{W}$) represent the coupling between the auxiliary function $Y(r)$ and certain structures of the source, while $\mathcal{N}$ and $\bar{\mathcal{M}}$ contain the information about the motion of the particle along the $n_{\mu}$ and $\bar{m}_{\mu}$ direction in the \acrlong{NP} tetrad, respectively.
The expressions for the $W$ terms can be found in Appendix~\ref{Appendix_A_and_W}.

\subsubsection{$Y(r)$ function}
\label{subsubsec:Y(r)}
A core ingredient of our new \gls{SN}-\gls{IBP} approach is the auxiliary functions $Y^{\rm in, up}$, which are constructed as the solutions to the second order \gls{ODE} in Eq.~\eqref{Eq.ODEforY} subjecting to different initial conditions, respectively.
Here, we give a prescription on how to solve for these functions.

For the $Y^{\rm in}$ function, we impose the initial conditions that $Y^{\rm in}(r \to \infty) = {Y^{\rm in}}'(r \to \infty) = 0$.\footnote{\label{footnote:IC_choice}Note that the initial conditions can be chosen arbitrarily at this stage, where different choices correspond to different particular solutions to Eq.~\eqref{Eq.ODEforY}.}
Unfortunately, Eq.~\eqref{Eq.ODEforY} needs to be integrated numerically.
To speed up the computation, we expand ${Y^{\rm in}}''$ near infinity as
\begin{multline}
\label{Eq.Yin''AsymptoticExpansion}
{Y^{\rm in}}''(r\to\infty) = \\ \frac{B_{\rm SN}^{\rm ref}}{r^3}\sum_{j=0}^\infty\frac{Y^\infty_{+,j}}{r^j}+\frac{B_{\rm SN}^{\rm inc}e^{4i\omega \ln 2-2i\omega r}}{r^{3+4i\omega}}\sum_{j=0}^\infty\frac{Y^\infty_{-,j}}{r^j},
\end{multline}
where the coefficients $Y^{\infty}_{\pm, j}$ are given in Appendix \ref{App:subsec:Y_in}.
This allows us to start the numerical integration for $Y^{\rm in}$ at a smaller outer boundary since we can analytically integrate Eq.~\eqref{Eq.Yin''AsymptoticExpansion} to evaluate the proper initial values to use at the outer boundary.

\begin{figure}[hpb]
    \centering
    
    \subfloat[]{\label{subfig:Y1}\includegraphics[width=1.0\linewidth]{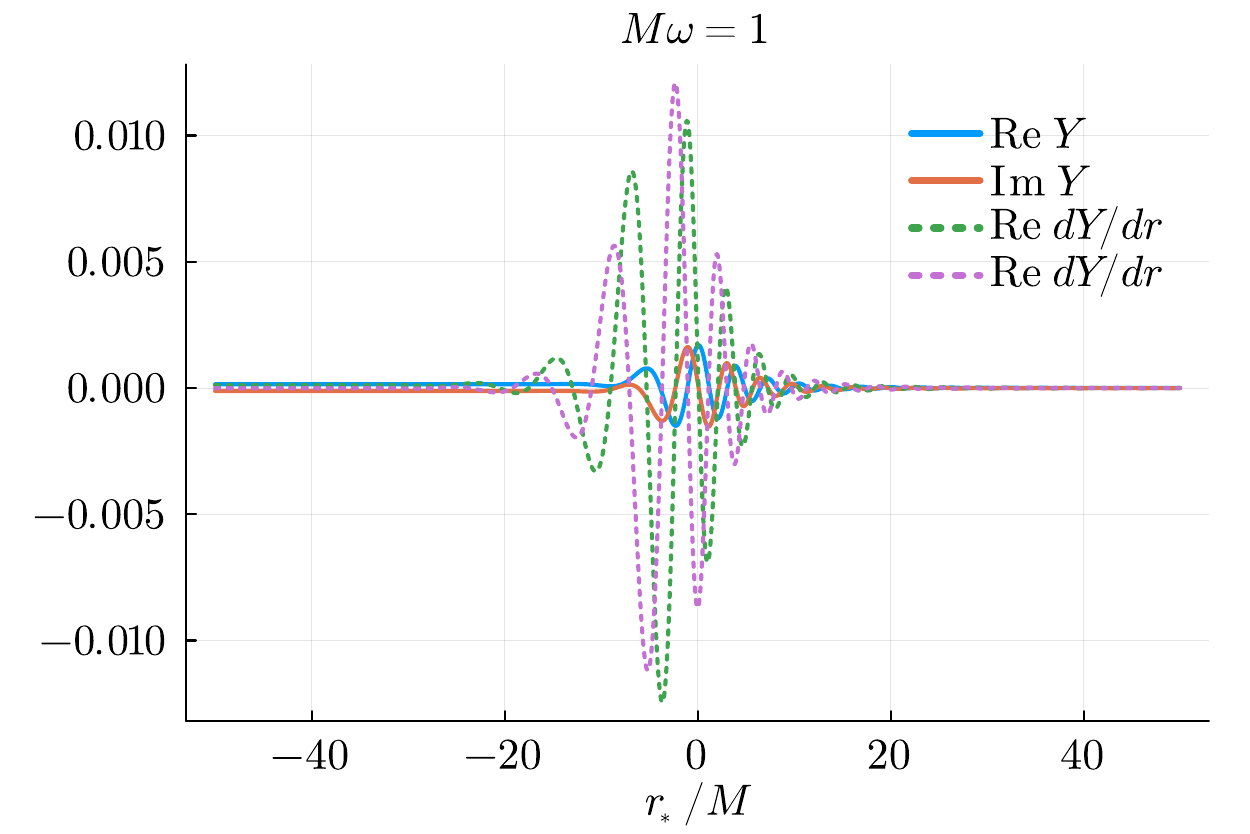}}

    \subfloat[]{\label{subfig:Y2}\includegraphics[width=1.0\linewidth]{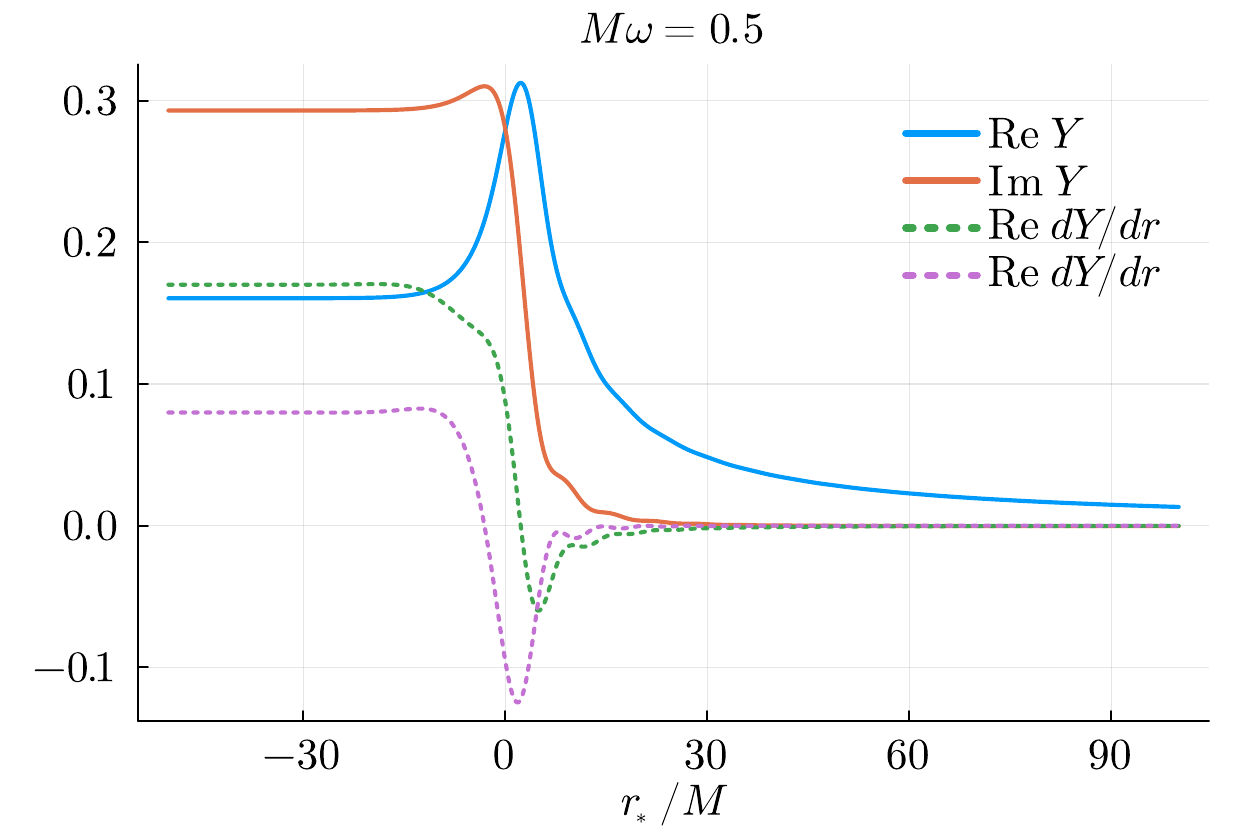}}

    \subfloat[]{\label{subfig:Y3}\includegraphics[width=1.0\linewidth]{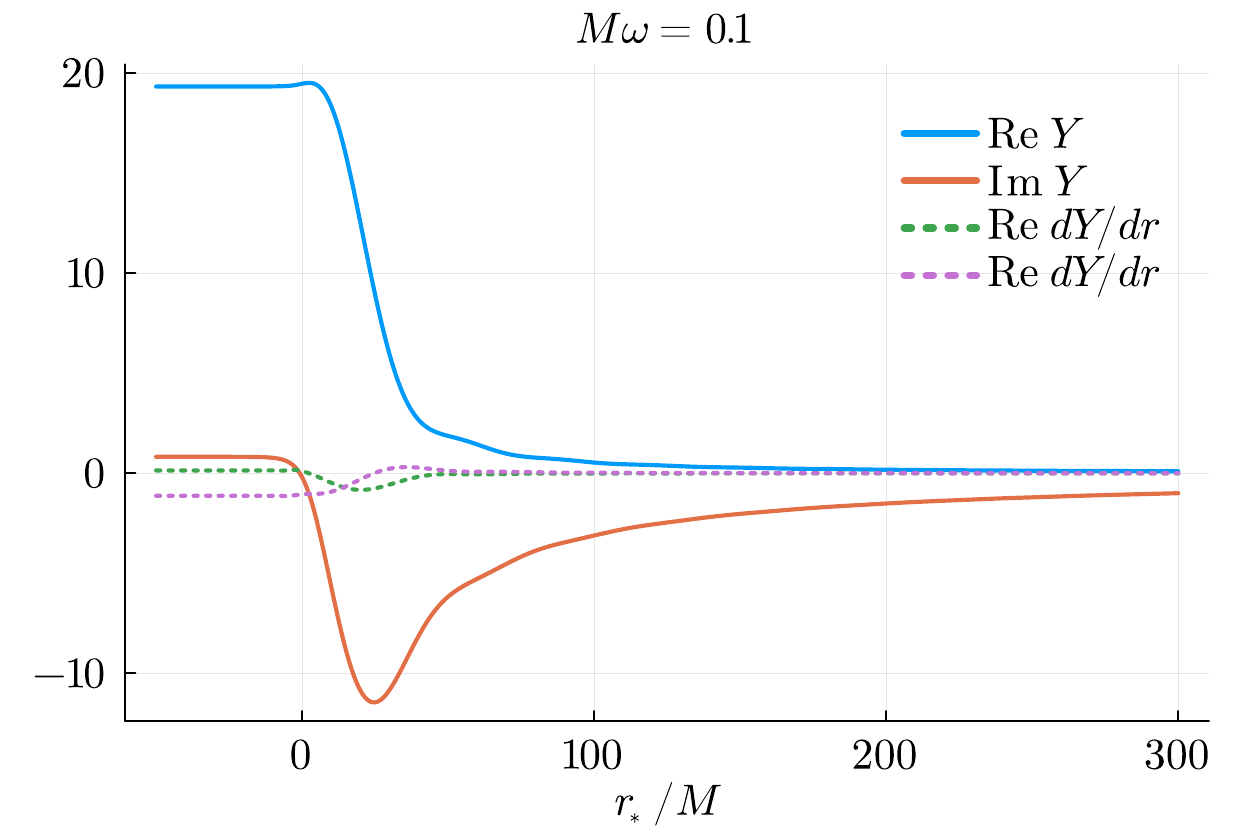}}

    \caption{The $Y^{\rm in}$ solutions for Eq.~\eqref{Eq.ODEforY} with boundary conditions $Y^{\rm in}(r\to\infty)={Y^{\rm in}}'(r\to\infty)=0$ and $\ell=m=2$, $a/M=0.9$. From the top to the bottom, the frequency is set to $M\omega=1$, $0.5$, and $0.1$, respectively.}
    \label{fig:Y}
\end{figure}

In addition, we found that it is easier to integrate the \gls{ODE} in $\rs$ instead, which is now given by
\begin{multline}
\label{Eq.Y_rs_ODE}
        \frac{\diff^2Y}{\diff \rs^2}= \frac{2(r^2-a^2)}{(r^2+a^2)^2}\frac{\diff Y}{\diff \rs}\\
        +\frac{\Delta^2X(\rs)}{r^2(r^2+a^2)^{5/2}}\exp\left(-i\int^r\frac{K}{\Delta}\diff \tilde{r}\right),
\end{multline}
where we have omitted the $\left\{\rm{in, up}\right\}$ superscript for simplicity.\footnote{Notice that the term $\exp\left(\pm i\int^r\frac{K}{\Delta}\diff \tilde{r}\right)$ can be evaluated analytically [cf Eq.~\eqref{eq:KoverDeltaIntegral}].}
As an example, Fig.~\ref{fig:Y} shows the solution for $Y^{\rm in}(\rs)$ and ${Y^{\rm in}}'(\rs)$ with $\ell=m=2$, $a/M=0.9$, and $M\omega=1$, $0.5$, and $0.1$.
We see that both $Y(\rs\to\infty)$ and $Y'(\rs\to\infty)$ converge to zero, while $Y(\rs\to-\infty)$ and $Y'(\rs\to-\infty)$ are constants.
Importantly, $Y$ and $Y'$ are nonoscillatory at both ends, unlike the Teukolsky variable $R$ or the \gls{SN} variable $X$.

Similarly, for the $Y^{\rm up}$ function, we impose the initial conditions that $Y^{\rm up}(r = r_+) = {Y^{\rm up}}'(r = r_+) = 0$.\textsuperscript{\ref{footnote:IC_choice}}
We expand ${Y^{\rm up}}''$ near the horizon as
\begin{multline}
\label{Eq.Yup''AsymptoticExpansion}
        {Y^{\rm up}}''(r\to\rp)=C_{\rm SN}^{\rm inc}\sum_{j=0}^\infty Y_{+,j}^{\rm H}\left(r-\rp\right)^j \\
        +C_{\rm SN}^{\rm ref}\left(r-\rp\right)^{i\frac{(a\rp m+2a^2\omega-4\rp\omega)}{\rp\sqrt{1-a^2}}}\sum_{j=0}^\infty Y_{+,j}^{\rm H}\left(r-\rp\right)^j,
\end{multline}
where the coefficients $ Y_{\pm,j}^{\rm H}$ are given in Appendix \ref{App:subsec:Y_up}.
Again, this allows us to start the numerical integration for $Y^{\rm up}$ at a finite inner boundary (in $\rs$) since we can analytically integrate Eq.~\eqref{Eq.Yup''AsymptoticExpansion} to obtain the proper initial values to use at the inner boundary.

\subsubsection{$\mathcal{W}(r)$ function}
\label{subsubsec:W(r)}
Another ingredient that is needed for the \gls{SN} formalism is the $\mathcal{W}$ function, which in turns gives the actual source term for the inhomogeneous \gls{SN} equation.
While we refer readers to Appendix A of Ref.~\cite{Watarai:2024huy} where the general solution of $\mathcal{W}(r)$ for a generic geodesic was presented, here we rederive these formulas using notations consistent with this paper for the sake of clarity.

The key to deriving the expression for $\mathcal{W}(r)$ lies in decoupling the terms related to $\mathcal{N}$ and $\bar{\mathcal{M}}$, respectively, and decomposing the expression using \gls{IBP}.
This improves the convergence of the integrand.
Meanwhile, the expressions outside the integral reflect the asymptotic behavior of $\mathcal{W}(r)$ at infinity.
Integrating the resultant expression inward from infinity then yields the final expression of $\mathcal{W}(r)$.

In addition, we also need to use two crucial identities. One is associated with $\mathcal{N}$ and the other with $\bar{\mathcal{M}}$.
In Ref.~\cite{Watarai:2024huy}, these derivations and identities are restricted to geodesic motions.
Here, we show that these identities and expressions remain valid in all cases including nongeodesics.

When solving for $\mathcal{W}(r)$ in Eq.~\eqref{Eq.d2W} [not to be confused with $W$ defined in Eq.~\eqref{Eq.I_SNIBP}], we generally partition it into three terms, namely,
\begin{equation}
    \mathcal{W}(r)=\mathcal{W}_{nn}(r)+\mathcal{W}_{n\bar{m}}(r)+\mathcal{W}_{\bar{m}\bar{m}}(r).
\end{equation}
In particular, we use $\mathcal{W}_{nn}$ as an example to present part of the derivation.
This is because $\mathcal{W}_{nn}$ will be used in our subsequent analysis of \glspl{GW} excited by particles falling radially along the spin axis (see Sec.~\ref{subsec:result_unbound}).
The results for $\mathcal{W}_{n\bar{m}}$ and $\mathcal{W}_{\bar{m}\bar{m}}$ will be given without detailed derivation.

It is not difficult to show from Eq.~\eqref{Eq.d2W} and Eq.~\eqref{Eq.T_definition} that $\mathcal{W}_{nn}$ satisfies the \gls{ODE}
\begin{widetext}
\begin{equation}\label{Eq.d2Wnn}
    \begin{aligned}
        \frac{\diff^2\mathcal{W}_{nn}}{\diff r^2} = &-\frac{\mathscr{A}\mu}{2}r^2\exp\left(i\int^r\frac{K}{\Delta}\diff\tilde{r}\right)\int_\gamma\diff\tau\ e^{i\omega t(\tau)-im\varphi(\tau)}\rho\bar{\rho}^2\mathcal{N}^2\mathscr{L}_1^\dagger\left[\rho^{-4}\mathscr{L}_2^\dagger\left(\rho^3S\right)\right]\delta(r-r(\tau))\\
        =&-\frac{\mathscr{A}\mu}{2}\sum_{j}\left\{\frac{1}{u^r}r^2\rho\bar{\rho}^2\mathcal{N}^2\mathscr{L}_1^\dagger\left[\rho^{-4}\mathscr{L}_2^\dagger\left(\rho^3S\right)\right]e^{i\chi(r)}\right\}_{r=r(\tau_j)},
    \end{aligned}
\end{equation}
\end{widetext}
where $\mathscr{L}^\dagger_s\equiv\partial_\theta-m/\sin\theta+a\omega\sin\theta+s\cot\theta$ is a differential operator on the angular sector and
\begin{equation}\label{Eq.chi}
    \chi(r)\equiv\omega t(r)-m\varphi(r)+\int^r\frac{K}{\Delta}\diff \tilde{r}=\omega v(r)-m\tilde{\varphi}(r),
\end{equation}
with $v=t+\rs$ and $\tilde{\varphi}=\varphi+\int^r\frac{a}{\Delta}\diff\tilde{r}$ defined as Kerr ingoing coordinates. 
Note that Eq.~\eqref{Eq.d2Wnn} is a more general version of Eq.~(A25) in Ref.~\cite{Watarai:2024huy}, where there was no $\sum_{j}^{r=r_j}$ summation in the expression.
The summation here is defined such that the particle is located at $r$ when $\tau = \tau_1, \tau_2, \cdots, \tau_j$.
For unidirectional trajectories, e.g., radial infalls and quasicircular plunges, we have $j=1$ and the summation can be omitted.
While for bound orbits, $j=\infty$, and for scattering orbits, $j=2$.
Most of the previous works have considered the case when $j=1$ only, i.e., a particle moves unidirectionally. 

For cases where $j>1$, we should divide the orbit into multi-unidirectional pieces.
Divisions are at the turning points where $u^r=0$.
At these turning points, the denominator becomes zero, making the expression singular.
However, this does not affect the subsequent integrations, as these singular points of the integrand can be transformed into a smooth form through changing the integration variable.
For details, see Ref.~\cite{10.1143/PTPS.90.1} for scattering orbits and Ref.~\cite{10.1143/ptp/90.3.595} for spherical-inclined bound orbits.

Here, we first suppose that the trajectory is unidirectional and omit the summation.
By taking the $r$ derivative of $\chi(r)$, one can show that
\begin{equation}
    \chi'(r)=\omega\frac{\mathcal{N}}{u^r}+\left(a\omega\sin^2\theta-m\right)\tilde{\varphi}'.
\end{equation}
Therefore, we can obtain an identity related to $\mathcal{N}$ that reads
\begin{equation}\label{Eq.N_identity}
    \begin{aligned}
        f(r)\frac{\mathcal{N}}{u^r}e^{i\chi(r)}=&\frac{1}{i\omega}\left\{\left[f(r)e^{i\chi(r)}\right]'\right.\\
        &\left.-\left[f'(r)+i\xi(r) f(r)\right]e^{i\chi(r)}\right\},
    \end{aligned}
\end{equation}
where
\begin{equation}
    \xi(r)=\left(a\omega\sin^2\theta-m\right)\tilde{\varphi}'(r)\sim\bigO\left(r^{-3/2}\right),
\end{equation}
and $f(r)$ is an arbitrary smooth function of $r$. 
By integrating Eq.~\eqref{Eq.d2Wnn} and using Eq.~\eqref{Eq.N_identity} twice, we obtain the expression of $\mathcal{W}_{nn}$ as a three-term form.
Similarly, one can obtain the expressions of $\mathcal{W}_{n\bar{m}}$ and $\mathcal{W}_{\bar{m}\bar{m}}$ with the help of the identity in Eq.~\eqref{eq:identity_with_Wprime_and_Jdagger}.
Schematically, they are
\begin{widetext}
\begin{subequations}\label{Eq.W_nn,nm,mm}
    \begin{align}
        &\frac{1}{\mu}\mathcal{W}_{nn}(r)=f_0(r)e^{i\chi(r)}+\int_{r}^{\infty} f_1(r_1)e^{i\chi(r_1)}\diff r_1+\int_{r}^\infty\diff r_1\int_{r_1}^\infty f_2(r_2)e^{i\chi(r_2)}\diff r_2,\label{Eq.W_nn}\\
        &\frac{1}{\mu}\mathcal{W}_{n\bar{m}}(r)=g_0(r)e^{i\chi(r)}+\int_r^\infty g_1(r_1)e^{i\chi(r_1)}\diff r_1+\int_r^\infty\diff r_1\int_{r_1}^\infty g_2(r_2)\ e^{i\chi(r_2)}\diff r_2,\label{Eq.W_nm}\\
        &\frac{1}{\mu}\mathcal{W}_{\bar{m}\bar{m}}(r)=h_0(r)e^{i\chi(r)}+\int_r^\infty h_1(r_1)e^{i\chi(r_1)}\diff r_1+\int_r^\infty\diff r_1\int_{r_1}^\infty h_2(r_2)\ e^{i\chi(r_2)}\diff r_2,\label{Eq.W_mm}
    \end{align}
\end{subequations}
\end{widetext}
where the expressions of $f_{0,1,2}$, $g_{0,1,2}$, and $h_{0,1,2}$ can be found in Appendix \ref{Appendix_W_ingredients}.

Moreover, for bound orbits, Eq.~\eqref{Eq.d2W} actually reads
\begin{equation}
    \begin{aligned}
        \mathcal{W}''=&-\frac{r^2}{\Delta^2}\mathcal{T}\exp\left(i\int^r\frac{K}{\Delta}\diff\tilde{r}\right)\\
        &\times\Theta(r-r_{\rm min})\Theta(r_{\rm max}-r),
    \end{aligned}
\end{equation}
where $\Theta(x)$ is the Heaviside step function, $r_{\rm min}$ and $r_{\rm max}$ are the inner and outer edges of the orbit.
Thus, we need to multiply all of the $f$, $g$, and $h$ functions in Eqs.~\eqref{Eq.W_nn,nm,mm} by $\Theta(r-r_{\rm min})\Theta(r_{\rm max}-r)$.
This implies that
\begin{equation}\label{Eq.W_bound_infinity}
    \mathcal{W}(r)=\mathcal{W}'(r)=0\qquad r>r_{\rm max}.
\end{equation}
In Sec.~\ref{subsubsec:bound orbits}, we will see that this result allows us to discard boundary terms when using the \gls{SN}-\gls{IBP} method.

\subsubsection{Boundary terms $Y'\mathcal{W}$ and $Y\mathcal{W}'$}
\label{subsubsec:boundary terms}
Recall from Eq.~\eqref{eq:Xinhomo_IBP} that for our \gls{SN}-\gls{IBP} approach, we need to evaluate $Y'(r)\mathcal{W}(r)$ and $Y(r)\mathcal{W}'(r)$ at both the horizon and infinity, respectively.
Here, we will discuss when it is justified to discard these boundary terms.

The general solutions of $Y(r)$ and $\mathcal{W}(r)$ can be written as
\begin{equation}
    \begin{aligned}
        Y(r) & =Y^{\rm part}(r)+y_1r+y_0,\\
        \mathcal{W}(r) & =\mathcal{W}^{\rm part}(r)+w_1r+w_0,
    \end{aligned}
\end{equation}
where $Y^{\rm part}$ and $\mathcal{W}^{\rm part}$ are the particular solutions to Eq.~\eqref{Eq.ODEforY} and Eq.~\eqref{Eq.d2W} that we gave earlier in the paper, respectively, and $y_{0,1}$, $w_{0,1}$ are some constants.
We can choose these constants to our advantages.
Moreover, we will refer to the particular solution of $Y(r)$ obtained in Sec.~\ref{subsubsec:Y(r)} [$Y^{\rm part}(r\to\infty)={Y^{\rm part}}'(r\to\infty)=0$] with $y_0=y_1=0$ as the canonical solution of Eq.~\eqref{Eq.ODEforY}, and similarly we will refer to the particular solution of $\mathcal{W}(r)$ obtained in Sec.~\ref{subsubsec:W(r)} with $w_0=w_1=0$ as the canonical solution of Eq.~\eqref{Eq.d2W}.

In general, the canonical solution $\mathcal{W}^{\rm canonical}(r)$ has the asymptotic behaviors
\begin{equation}
    \mathcal{W}^{\rm canonical}(r)\sim\begin{cases}
        \bigO(1), & r\to\rp\\
        \bigO(r^{1/2}), & r\to\infty
    \end{cases},
\end{equation}
and
\begin{equation}
    {\mathcal{W}^{\rm canonical}}'(r)\sim\begin{cases}
        \bigO(1),&r\to\rp\\
        \bigO(1),&r\to\infty
    \end{cases}.
\end{equation}
If one chooses the canonical solution as the particular solution and set
\begin{subequations}\label{Eq.w0w1}
    \begin{align}
        &w_0=-\mathcal{W}^{\rm part}(\rp)+\rp{\mathcal{W}^{\rm part}}'(\rp),\\
        &w_1=-{\mathcal{W}^{\rm part}}'(\rp),
    \end{align}
\end{subequations}
then the boundary terms of $\mathcal{W}(r)$ at the horizon vanish \cite{10.1093/ptep/ptaa149}.
However, this assumes that the limits $\mathcal{W}^{\rm part}(r\to\rp)$ and ${\mathcal{W}^{\rm part}}'(r\to\rp)$ exist. From the analysis in Sec.~\ref{subsubsec:W(r)}, this requirement translates into the condition that the limit $\chi(r\to\rp)$ exists. 

When a particle is close enough to the horizon of a \gls{BH}, all external forces will be negligible compared to the influence of spacetime curvature itself.
This means that the particle moves along a geodesic when approaching the horizon. Therefore, we can use the geodesic equations and obtain
\begin{equation}
    \frac{\diff \chi}{\diff\rs}\sim\bigO(\Delta), \qquad \rs\to-\infty.
\end{equation}
As a result, we have
\begin{equation}
    \chi(\rs\to-\infty)={\rm const}.
\end{equation}
With this choice of $w_{0, 1}$, we have the following asymptotic behaviors for $\mathcal{W}$ as
\begin{equation}
    \mathcal{W}(r)\sim\begin{cases}
        \bigO(\Delta^2),&r\to\rp\\
        \bigO(r),&r\to\infty
    \end{cases},
\end{equation}
and
\begin{equation}
    \mathcal{W}'(r)\sim\begin{cases}
        \bigO(\Delta),&r\to\rp\\
        \bigO(1),&r\to\infty
    \end{cases},
\end{equation}
at the expense that $\mathcal{W}(r\to\infty)$ becomes less convergent.\footnote{Note that $\mathcal{S}$ is still convergent when $r \to \infty$.} 

Similarly, the canonical solution $Y^{\rm canonical}(r)$ has the asymptotic behaviors (which can also be seen in Fig.~\ref{fig:Y}) as 
\begin{equation}
Y^{\rm canonical}(r) \sim\begin{cases}
        \bigO(1), & r\to\rp\\
        \bigO(1/r), & r\to\infty
    \end{cases},
\end{equation}
and
\begin{equation}
{Y^{\rm canonical}}'(r) \sim\begin{cases}
        \bigO(1),&r\to\rp\\
        \bigO(1/r^2),&r\to\infty
    \end{cases}.
\end{equation}
By setting
\begin{subequations}\label{Eq.y0y1}
    \begin{align}
        y_0 & =-Y^{\rm part}(\rp)+\rp{Y^{\rm part}}'(\rp),\\
        y_1 & =-{Y^{\rm part}}'(\rp),
    \end{align}
\end{subequations}
we can obtain a solution with the asymptotic behaviors where
\begin{equation}
    Y(r) \sim 
    \begin{cases}
        \bigO(\Delta^2),&r\to\rp\\
        \bigO(r),&r\to\infty
    \end{cases},
\end{equation}
and
\begin{equation}
    Y'(r) \sim
    \begin{cases}
        \bigO(\Delta),&r\to\rp\\
        \bigO(1),&r\to\infty
    \end{cases}.
\end{equation}

\begin{figure*}[ht!]
\begin{center}
\includegraphics[width=2\columnwidth]{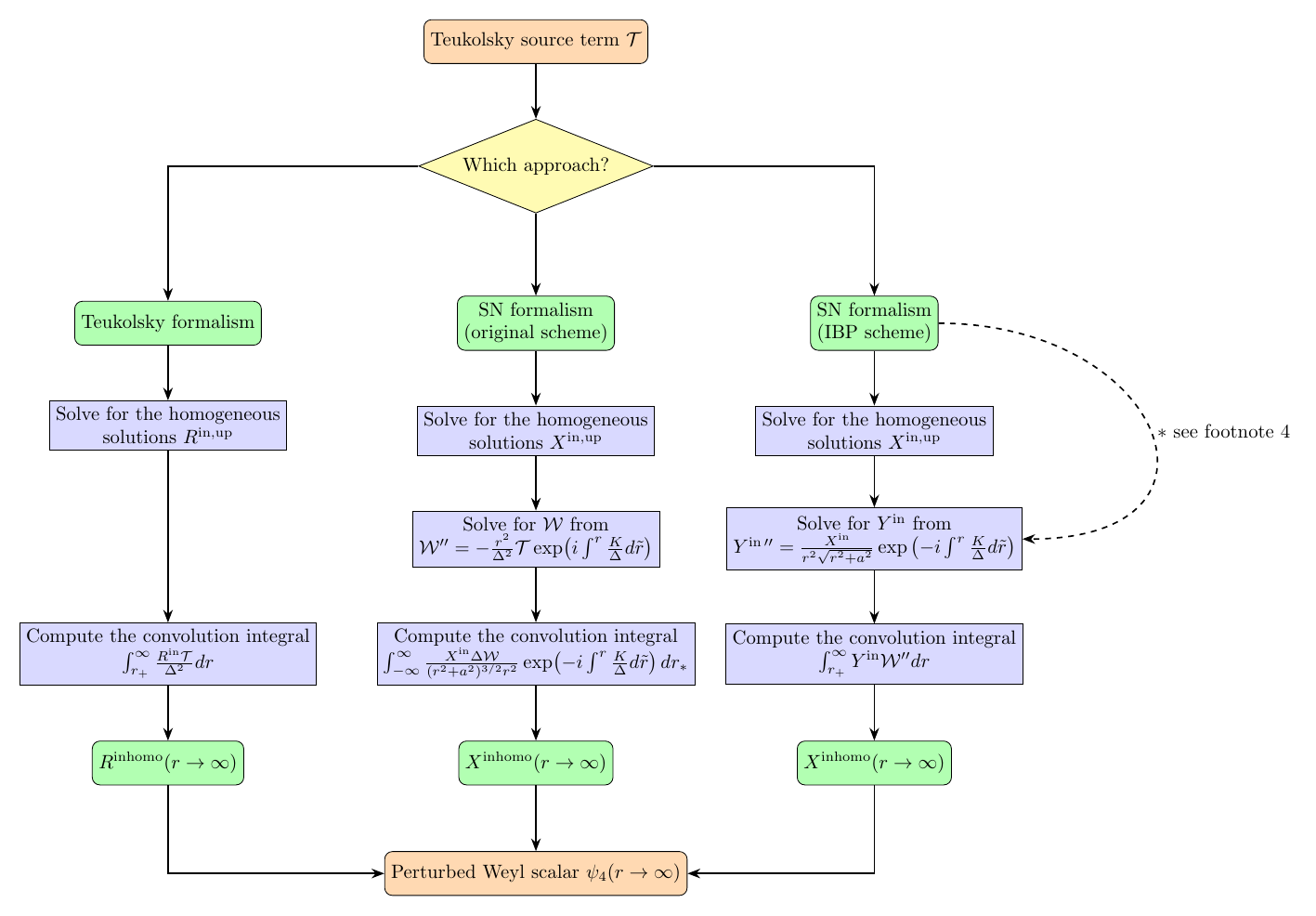}
\end{center}
\caption{\label{fig:flowchart}A flowchart summarizing the three different approaches described in this paper for computing the asymptotic value of the perturbed Weyl scalar $\psi_4(r \to \infty)$ given the Teukolsky source term $\mathcal{T}$, namely, the Teukolsky formalism, the \gls{SN} formalism using the original scheme, and the \gls{SN} formalism using the \gls{IBP} scheme (this work).}
\end{figure*}

We summarize the three different approaches for computing $\psi_4(r \to \infty)$, and in turn gravitational waveforms and fluxes at infinity, that are described in this paper using a flowchart in Fig.~\ref{fig:flowchart}. The new approach introduced in this subsection is shown in the rightmost column. In the next subsection, we give the recipes to calculate gravitational waveforms and fluxes at infinity using the \gls{SN} formalism.

\subsection{Recipes for calculating waveforms and fluxes at infinity}
\label{subsec:recipes}
\subsubsection{Bound orbits}
\label{subsubsec:bound orbits}

We have seen from Eq.~\eqref{Eq.W_bound_infinity} that the canonical solution of $\mathcal{W}(r)$ for bound orbits, i.e., $w_0=w_1=0$,  vanishes at infinity.
Therefore, we only need to
\begin{enumerate}
    \item Solve for $Y(r)$ following the scheme introduced in Sec.~\ref{subsubsec:Y(r)}.
    \item Extract the boundary values of $Y(r)$ and $Y'(r)$ at the horizon and calculate $y_0$ and $y_1$ following Eqs.~\eqref{Eq.y0y1}.
\end{enumerate}
With these choices, all four of the boundary terms in Eq.~\eqref{Eq.Xinf after IBP} vanish.
We can then calculate the inhomogeneous solution using Eq.~\eqref{Eq.I_SNIBP}, and therefore the gravitational waveform and fluxes at infinity.

Here, we briefly introduce a procedure for calculating the amplitude for each harmonic of an \gls{EMRI} waveform on a generic bound geodesic.
The derivation is analogous to the one in Ref.~\cite{Drasco:2005kz}, but under the \gls{SN} formalism. 
Then, in Sec.~\ref{subsec:result_bound}, we show some examples of \glspl{EMRI} waveforms on generic (eccentric-inclined) orbits using our \gls{SN}-\gls{IBP} scheme.

A generic bound geodesic orbit in the \gls{BL} coordinates can be decoupled into harmonics of $r$ and $\theta$.
This is because the Kerr metric components have no dependence on $t$ and $\varphi$.
The general solutions to the timelike bound geodesic equation can be expressed as
\begin{subequations}\label{Eq.KerrGeoOrbit}
    \begin{align}
        &t(\lambda)=\Gamma\lambda+\Delta t[r(\lambda),\theta(\lambda)],\\
        &r(\lambda)=\sum_{n=-\infty}^\infty r_ne^{-in\Upsilon_r\lambda},\\
        &\theta(\lambda)=\sum_{k=-\infty}^\infty\theta_ke^{-ik\Upsilon_\theta\lambda},\\
        &\varphi(\lambda)=\Upsilon_\varphi\lambda+\Delta \varphi[r(\lambda),\theta(\lambda)],
    \end{align}
\end{subequations}
where $\Gamma$, $\Upsilon_r$, $\Upsilon_\theta$, $\Upsilon_\varphi$ are frequencies parametrized by the Mino time $\lambda$ which is defined by $\diff\tau=\Sigma\diff\lambda$.
To help with our calculations, we also introduce an open source \texttt{julia} package \texttt{KerrGeodesics.jl} for solving timelike Kerr geodesics, see Appendix \ref{Appendix_geodesic_motions} for details.

Therefore, we can write the Green's function integral Eq.~\eqref{Eq.I_SNIBP} as
\begin{equation}
    I=-\mu\int_{\gamma} J_{\ell m\omega}\left[r(\lambda),\theta(\lambda)\right]e^{i(\omega\Gamma-m\Upsilon_\varphi)\lambda}\diff\lambda.
\end{equation}
The integrand kernel is defined by
\begin{equation}
    \begin{aligned}
        J_{\ell m\omega} & = \frac{\diff\tau}{\diff\lambda}\left(W_{nn}\mathcal{N}^2+W_{n\bar{m}}\mathcal{N}\mathcal{M}+W_{\bar{m}\bar{m}}\mathcal{M}^2\right)\\
        & = \sum_{k=-\infty}^\infty\sum_{n=-\infty}^\infty J_{\ell mkn}(\omega)e^{-i(k\Upsilon_\theta+n\Upsilon_r)\lambda},
    \end{aligned}
\end{equation}
where
\begin{equation}\label{Eq.J_double_integral}
    J_{\ell mkn}=\int_0^{2\pi}\int_0^{2\pi}e^{i(k\phi_\theta+n\phi_r)}J_{\ell m\omega}(\phi_r,\phi_\theta)\frac{\diff\phi_\theta\diff\phi_r}{(2\pi)^2},
\end{equation}
with $\phi_r=\Upsilon_r\lambda$, $\phi_\theta=\Upsilon_\theta\lambda$ defined as the decoupled phases.
Finally, we can rewrite the integral, with $\gamma=(-\infty,\infty)$, as
\begin{equation}
    \begin{aligned}
        I =&\int_{-\infty}^\infty e^{i(\omega\Gamma-m\Upsilon_\varphi-k\Upsilon_\theta-n\Upsilon_r)\lambda}\sum_{k=-\infty}^\infty\sum_{n=-\infty}^\infty J_{\ell mkn}(\omega)\diff\lambda\\
        =&\sum_{k=-\infty}^\infty\sum_{n=-\infty}^\infty2\pi\delta(\omega\Gamma-m\Upsilon_\varphi-k\Upsilon_\theta-n\Upsilon_r) J_{\ell mkn}(\omega).
    \end{aligned}
\end{equation}
Then, we insert it into Eq.~\eqref{Eq.h} and Eq.~\eqref{Eq.Z} and obtain the gravitational waveform at infinity as
\begin{equation}
    \begin{aligned}
        h&=h_+-ih_\times\\
        &=\frac{8}{r}\sum_{\ell m}\int_{-\infty}^\infty\frac{I}{2i\omega B^{\rm inc}_{\rm SN}}{_{-2}}S^{a\omega}_{\ell m}(\theta)e^{-i\omega(t-\rs)+im\varphi}\diff\omega\\
        &=\sum_{\ell mnk}h_{\ell mnk},
    \end{aligned}
\end{equation}
where
\begin{equation}
    \omega_{mnk}=m\frac{\Upsilon_{\varphi}}{\Gamma}+n\frac{\Upsilon_r}{\Gamma}+k\frac{\Upsilon_\theta}{\Gamma}
\end{equation}
and 
\begin{equation}\label{Eq.hlmnk_bound}
    h_{\ell mnk}=-\frac{2\mu}{r}\frac{Z_{\ell mnk}^\infty}{\omega_{mnk}^2}{_{-2}}S^{a\omega_{mnk}}_{\ell m}(\theta)e^{-i\omega_{mnk}(t-\rs)+im\varphi},
\end{equation}
where
\begin{equation}\label{Eq.Zlmnk_bound}
    Z_{\ell mnk}^{\infty}=-\frac{4i\pi\omega_{mnk}}{B_{\rm SN}^{\rm inc}\Gamma}J_{\ell mnk}.
\end{equation}
The averaged energy flux, angular momentum flux, and Carter constant flux at infinity are given by
\begin{subequations}
    \begin{align}
        \left\langle \dot{\mathcal{E}} \right\rangle^\infty=&\sum_{\ell mnk}\frac{\left|Z_{\ell mnk}^{\infty}\right|^2}{4\pi\omega_{mnk}^2},\label{Eq.energy_flux}\\
        \left\langle \dot{\mathcal{L}_{z}} \right\rangle^{\infty} = &\sum_{\ell mnk}\frac{m\left|Z_{\ell mnk}^{\infty}\right|^2}{4\pi\omega_{mnk}^3},\label{eq:Lz_flux}\\
        \left\langle \dot{\mathcal{Q}} \right\rangle^{\infty} = &\sum_{\ell mnk}\frac{\left(\mathcal{L}_{mnk}+k\Upsilon_\theta\right)\left|Z_{\ell mnk}^{\infty}\right|^2}{2\pi\omega_{mnk}^3},\label{eq:Q_flux}
    \end{align}
\end{subequations}
where
\begin{subequations}
    \begin{align}
        \mathcal{L}_{mnk} & = m\langle\cot^2\theta\rangle\mathcal{L}_z-a^2\omega_{mnk}\langle\cos^2\theta\rangle\mathcal{E},\\
        \langle\cot^2\theta\rangle & =\frac{1}{\pi}\int_0^{\pi}\left[\cot\theta(\phi_\theta)\right]^2\diff\phi_\theta,\\
        \langle\cos^2\theta\rangle & =\frac{1}{\pi}\int_0^{\pi}\left[\cos\theta(\phi_\theta)\right]^2\diff\phi_\theta.
    \end{align}
\end{subequations}

\subsubsection{Unbound orbits}
\label{subsubsec:unbound orbits}
Unlike bound orbits, $\mathcal{W}(r \to \infty)$ and $\mathcal{W}'(r \to \infty)$ may not vanish for unbound orbits.
An example of this would be the radial infall of a particle from infinity.
In this case, the natural thing to do with our \gls{SN}-\gls{IBP} approach would be choosing $y_0 = y_1 = 0$ such that the boundary terms at infinity in Eq.~\eqref{Eq.Xinf after IBP} vanish, but one still needs to solve Eq.~\eqref{Eq.d2W} to evaluate $\mathcal{W}(r = r_+)$ and $\mathcal{W}(r = r_+)$.
While for the original formulation (without using \gls{IBP}), we also need to solve for $\mathcal{W}(r)$ and integrate Eq.~\eqref{Eq.X^Infty}. 
Therefore, one needs to solve for $\mathcal{W}(r)$ either way, and our \gls{IBP} approach does not have any advantage over the original formulation.
A question naturally arises as to which method should be used for unbound orbits?
We can answer this question by analyzing the convergence of the integrands in both formulations.

As an example, we derive how one calculates the waveform induced by a particle falling radially into a Kerr \gls{BH} along its spin axis without using \gls{IBP} in the original \gls{SN} formulation.
The 4-velocity is given by
\begin{subequations}
    \begin{align}
        u^t&=\mathcal{E}\frac{r^2+a^2}{\Delta},\\
        u^r&=-\frac{\sqrt{\mathcal{E}^2(r^2+a^2)^2-\Delta(r^2+a^2\mathcal{E}^2)}}{r^2+a^2},\\
        u^\theta&=u^\varphi=0,
    \end{align}
\end{subequations}
where $\mathcal{E}$ is the orbital energy per mass.
One instantly find that $\bar{\mathcal{M}}=0$ from the definition in Eq.~\eqref{Eq.N_and_M} and therefore only $\mathcal{W}_{nn}$ is nonvanishing.

Specifically, we see that when $\mathcal{E}=1$, we have $u^r(r\to\infty)\sim\bigO(r^{-1/2})$ and therefore $\mathcal{W}(r\to\infty)\sim f_0\sim\bigO(1/r^{1/2})$.
When $\mathcal{E}>1$, we have $u^r(r\to\infty)\sim\bigO(1)$ as $r\to\infty$ and therefore $\mathcal{W}(r\to\infty)\sim f_0\sim\bigO(1)$.\footnote{For $\mathcal{E}<1$, the particle cannot escape to infinity, which reduces to the bound case.}
By setting $y_0=y_1=0$, we have $Y(r\to\infty)\sim\bigO(1/r)$. From the definition in Eq.~\eqref{Eq.ODEforY} and Eq.~\eqref{Eq.d2W}, we already know that $Y''(r\to\infty)\sim\bigO(1/r^3)$, $\mathcal{W}''(r\to\infty)\sim\bigO(r^{1/2})$ for $\mathcal{E}=1$, and $\mathcal{W}''(r\to\infty)\sim\bigO(1)$ for $\mathcal{E}>1$.
As a result, we obtain the convergence of the integrands in the \gls{SN}-\gls{IBP} and the original \gls{SN} method, respectively, as
\begin{equation}\label{Eq.IBP/non-IBP_convergence}
    \begin{aligned}
        &\text{\gls{SN}-\gls{IBP}:}\quad Y(r)\mathcal{W}''(r)\sim\begin{cases}
        \bigO(1/r^{1/2}),&\mathcal{E}=1\\
        \bigO(1/r),&\mathcal{E}>1
    \end{cases},\\
        &\text{original \gls{SN}:}\quad Y''(r)\mathcal{W}(r)\sim\begin{cases}
        \bigO(1/r^{7/2}),&\mathcal{E}=1\\
        \bigO(1/r^3),&\mathcal{E}>1
    \end{cases}.
    \end{aligned}
\end{equation}
From Eq.~\eqref{Eq.IBP/non-IBP_convergence}, we conclude that the integrand of the non-\gls{IBP} method converge way faster than that of the \gls{IBP} method and suggest using the original \gls{SN} formulation for unbound orbits. We will show the convergence speed clearly in Sec.~\ref{subsec:result_unbound}. 

For an unbound orbit, there is no discrete frequency spectrum as in the case for a bound orbit.
The frequency domain waveform $\tilde{h}(\omega)$ and the now-continuous energy spectrum $\diff \mathcal{E}/d\omega$, in the case of radial infall, can be expressed as
\begin{subequations}\label{Eq.amp_spectra_unbound}
    \begin{align}
        \tilde{h}_{\ell}(\omega)&=-\frac{2\mu}{r}\frac{Z^\infty_{\ell0\omega}}{\omega^2}{_{-2}}S^{a\omega}_{\ell 0}(\theta)=\frac{8\mu}{r}\frac{X_{\ell 0\omega}^\infty}{c_0}{_{-2}}S^{a\omega}_{\ell 0}(\theta),\\
        \left(\frac{\diff \mathcal{E}}{\diff \omega}\right)_\ell^\infty &= \frac{\mu^2}{2\omega^2}\left(\left|Z_{\ell 0\omega}^\infty\right|^2+\left|Z_{\ell 0-\omega}^\infty\right|^2\right)\notag\\
        & =8\omega^2\mu^2\left(\left|\frac{X_{\ell 0\omega}^\infty}{c_0}\right|^2+\left|\frac{X_{\ell 0-\omega}^\infty}{c_0}\right|^2\right).
    \end{align}
\end{subequations}
The corresponding time-domain waveform is given by
\begin{equation}\label{Eq.Waveform_time_domain_unbound}
    h_+-ih_\times=\sum_{\ell}\int_{-\infty}^\infty\tilde{h}_\ell(\omega)e^{-i\omega u}\diff \omega,
\end{equation}
where $u=t-\rs$ is the retarded time.

\section{Results}

In this section, we present some example waveforms and energy flux calculations for both bound and unbound orbits.
Specifically, for bound orbits, we use the \gls{SN}-\gls{IBP} approach introduced in this paper to compute the \gls{EMRI} waveform snapshot (\textit{à la} Ref.~\cite{Drasco:2005kz}) for a generic timelike geodesic.
For unbound orbits, we consider particles falling radially from infinity along the spin axis and covering two cases---the rest limit ($\mathcal{E}=1$) and the ultrarelativistic limit ($\mathcal{E}\to\infty$).

\subsection{Generic bound stable orbits}
\label{subsec:result_bound}
Numerous studies have already calculated the gravitational radiation from particles on bound Kerr geodesic orbits using the Teukolsky formalism, including eccentric-equatorial orbits \cite{PhysRevD.66.044002}, inclined-spherical orbits \cite{PhysRevD.61.084004}, and generic orbits \cite{Drasco:2005kz}.
Prior to this work, there were also calculations using the \gls{SN} formalism on circular-equatorial orbits \cite{PhysRevD.48.663}, eccentric-equatorial orbits \cite{PhysRevD.50.6297}, and inclined-spherical orbits \cite{10.1143/ptp/90.3.595}.
No calculation for generic orbits has been done with the \gls{SN} formalism.
We present our results and compare them with the literature and codes using the Teukolsky formalism, namely the \texttt{Teukolsky} package from \texttt{BHPToolkit} \cite{BHPToolkit} and \texttt{pybhpt} \cite{PhysRevD.106.064042, PhysRevD.109.044020}.

Here, we show the results of Eq.~\eqref{Eq.hlmnk_bound}--\eqref{Eq.energy_flux}.
We set $a=0.9M$, $p=6M$, $e=0.7$, $x=\cos\pi/4$ as our fiducial parameters for a generic geodesic orbit.\footnote{The trajectory is also visualized in Fig.~\ref{fig:generic_trjectory} in Appendix \ref{Appendix_geodesic_motions}.}
For higher values of $n$ and $k$, Eq.~\eqref{Eq.J_double_integral} becomes a highly oscillatory double integral, which is hard to integrate numerically.
To achieve a better precision and speed, we employ Levin's method, which converts a quadrature problem into an \gls{ODE} problem.
The algorithm is introduced in Appendix \ref{Appendix_Levin}. 

\begin{figure}[htpb!]
    \centering
    \includegraphics[width=1.0\linewidth]{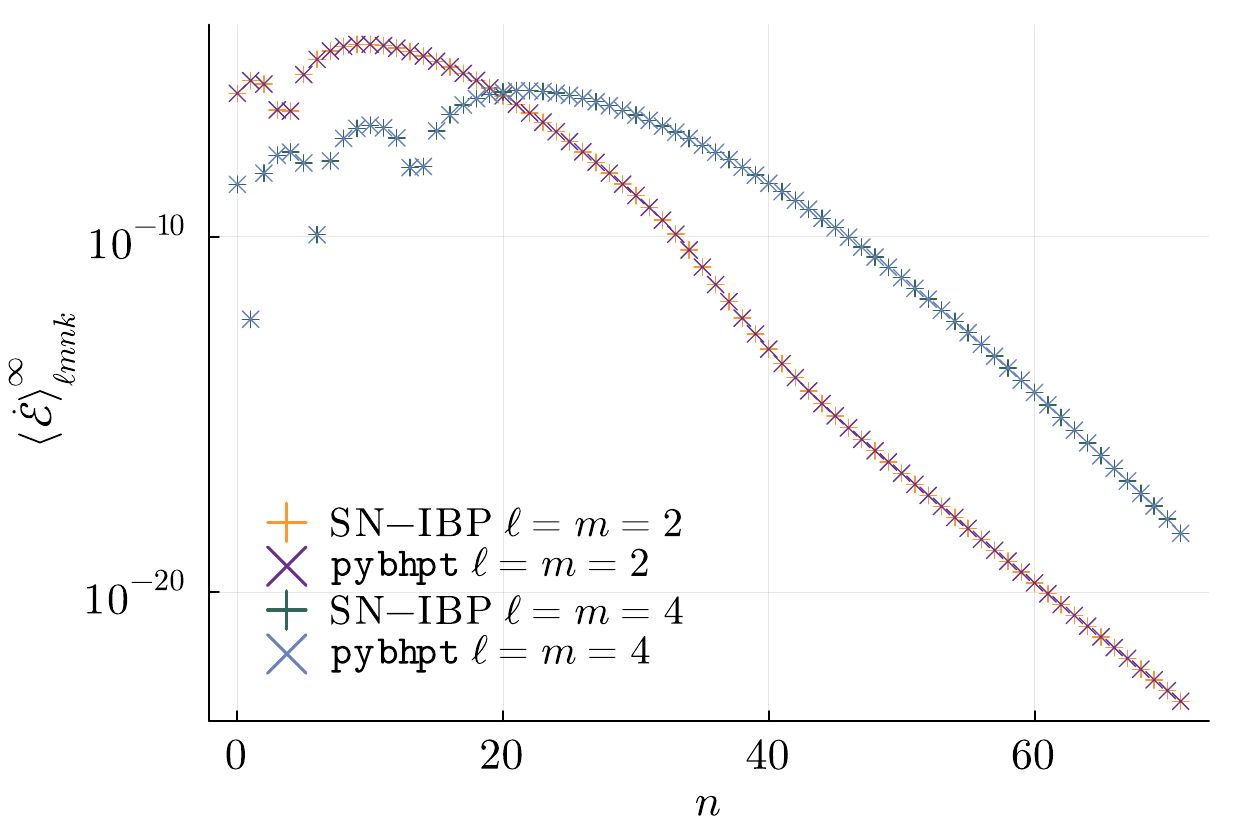}
    \caption{The energy flux at infinity for $a=0.9M$, $p=6M$, $e=0.7$, $x=\cos\pi/4$. The mode indexes are $\ell=m=2$ and $\ell=m=4$ with polar index $k=0$ and radial index $n=0$ to $n=70$. The two approaches agree very well.}
    \label{fig:bound_flux}
\end{figure}

To verify our codes, we calculate the energy flux using the \gls{SN}-\gls{IBP} method in this work (implemented in \texttt{GeneralizedSasakiNakamura.jl}\footnote{\url{https://github.com/ricokaloklo/GeneralizedSasakiNakamura.jl} from v0.7.0 onwards.}) and \texttt{pybhpt} for the $\ell=m=2$ and $\ell=m=4$ modes with $k=0$ and $n=0$ to $n=70$ in Fig.~\ref{fig:bound_flux}.
In addition, we tabulate the total energy flux for each $\ell$ mode from the two codes, which is defined as
\begin{equation}\label{Eq.flux_l_mode}
    \left\langle\dot{\mathcal{E}}\right\rangle^\infty_\ell=\sum_{mnk}\left\langle\dot{\mathcal{E}}\right\rangle^\infty_{\ell mnk}.
\end{equation}
The truncation rules\footnote{Note that we do not claim this set of truncation rules to be optimal.} for the summation in Eq.~\eqref{Eq.flux_l_mode} are specified as follows:
\begin{enumerate}
    \item For each $\ell$ mode, we manually set the truncation limits as $n_{\rm max}^\ell=80+20\ell$ and $k_{\rm max}^\ell=8+2\ell$.
    \item For fixed $\ell$, $m$, and $n=0$, if three consecutive values of $\langle\dot{\mathcal{E}}\rangle^\infty_{\ell mnk}$ are smaller than $10^{-6}\times\langle\dot{\mathcal{E}}\rangle^\infty_{\ell}$ (i.e., the current value of the summation of that $\ell$ mode), then we truncate the $k$ summation.
    \item For fixed $\ell$, $m$, and $k$, if three consecutive values of $\langle\dot{\mathcal{E}}\rangle^\infty_{\ell mnk}$ are smaller than current $10^{-6}\times\langle\dot{\mathcal{E}}\rangle^\infty_{\ell}$ (i.e., current value of the summation of this $\ell$ mode), then we truncate the $n$ summation.
\end{enumerate}

\begin{figure}[htpb!]
    \centering
    \includegraphics[width=1.0\linewidth]{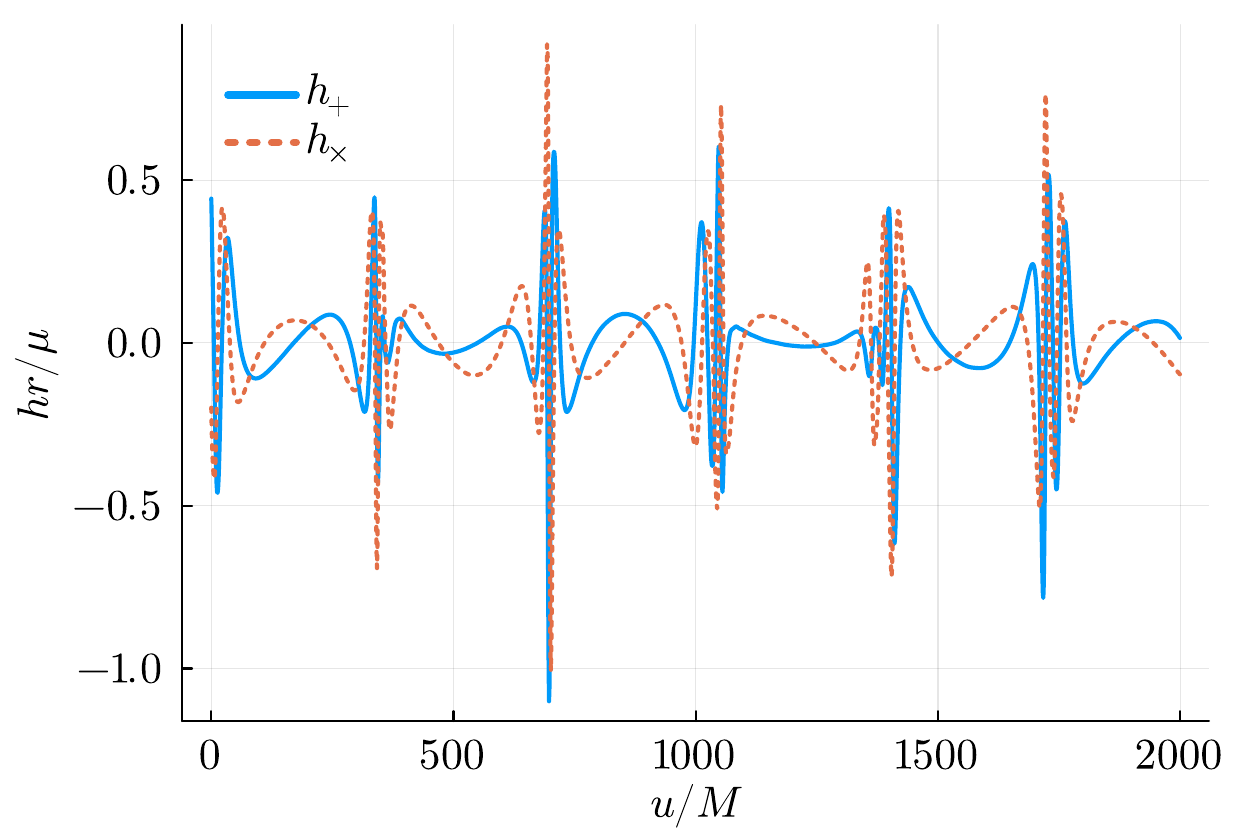}
    \caption{The \gls{GW} waveform snapshot for $a=0.9M$, $p=6M$, $e=0.7$, $x=\cos\pi/4$ viewing at $\theta=\pi/2$ and $\varphi=0$.}
    \label{fig:wavform_bound}
\end{figure}

Following these rules, we calculate the energy fluxes for the $\ell=2$, $3$, $4$, $5$, and $6$ modes. These values are tabulated Table~\ref{tab:flux_values}, together with the total number of modes summed in those calculations.\footnote{Note that in these calculations, both code use the same truncation strategy presented above.}
The two sets of numbers agree to the twelve digit, and disagreement only appears after the thirteenth digit (indicated by the brackets in Table~\ref{tab:flux_values}).
Moreover, in Fig.~\ref{fig:wavform_bound}, we show the waveform snapshot with the fiducial parameters, using the amplitude data from Table~\ref{tab:flux_values}. In total, $5874$ modes were used for the generation of the waveform.

\begin{table}[h]
    \caption{The energy fluxes of different $\ell$ modes for $a=0.9M$, $p=6M$, $e=0.7$, $x=\cos\pi/4$. The last column is the total number of modes in the summation.}
    \label{tab:flux_values}
    \centering
    \begin{ruledtabular}
    \begin{tabular}{cccc}
    $\langle\dot{\mathcal{E}}\rangle^\infty_\ell$ & \gls{SN}-\gls{IBP}($\times 10^{-4}$) & \texttt{pybhpt}($\times 10^{-4}$) & modes\\ 
    \hline
    $\ell=2$ & $6.2645935(4855)$ & $6.2645935(8421)$ & $860$\\
    $\ell=3$ & $1.7855172(0137)$ & $1.7855172(0344)$ & $1053$\\
    $\ell=4$ & $0.6318417(2096)$ & $0.6318417(2469)$ & $1237$\\
    $\ell=5$ & $0.2441166(3865)$ & $0.2441166(4101)$ & $1324$\\
    $\ell=6$ & $0.0966104(0775)$ & $0.0966104(0763)$ & $1400$\\
    \end{tabular}
    \end{ruledtabular}
\end{table}

\subsection{Radial infalls}
\label{subsec:result_unbound}

As discussed in Sec.~\ref{subsubsec:unbound orbits}, there are two cases---$\mathcal{E}=1$, where the particle has no initial velocity at infinity (also referred to as the rest limit) and $\mathcal{E}>1$.
In addition, $\mathcal{E} \gg 1$ or the ultrarelativistic limit corresponds to a particle moving nearly at the speed of light and hitting a Kerr \gls{BH} along its spin axis.

\begin{figure}[htpb]
    \centering
    
    \subfloat[]{\label{subfig:f_rest}\includegraphics[width=1.0\linewidth]{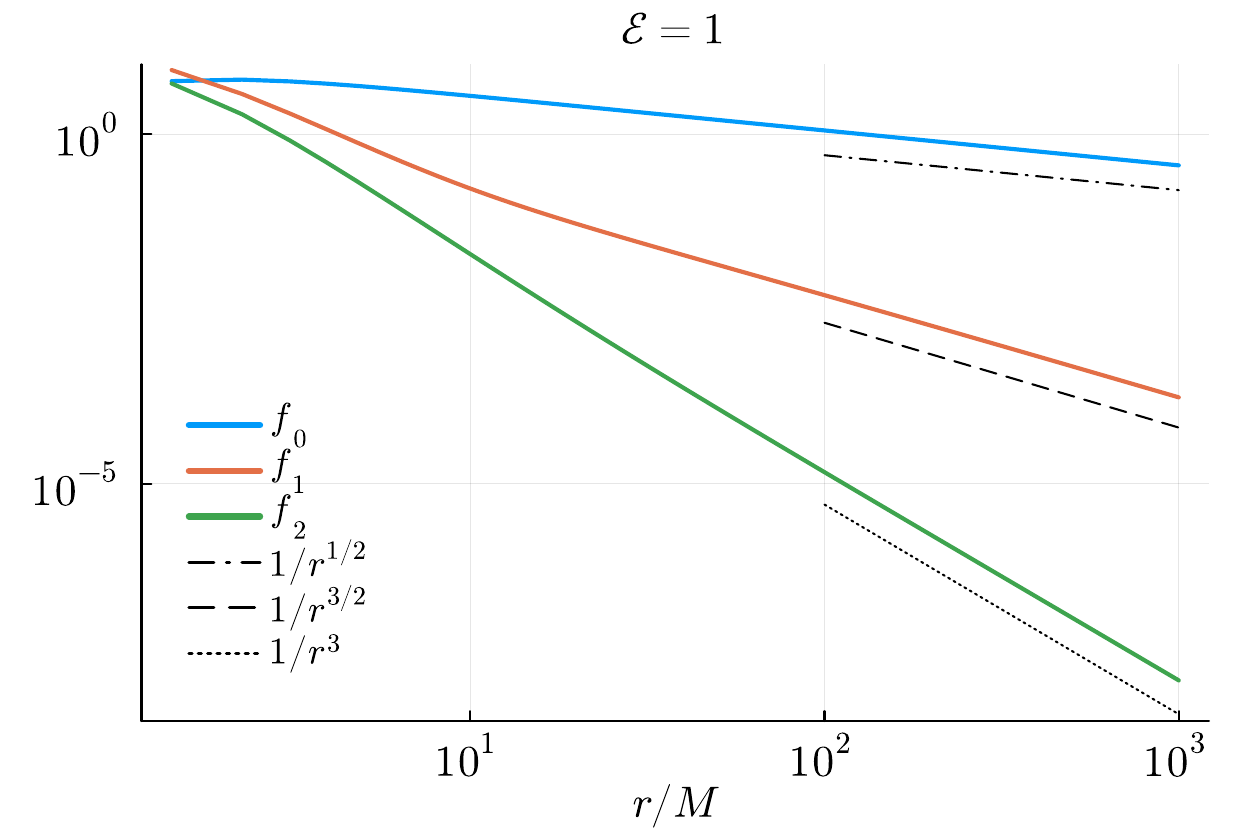}}

    \subfloat[]{\label{subfig:W_1}\includegraphics[width=1.0\linewidth]{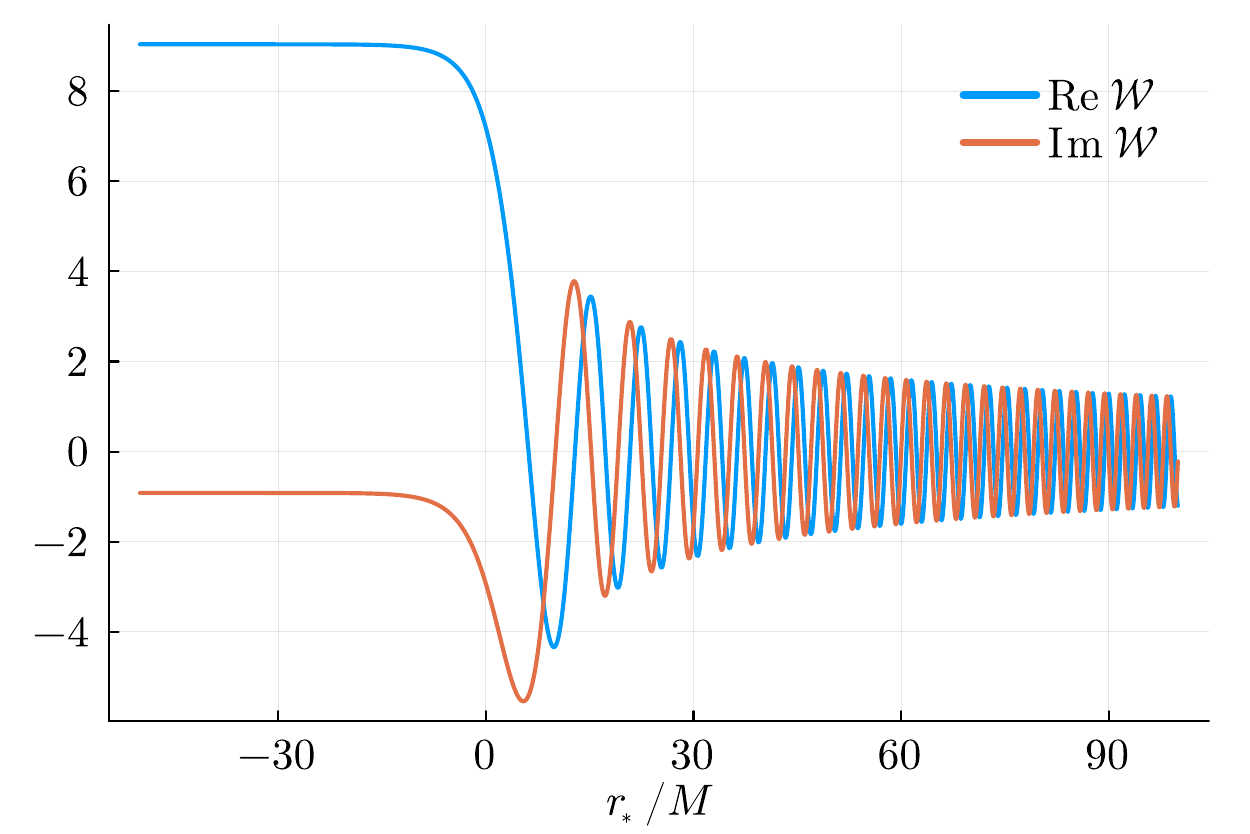}}

    \subfloat[]{\label{subfig:unbound_comparison_rest}\includegraphics[width=1.0\linewidth]{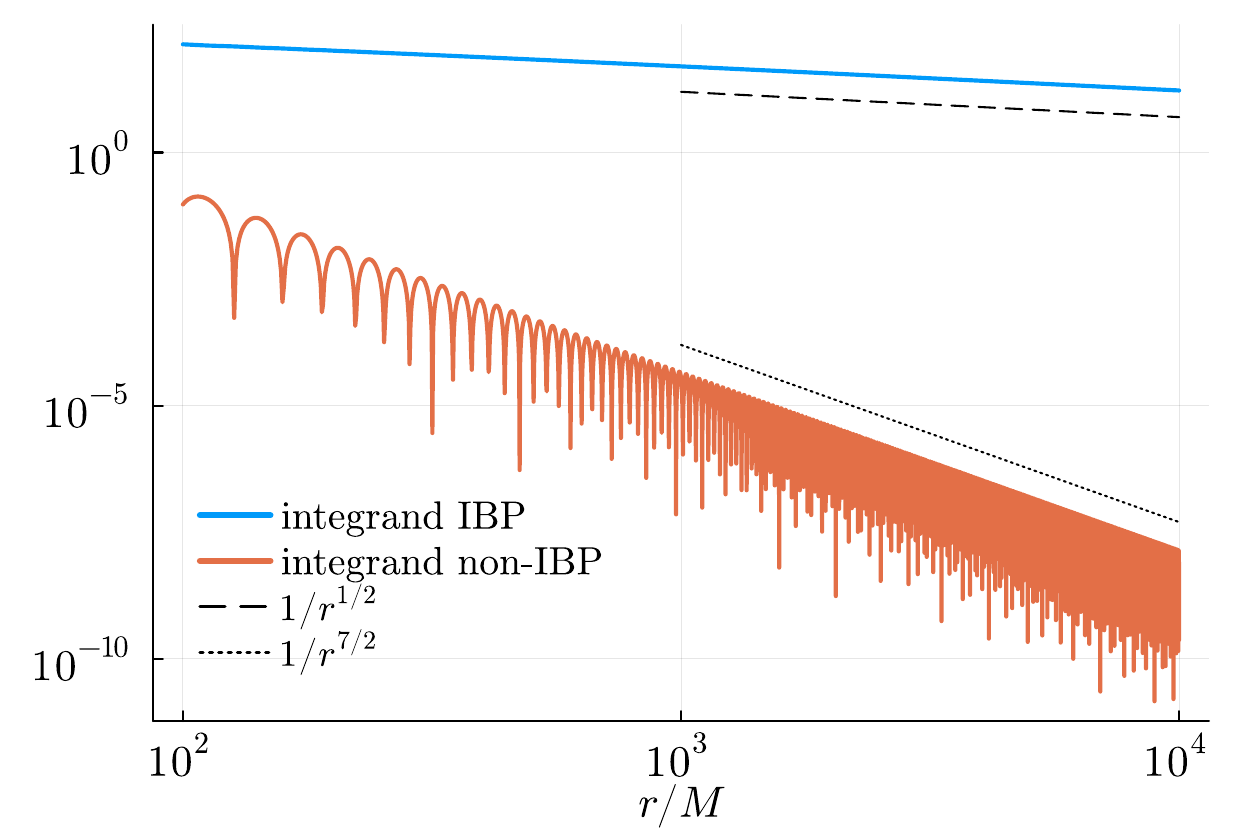}}

    \caption{The $\mathcal{E}=1$ case. Panel (a) illustrates the variation of $f_0$, $f_1$, and $f_2$ with $r$. As $r\to\infty$, $f_0$ converges at a rate of $1/r^{1/2}$, $f_1$ converges at a rate of $1/r^{3/2}$, and $f_2$ converges at a rate of $1/r^3$. Panel (b) shows the variation of the $\mathcal{W}(\rs)$ function. It converges at the same rate as $f_0$, i.e., $1/r^{1/2}$, and its oscillation frequency increases with increasing $\rs$. Panel (c) presents the magnitudes of the integrands in the Green's function integrals for the \gls{IBP} and non-\gls{IBP} methods. The \gls{IBP} method defined in Eq.~\eqref{Eq.I_SNIBP} exhibits a convergence rate of $1/r^{1/2}$, while the non-\gls{IBP} (i.e., the original \gls{SN}) method defined in Eq.~\eqref{Eq.X^Infty} converges faster as $1/r^{7/2}$. Other parameters are $\ell=2$, $m=0$, $a/M=0.9$, and $M\omega=0.5$. }
    \label{fig:unbound_rest}
\end{figure}
\begin{figure}[htpb]
    \centering
    
    \subfloat[]{\label{subfig:f_ultra}\includegraphics[width=1.0\linewidth]{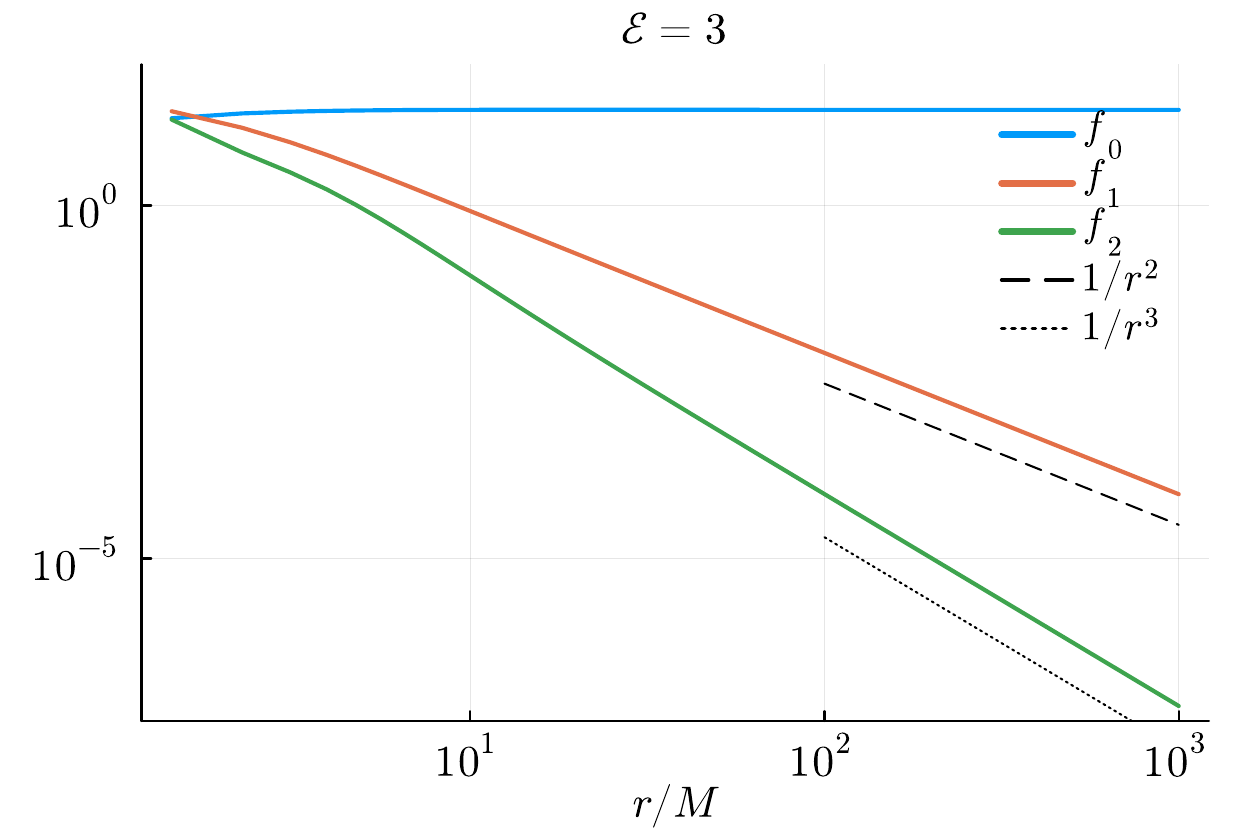}}

    \subfloat[]{\label{subfig:W_3}\includegraphics[width=1.0\linewidth]{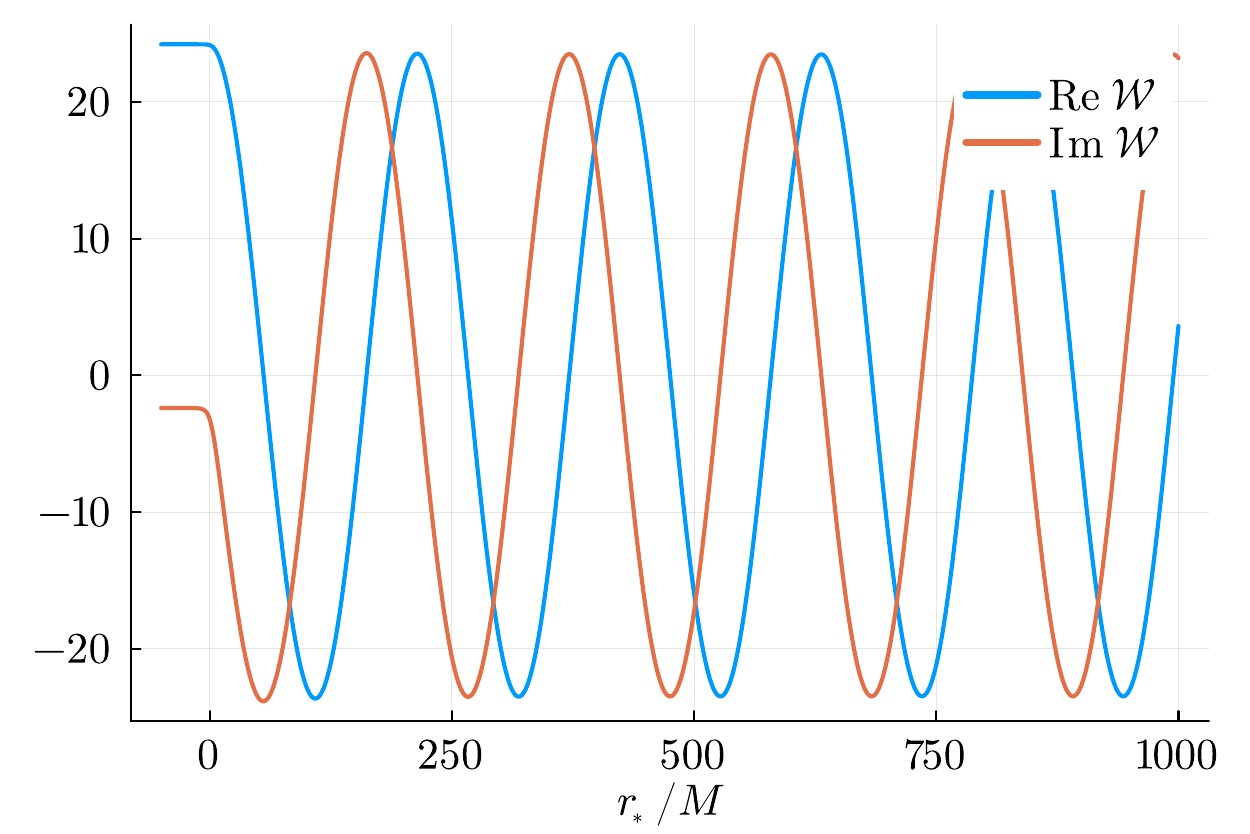}}

    \subfloat[]{\label{subfig:unbound_comparison_ultra}\includegraphics[width=1.0\linewidth]{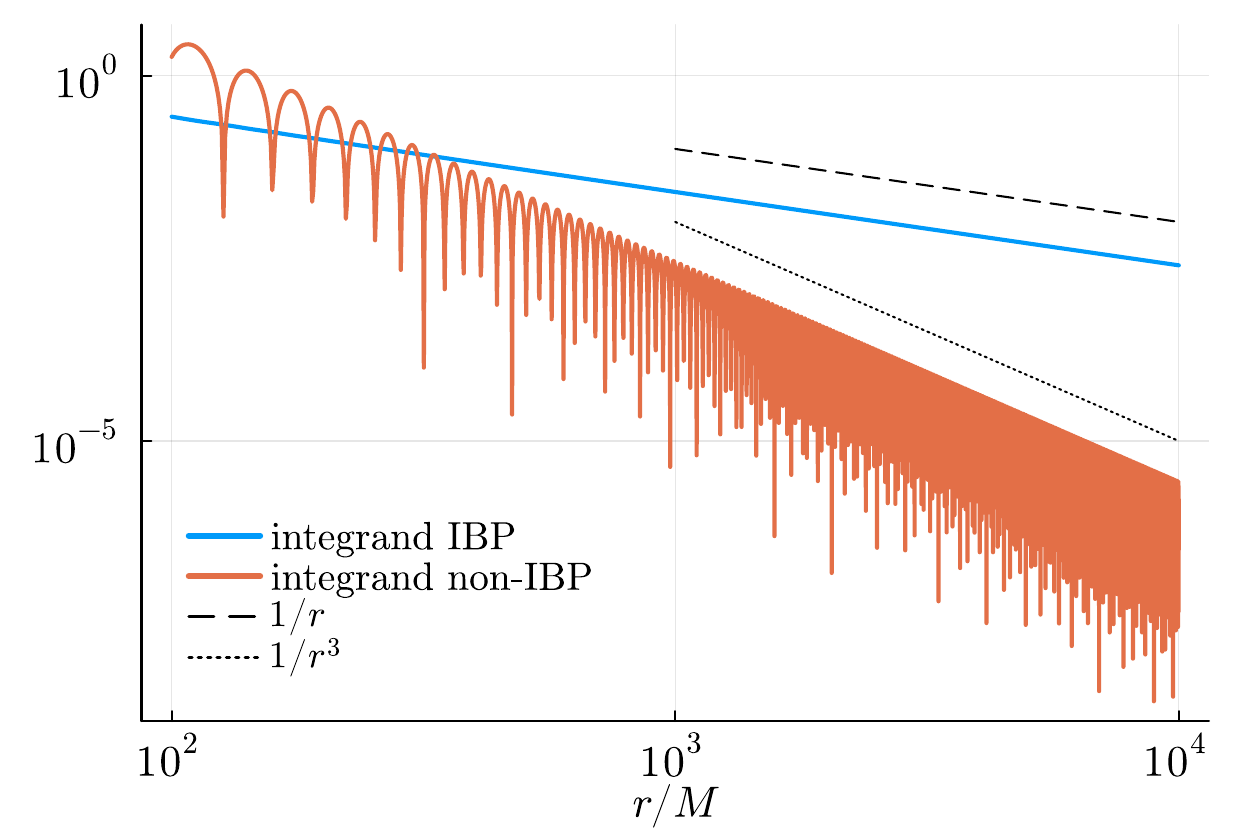}}

    \caption{The same as Fig.~\ref{fig:unbound_rest}, but with $\mathcal{E}=3$.}
    \label{fig:unbound_ultra}
\end{figure} 

The asymptotic behavior of $u^r$ at infinity differs in these two cases, leading to distinct asymptotic behaviors of $f_{0,1,2}$ in Eq.~\eqref{Eq.W_nn}.
This further results in differences in the asymptotic behavior of $\mathcal{W}(r)$ and the integrand within the Green's function integral involved in the calculations.
Consequently, the results obtained for these two cases are also significantly different. 
The behaviors of $f_{0,1,2}$, $\mathcal{W}$ and the Green's function integrand are shown in Fig.~\ref{fig:unbound_rest} for $\mathcal{E}=1$ and Fig.~\ref{fig:unbound_ultra} for $\mathcal{E}=3$.
One can see that $f_0\sim\bigO(1/r^{1/2})$, $f_1\sim\bigO(1/r^{3/2})$, $f_2\sim\bigO(1/r^{3})$ for $\mathcal{E}=1$ and $f_0\sim\bigO(1)$, $f_1\sim\bigO(1/r^{2})$, $f_2\sim\bigO(1/r^{3})$ for $\mathcal{E}>1$.

Since the convergence of $\mathcal{W}$ is controlled by $f_0$ following Eq.~\eqref{Eq.W_nn}, therefore the overall convergence of the integrand for the original \gls{SN} formulation [cf. Eq.~\eqref{Eq.X^Infty}] is $\sim\bigO(1/r^{7/2})$ for $\mathcal{E}=1$ and $\sim\bigO(1/r^{3})$ for $\mathcal{E}>1$.
However, the integrand in the \gls{SN}-\gls{IBP} approach in Eq.~\eqref{Eq.I_SNIBP} behaves as $\sim\bigO(1/r^{1/2})$ for $\mathcal{E}=1$ and $\sim\bigO(1/r)$ for $\mathcal{E}>1$. All of the asymptotic behaviors above agree with our theoretical analyses in Sec.~\ref{subsubsec:unbound orbits}.
\begin{figure}[htpb!]
    \centering

    \subfloat[]{\label{subfig:amp_rest}\includegraphics[width=1.0\linewidth]{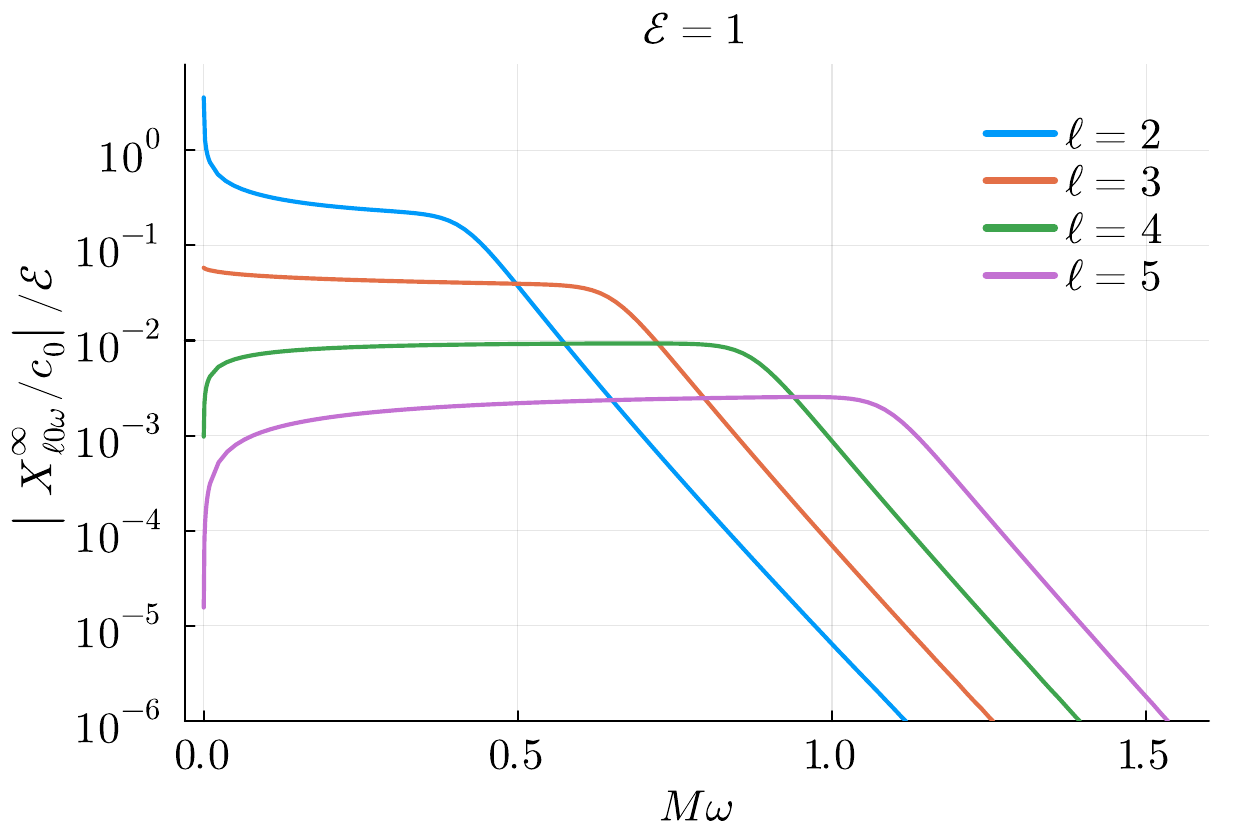}}

    \subfloat[]{\label{subfig:spectrum_rest}\includegraphics[width=1.0\linewidth]{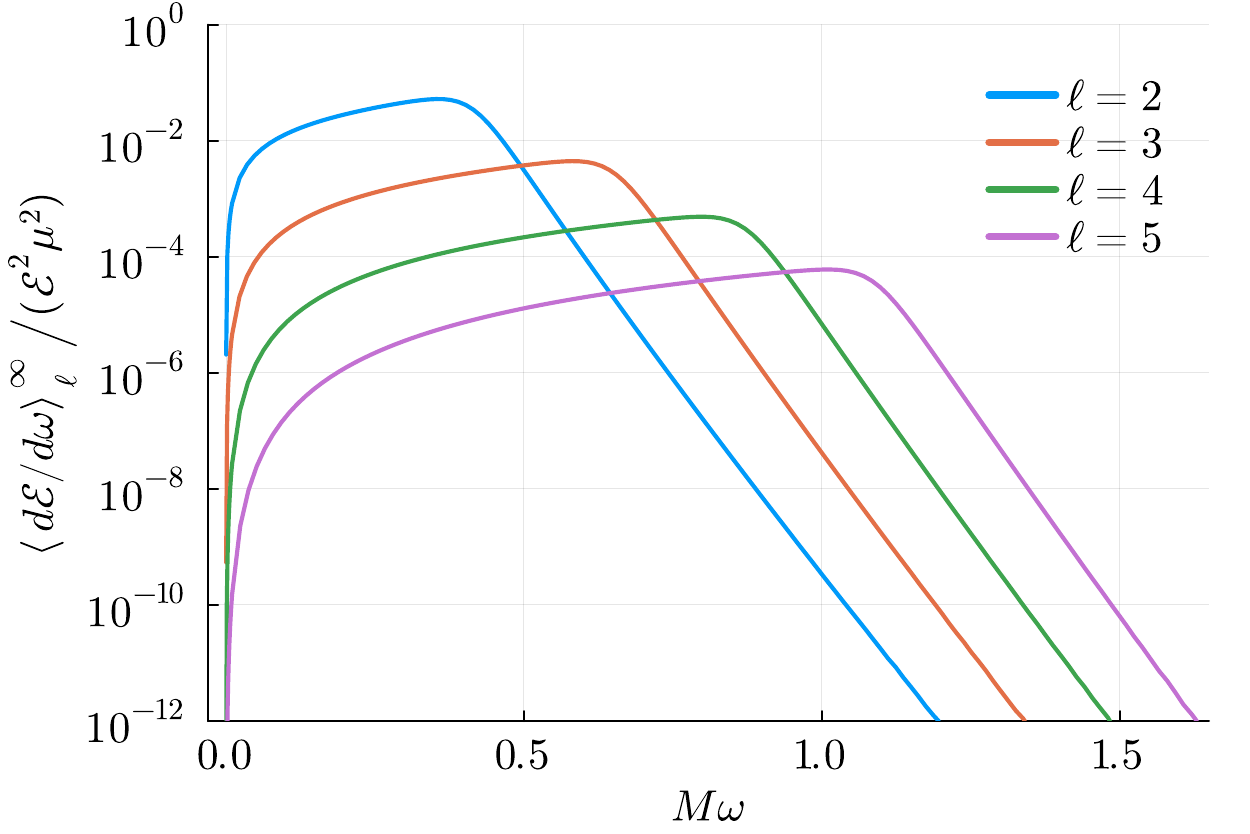}}
    
    \caption{The amplitude and energy spectrum of \gls{GW} induced by a particle falling radially into a Kerr \gls{BH} along its spin axis with zero initial velocity ($\mathcal{E}=1$) at infinity. The amplitude is normalized by $\mu$ and the energy spectrum is normalized by $\mu^2$. Other parameters are $a=0.9M$ and $m=0$.}
    \label{fig:Amp_spectrum_head_on_rest}
\end{figure}
\begin{figure}[htpb!]
    \centering

    \subfloat[]{\label{subfig:amp_ultra}\includegraphics[width=1.0\linewidth]{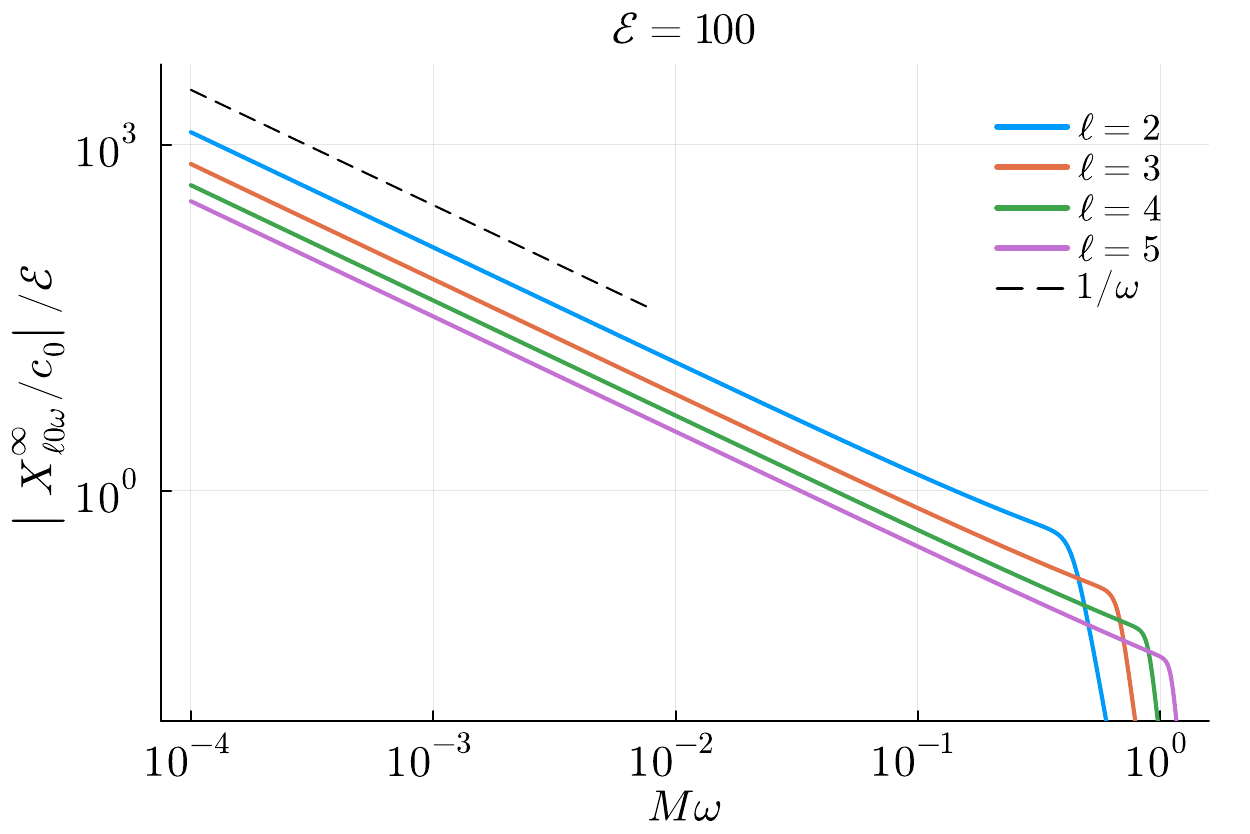}}

    \subfloat[]{\label{subfig:spectrum_ultra}\includegraphics[width=1.0\linewidth]{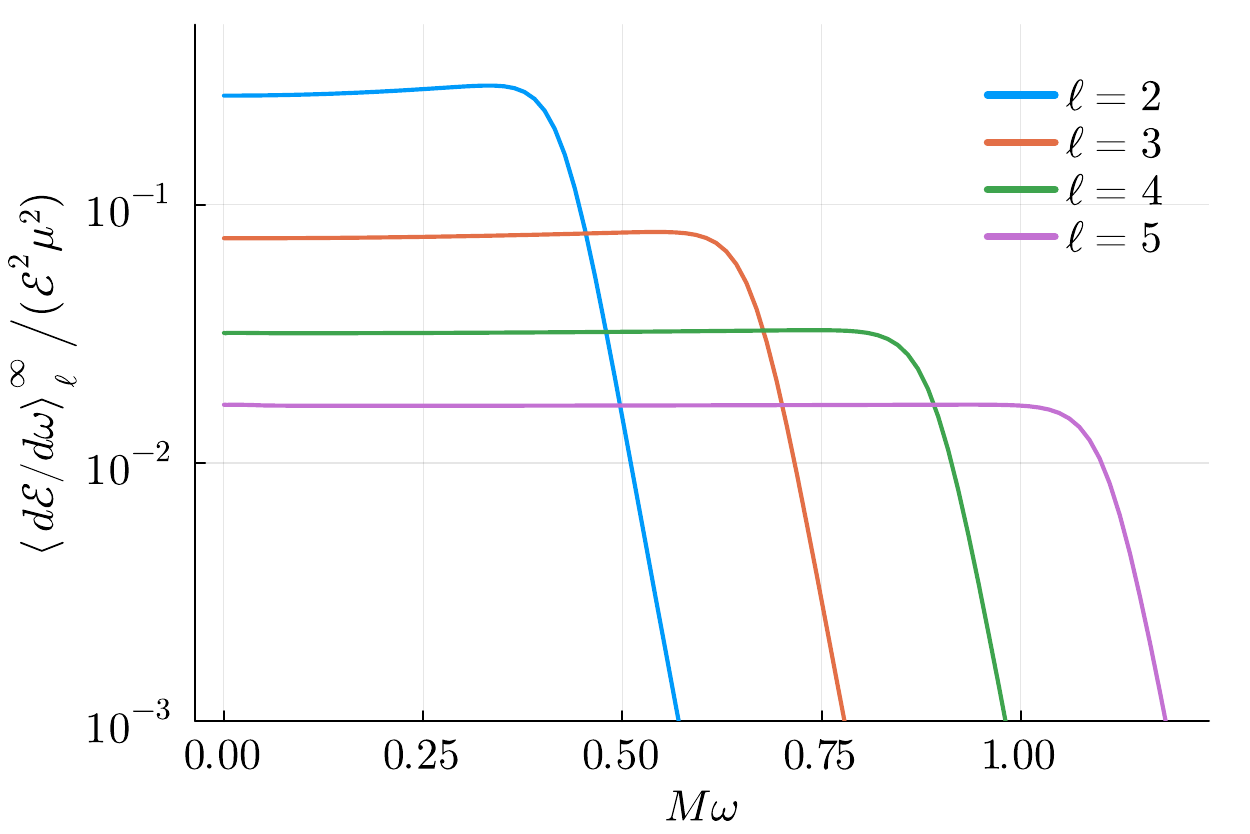}}
    
    \caption{The amplitude and energy spectrum of \gls{GW} induced by a particle falling radially into a Kerr \gls{BH} along its spin axis in the ultrarelativistic limit (we set $\mathcal{E}=100$). The amplitude is normalized by $\mu\mathcal{E}$ and the energy spectrum is normalized by $\mu^2\mathcal{E}^2$. Other parameters are $a=0.999M$ and $m=0$.}
    \label{fig:Amp_spectrum_head_on_ultra}
\end{figure}

Figures \ref{fig:Amp_spectrum_head_on_rest} and \ref{fig:Amp_spectrum_head_on_ultra} show the amplitude $\left|X_{\ell0\omega}^\infty/c_0\right|$ and energy spectrum $\left(\diff\mathcal{E}/\diff\omega\right)_\ell^\infty$ in the rest limit and the ultrarelativistic limit (using $\mathcal{E} = 100$ as an approximation), respectively.
We can see that Fig.~\ref{fig:Amp_spectrum_head_on_rest} agrees well with Fig.~1 in Ref.~\cite{PhysRevD.83.044039}, which shows the amplitude and the energy spectrum for a radial infall into a Schwarzchild \gls{BH} in the rest limit. We also find the same power-law behavior of the amplitudes as
\begin{equation}
    \left|\frac{X_{\ell0\omega}^\infty}{c_0}\right|\sim \omega^{(\ell-3)/3}
\end{equation}
at the \gls{ZFL}, i.e. $\omega\to0$, consistent with the result reported in Ref.~\cite{PhysRevD.83.044039}.
In the ultrarelativistic limit, one can see from Fig.~\ref{fig:Amp_spectrum_head_on_ultra} that the power-law behavior of all the $\ell$ modes are
\begin{equation}
    \left|\frac{X_{\ell0\omega}^\infty}{c_0}\right|\sim 1/\omega
\end{equation}
at the \gls{ZFL}. This makes the energy spectra nonvanishing at the \gls{ZFL} by definition in Eq.~\eqref{Eq.amp_spectra_unbound}. 

Therefore, we can extract the value for the energy spectrum values in the \gls{ZFL} per $\ell$ mode numerically from our calculations.
Theoretically, the total energy spectrum (summed over all $\ell$ modes) in the \gls{ZFL} was derived in Ref.~\cite{PhysRevD.15.2069}, which is given by
\begin{equation}
    \left(\frac{\diff\mathcal{E}}{\diff\omega}\right)^{\rm ZFL}=\sum_{\ell=2}^\infty\left(\frac{\diff\mathcal{E}}{\diff\omega}\right)_\ell^{\rm ZFL}=\frac{4}{3\pi}\mathcal{E}^2\mu^2,
\end{equation}
while the per-$\ell$ mode value was also given in Ref.~\cite{PhysRevD.81.104048} as
\begin{equation}\label{Eq.ZFL_l}
    \left(\frac{\diff\mathcal{E}}{\diff\omega}\right)^{\rm ZFL}_\ell=\frac{4\mathcal{E}^2\mu^2}{\pi}\frac{(2\ell+1)(\ell-2)!}{(\ell+2)!}.
\end{equation}

Here we show the numerical values extracted from Fig.~\ref{subfig:spectrum_ultra} and the theoretical predictions using Eq.~\eqref{Eq.ZFL_l}.
The results are tabulated in Table~\ref{tab:ZFL_values}.
Our numerical results match the theoretical results within an error of $0.1\%$ and are consistent with those shown in Ref.~\cite{Cardoso:2002jr}.
\begin{table}[htpb!]
    \caption{The numerical and theoretical \gls{ZFL} values of the energy spectrum in the ultrarelativistic limit, normalized by $\mathcal{E}^2\mu^2$.}
    \label{tab:ZFL_values}
    \centering
    \begin{ruledtabular}
    \begin{tabular}{ccc}
    $\langle\diff\mathcal{E}/\diff\omega\rangle^{\rm ZFL}_\ell$ & Numerical result & Theoretical prediction \\ 
    \hline
    $\ell=2$ & $0.26524876$ & $0.26525824$ \\
    $\ell=3$ & $0.07422927$ & $0.07427231$ \\
    $\ell=4$ & $0.03180048$ & $0.03183099$ \\
    $\ell=5$ & $0.01668822$ & $0.01667338$ \\
    \end{tabular}
    \end{ruledtabular}
\end{table}

\begin{figure*}[htpb!]
    \centering
    \includegraphics[width=1.0\linewidth]{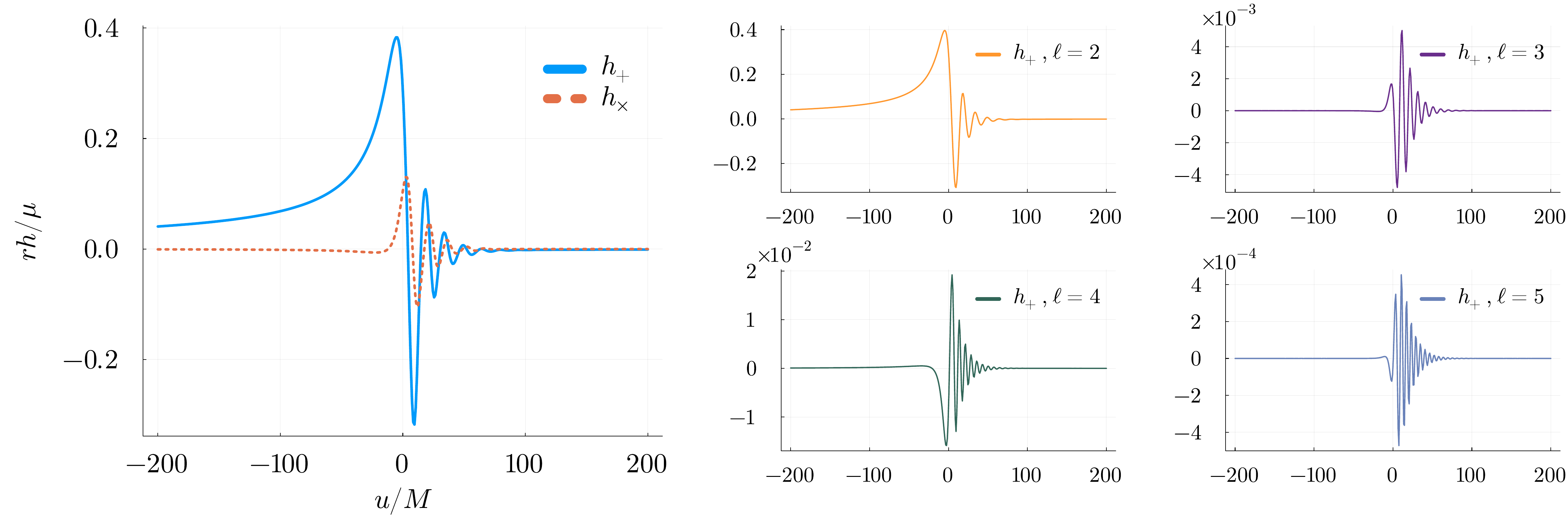}
    \caption{The \gls{GW} waveform of the radial infall for $\mathcal{E}=1$ viewing at $\theta=\pi/2$ and $\varphi=0$. The left panel shows the summation $h_+-ih_\times=h_2+h_3+h_4+h_5$. The right panel show the waveforms of the separate modes.}
    \label{fig:waveform_unbound}
\end{figure*}

Finally, we show in Fig.~\ref{fig:waveform_unbound} the time-domain waveform in the rest limit by performing an inverse \gls{FT} in Eq.~\eqref{Eq.Waveform_time_domain_unbound}.

\section{Discussions}
\label{sec:discussions}

\subsection{Waveform modeling for extreme mass ratio inspirals}
\label{subsec:applications}
One obvious application of our \gls{SN}-\gls{IBP} approach would be computing \gls{EMRI} waveforms, which we have already demonstrated in Sec.~\ref{subsec:result_bound} (and Fig.~\ref{fig:wavform_bound}).
However, such waveform generation requires many---around thousands of---modes to be calculated and summed up, which can take upwards of seconds per waveform and thus too slow for the purpose of \gls{LISA} data analysis.

Fortunately, the \textsc{FastEMRIWaveforms} framework \cite{Chua:2020stf, Katz:2021yft, Speri:2023jte, Chapman-Bird:2025xtd} solves this problem by generating \gls{EMRI} waveforms using precomputed waveform amplitude and flux data and thus decouples the waveform generation for data analysis from the relatively expensive waveform calculation.
As mentioned in Ref.~\cite{Chapman-Bird:2025xtd}, the framework can be easily extended handle eccentric and inclined orbits around a Kerr \gls{BH} once the corresponding amplitude and flux data are available, which we can easily generate with the \gls{SN}-\gls{IBP} approach.

Since our formalism and code implementation are independent from the Teukolsky formalism, one can also use our code to cross-validate the adiabatic---or 0PA---amplitude and flux data in the literature.
In terms of performance, our implementation is comparable with the state-of-the-art \texttt{pybhpt}.
We benchmark the performance of three different codes on calculating the waveform amplitude at infinity using the same set of fiducial parameters, i.e., $a=0.9M$, $p=6M$, $e=0.7$, $x=\cos\pi/4$, that are used throughout the paper, namely, ours, the \texttt{Teukolsky} package from \texttt{BHPToolkit}, and \texttt{pybhpt}. The single-core CPU times are tabulated in Table~\ref{tab:speed_comparison}.\footnote{The benchmarking was done with an Apple M2 chip. Specifically, \texttt{Teukolsky v1.1.1} with \texttt{Mathematica 14.0} and \texttt{pybhpt v0.9.10} with \texttt{Python 3.12} were used. Default machine precision is used in all calculations.}
Note that no attempt was made to optimize our current implementation, and there is still room for improvement.
For example, the current bottleneck of the calculation is actually in solving the homogeneous solutions $X^{\rm in, up}$.
Optimization of our implementation is planned but it is outside the scope of this paper.

\begin{table}[h]
    \caption{Runtime comparison using three different codes calculating the waveform amplitude at infinity with the same set of fiducial parameters $a=0.9M$, $p=6M$, $e=0.7$, $x=\cos\pi/4$ that are used throughout the paper. Runtimes are rounded to the nearest millisecond.}
    \label{tab:speed_comparison}
    \centering
    \begin{ruledtabular}
    \begin{tabular}{cccc}
    $(\ell, m, n, k)$ & this work [ms] & \texttt{BHPToolkit} [ms] & \texttt{pybhpt} [ms] \\ 
    \hline
    $(2,2,0,0)$ & $54$ & $1890$ & $45$ \\
    $(2,2,0,5)$ & $53$ & $7636$ & $63$ \\
    $(2,2,10,0)$ & $58$ & $5464$ & $48$ \\
    $(2,2,50,0)$ & $83$ & $39740$ & $389$ \\
    $(4,4,0,0)$ & $68$ & $1840$ & $53$ \\
    $(4,4,0,5)$ & $73$ & $4672$ & $47$ \\
    $(4,4,10,0)$ & $70$ & $6488$ & $54$ \\
    $(4,4,50,0)$ & $205$ & $33994$ & $398$ \\
    \end{tabular}
    \end{ruledtabular}
\end{table}

\subsection{Current limitations and future extensions}
\label{subsec:limitations_and_extensions}
While the \gls{IBP} approach we presented here drastically simplifies the computation of the waveform amplitude and fluxes at infinity for bound orbits when using the \gls{SN} formalism, there are still some limitations to our formulation.
For instance, the \gls{IBP} approach has no advantage in computing those quantities near the \gls{BH} horizon over the original formulation.
This is because, when near the horizon (or $r \to \rp$), the inverse transformation from the \gls{SN} variable to the Teukolsky variable [cf. Eq.~\eqref{Eq.SNtransformationRtoX}] is, in fact, dominated by the contribution coming from $\mathcal{S}$ when using the canonical solution $\mathcal{W}^{\rm canonical}$ (and by extension $\mathcal{S}^{\rm canonical}$).
Therefore, we still need to compute $\mathcal{W}(r = \rp)$ when computing fluxes down the horizon.

Note that the \gls{SN} formalism itself is perfectly valid in this case.
In fact, one can choose $w_{0, 1}$ [cf. Eq.~\eqref{Eq.w0w1} in Sec.~\ref{subsubsec:boundary terms}] such that it is the contribution coming from the $\Lambda^{-1}$ operator acting on the inhomogeneous \gls{SN} solution that dominates the transformation \cite{10.1093/ptep/ptaa149}.
However, we have already imposed the boundary conditions that $w_{0,1} = 0$ to make the boundary terms in Eq.~\eqref{Eq.Xinf after IBP} at infinity vanish.
A workaround to this issue is to solve for the inhomogeneous solution with a spin weight of $s = +2$ instead, which allows us to still use the \gls{IBP} approach to simplify calculations.
We leave this for a separate publication \cite{Lo:2025lpo}.

Another extension to our work here is to consider also generic bound plunge orbits into a Kerr \gls{BH}. We believe that the \gls{IBP} approach is still advantageous over the original \gls{SN} formulation.
These kinds of problems also serve as an analytical model for studying and understanding more about the physics and mechanism of the quasinormal mode excitation in binary black hole mergers using the \gls{SN} formalism \cite{Zhang:2013ksa, Watarai:2024huy, Lo:2025njp}.
We again leave this for a future publication.
An interesting avenue to employ the \gls{SN} formalism is computing gravitational radiation from scattered orbits around a Kerr \gls{BH}. This is particularly exciting since the same calculation can also be done with scattering amplitude techniques with post-Minkowskian expansions \cite{Jakobsen:2023hig}. The calculation from the \gls{BH} perturbation theory has only be done in the nonspinning limit (e.g., Ref.~\cite{Warburton:2025ymy}). However, due to their unbound nature, we expect that our \gls{SN}-\gls{IBP} approach will not be advantageous over the original \gls{SN} formulation for those scattered orbits (cf. Sec.~\ref{subsubsec:unbound orbits}).

\section{Conclusions}
In this work, we introduce a new scheme for solving the inhomogeneous \gls{SN} equation using integration by parts. %
When computing gravitational waveforms and fluxes at infinity coming from Kerr \glspl{BH} perturbed by particles in bound orbits, this simple trick eliminates the need for performing yet another radial integration to obtain the source term that needs to be convolved with a Green's function as in the original \gls{SN} formulation.
Our approach enables the efficient computation of gravitational waveforms within the \gls{SN} formalism now in \emph{all} cases, from bound to unbound orbits, without having to transform between the Teukolsky and \gls{SN} formalisms in intermediate steps.

Specifically, we define a new auxiliary variable $Y$ in place of the \gls{SN} variable $X$ that we convolve with the source term $\mathcal{T}$ that one would use in the Teukolsky formalism.
This new variable $Y$ is independent of the source term and therefore only needs to be computed once per frequency.
Furthermore, it is nonoscillatory and regular at the \gls{BH} horizon and spatial infinity, thus allowing for easy numerical calculations.
As a byproduct of this work, we also derive a source term for the \gls{SN} formalism that is valid for arbitrary motion and not just for geodesic motions.

We demonstrate that our approach and code implementation yield waveform amplitude and flux data that are consistent with the literature, while already achieving comparable speed without any optimization attempt.
Getting these amplitude and flux data accurately and efficiently is crucial as they enable the rapid generation of waveforms for future \gls{LISA} data analysis, especially for \gls{EMRI} waveforms with generic (eccentric and inclined) bound orbits.

\begin{acknowledgments}
The Center of Gravity is a Center of Excellence funded by the Danish National Research Foundation under Grant No.~DNRF184.
This work was supported by the research Grants No.~VIL37766 and No.~VIL53101 from Villum Fonden, and the DNRF Chair program Grant No. DNRF162 by the Danish National Research Foundation.
This work has received funding from the European Union's Horizon 2020 research and innovation program under the Marie Sklodowska-Curie Grant Agreement No.~101131233.
This work received no direct support from the National Natural Science Foundation of China. X.C.~acknowledges support from the NSFC Grant No.~12473037.
Additionally, R.K.L.L.~would like to thank KIAA at Peking University for their hospitality during his visit and Norichika Sago for his help in the early stage of this work.
\end{acknowledgments}

\section*{Data availability}
The data that support the findings of this article are openly available \cite{10.5281/zenodo.17574059}.

\appendix
\section{Deriving the relation between the Teukolsky and Sasaki-Nakamura source terms}
\label{App:Derivation_of_inhomogeneous_SN}
In this Appendix, we rederive the relation between the Teukolsky source term $\mathcal{T}$ and the \gls{SN} source term $\mathcal{S}$ following Refs.~\cite{SASAKI198268, 10.1143/PTP.67.1788}.

We first consider the variable $\mathcal{X}$, which is related to the \gls{SN} variable $X$ by $X(r) = \sqrt{\left(r^2 + a^2 \right)/\Delta^{2}} \mathcal{X}$ (cf. Ref.~\cite{Lo:2023fvv}). It satisfies a Regge-Wheeler-like equation given by
\begin{equation}
\label{eq:inhomogeneous_RWlikeEqn}
	\Delta^{2} \left( \frac{1}{\Delta} \mathcal{X}' \right)' - \Delta F_{1} \mathcal{X}' - U_{1} \mathcal{X} = \mathscr{S},
\end{equation}
where $\mathscr{S}$ is the source term for the $\mathcal{X}$ variable.
Note that this equation reduces to the usual Regge-Wheeler equation when $a = 0$ since in this case $\eta(r) = c_0$ is just a constant.
We use Eq.~\eqref{eq:inhomogeneous_RWlikeEqn} to write $\mathcal{X}''$ in terms of $\mathcal{X}$ and $\mathcal{X}'$, which is
\begin{equation}
	\mathcal{X}'' = \dfrac{\mathscr{S}}{\Delta} + \dfrac{U_1}{\Delta} \mathcal{X}	+ \left[ F_1 + \dfrac{\Delta'}{\Delta} \right] \mathcal{X}',
\end{equation}
where setting $\mathscr{S} = 0$ recovers the sourceless case.

Following Refs.~\cite{SASAKI198268, 10.1143/PTP.67.1788}, we \emph{modify} the inverse transformation from \gls{SN} variables $X$ to Teukolsky variables $R$ as
\begin{equation}
	R = \dfrac{1}{\eta} \left[ \left( \alpha + \dfrac{\beta'}{\Delta} \right)\mathcal{X} - \dfrac{\beta}{\Delta} \mathcal{X}' \right] + \dfrac{\mathscr{S}}{\eta},
\end{equation}
and setting $\mathscr{S} = 0$ recovers the homogeneous case.

We then evaluate $R'$ in terms of $\mathcal{X}$ and $\mathscr{S}$ (and their derivatives) and substitute them back to the inhomogeneous Teukolsky equation in Eq.~\eqref{Eq.TeukolskyRadialEquation}.
As a result, we obtain an \gls{ODE} for $\mathscr{S}$, which is given by
\begin{equation}
\label{eq:inhomo_Teukolsky_like_eqn}
\begin{aligned}
	& \Delta^{2} \left[ \dfrac{1}{\Delta} \left( \dfrac{\mathscr{S}}{\eta} \right)' \right]' \\
	& + \Delta^{2} \left[ -\dfrac{\beta}{\Delta^{3}} \left(\dfrac{\mathscr{S}}{\eta}\right) \right]' + \left( \alpha - V_{\rm T}\right) \dfrac{\mathscr{S}}{\eta} = -\mathcal{T}.
\end{aligned}
\end{equation}

Our goal is to solve for $\mathscr{S}$ in terms of $\mathcal{T}$. Note that we can rewrite Eq.~\eqref{eq:inhomo_Teukolsky_like_eqn} into a much more compact form as
\begin{equation}
\label{eq:compact_form_for_inhomo_Teukolsky_like_eqn}
\mathscr{J}^{\dagger} \left[ \mathscr{J}^{\dagger} \left( \dfrac{r^2}{\Delta} \dfrac{\mathscr{S}}{\eta} \right) \right] = - \dfrac{r^2}{\Delta^2} \mathcal{T},
\end{equation}
where $\mathscr{J}^\dagger \equiv \partial_r+iK/\Delta$ is a differential operator.\footnote{The $\mathscr{J}^\dagger$ and $\mathscr{J}$ operators are identical to the $J_{+}$ and $J_{-}$ defined in Ref.~\cite{Lo:2023fvv}, respectively.}
If we introduce an auxiliary variable $\mathcal{W}$ such that
\begin{equation}
	\mathcal{W}(r) = f(r) \exp \left( \int^{r} i\dfrac{K}{\Delta} d\tilde{r} \right),
\end{equation}
for any differentiable function $f(r)$, then $\mathcal{W}'$ can be written as
\begin{equation}
\label{eq:identity_with_Wprime_and_Jdagger}
	\mathcal{W}'(r) = \exp \left( \int^{r} i\dfrac{K}{\Delta} d\tilde{r} \right) \mathscr{J}^{\dagger} \left[ f(r) \right].
\end{equation}
If we define
\begin{equation}
	\mathcal{W}(r) = \dfrac{r^2}{\Delta} \dfrac{\mathscr{S}}{\eta} \exp \left( \int^{r} i\dfrac{K}{\Delta} d\tilde{r} \right),
\end{equation}
then by using the identity in Eq.~\eqref{eq:identity_with_Wprime_and_Jdagger} twice, we have
\begin{equation}
	\left(\mathcal{W}' \right)' = \exp \left( \int^{r} i\dfrac{K}{\Delta} d\tilde{r} \right) \mathscr{J}^{\dagger} \left[ \mathscr{J}^{\dagger}\left( \dfrac{r^2}{\Delta} \dfrac{\mathcal{S}}{\eta} \right) \right].
\end{equation}
Using Eq.~\eqref{eq:compact_form_for_inhomo_Teukolsky_like_eqn}, we have
\begin{equation}
\tag{\ref{Eq.d2W}}
	\mathcal{W}'' = -\dfrac{r^2}{\Delta^2} \mathcal{T} \exp \left( \int^{r} i\dfrac{K}{\Delta} d\tilde{r} \right),
\end{equation}
which is the \gls{ODE} that one needs to solve to obtain $\mathscr{S}$ from $\mathcal{T}$.

Given the source term $\mathscr{S}$ for the variable $\mathcal{X}$, we can convert that to the source term needed the \gls{SN} equation $\mathcal{S}$ simply with
\begin{equation}\label{Eq.SN_source_term}
	\mathcal{S} = \dfrac{1}{(r^2 + a^2)^{3/2}} \mathscr{S},
\end{equation}
as $\mathcal{X}$ solutions are related to the corresponding $X$ solutions by $X = \sqrt{(r^2 + a^2)/\Delta^{2}} \mathcal{X}$. Putting everything together and the subscript back, we have
\begin{equation}
\tag{\ref{Eq.Source_S}}
    \mathcal{S}_{\ell m\omega}=\frac{\eta\Delta\mathcal{W}}{(r^2+a^2)^{3/2}r^2}\exp\left(-i\int^r\frac{K}{\Delta}\diff\tilde{r}\right).
\end{equation}

\section{$A$ terms and $W$ terms\label{Appendix_A_and_W}}
The source term components $A$ in the Teukolsky formalism are given by
\begin{subequations}
    \begin{align}
        &A_{nn0}=\frac{\mathscr{A}}{2}\rho\bar{\rho}^2\mathcal{N}^2\mathscr{L}_1^\dagger\left[\rho^{-4}\mathscr{L}_2^\dagger\left(\rho^3S\right)\right],\\
        &A_{n\bar{m}0}=\mathscr{A}\bar{\rho}^2\mathcal{N}\bar{\mathcal{M}}\left[\left(\mathscr{L}_2^{\dagger}S\right)\left(\frac{i K}{\Delta}-\rho-\bar{\rho}\right)\right.\\
        &\qquad\qquad\qquad\qquad\qquad \left.-a\sin\theta S\frac{K}{\Delta}\left(\rho-\bar{\rho}\right)\right],\notag\\
        &A_{\bar{m}\bar{m}0}=\frac{\mathscr{A}}{2}\bar{\rho}^2\bar{\mathcal{M}}^2S\left[-i\left(\frac{K}{\Delta}\right)_{,r}-\frac{K^2}{\Delta^2}-2i\rho\frac{K}{\Delta}\right],\\
        &A_{n\bar{m}1}=\mathscr{A}\bar{\rho}^2\mathcal{N}\bar{\mathcal{M}}\left[\mathscr{L}_2^{\dagger}S+i a\sin\theta\left(\rho-\bar{\rho}\right)S\right],\\
        &A_{\bar{m}\bar{m}1}=\mathscr{A}\bar{\rho}^2\bar{\mathcal{M}}^2S\left(i\frac{K}{\Delta}-\rho\right),\\
        &A_{\bar{m}\bar{m}2}=\frac{\mathscr{A}}{2}\bar{\rho}^2\bar{\mathcal{M}}^2S,
    \end{align}
\end{subequations}
where $S$ are the \gls{SWSH} with all of its subscripts suppressed to avoid confusion.

The source term components in the \gls{SN} formalism are given by
\begin{subequations}
    \begin{align}
        W_{nn}=&\mathscr{A}\frac{\rho\bar{\rho}^2}{2}\mathscr{L}_1^\dagger\left[\rho^{-4}\mathscr{L}_2^\dagger\left(\rho^3 S\right)\right]r^2Y\mathrm{phase},\\
        W_{n\bar{m}}=&-\mathscr{A}r\bar{\rho}^2\left\{\left(\mathscr{L}_2^\dagger S\right)\left(\rho+\bar{\rho}\right)rY\right.\\
        &\left.+\left[\mathscr{L}_2^\dagger S+i a\sin\theta\left(\rho-\bar{\rho}\right)S\right]\left(2Y+rY'\right)\right\}\mathrm{phase},\notag\\
        W_{\bar{m}\bar{m}}=&\mathscr{A}S\bar{\rho}^2\biggl[\frac{X}{2\sqrt{r^2+a^2}}+\left(Y+2rY'\right)\mathrm{phase}\\
        &+\rho r\left(2Y+rY'\right)\mathrm{phase}\biggr],\notag\\
        \text{phase}=&\exp\left(i \int^r\frac{K}{\Delta}\diff\tilde{r}\right) \label{eq:KoverDeltaIntegral}\\
        =&\exp\left(i \omega \rs-\frac{i am}{2\sqrt{1-a^2}}\ln\frac{r-\rp}{r-\rmi}\right)\notag.
    \end{align}
\end{subequations}

\subsection{Normalization conventions}
\label{subsec:normalization_conventions}
The value of the constant $\mathscr{A}$ above depends on the normalization conventions on the \gls{FT} and \glspl{SWSH} adopted.
Specifically for the \gls{FT}, there are canonically two conventions for the normalization, namely the unitary \gls{FT} convention where
\begin{equation}
    \begin{aligned}
        F(\omega)&=\frac{1}{\sqrt{2\pi}}\int_{-\infty}^\infty f(t)e^{i\omega t}\diff t,\\
        f(t)&=\frac{1}{\sqrt{2\pi}}\int_{-\infty}^\infty F(\omega)e^{-i\omega t}\diff\omega,
    \end{aligned}
\end{equation}
and the non-unitary \gls{FT} convention where
\begin{equation}
    \begin{aligned}
        F(\omega) & =\frac{1}{2\pi}\int_{-\infty}^\infty f(t)e^{i\omega t}\diff t,\\
        f(t) & =\int_{-\infty}^\infty F(\omega)e^{-i\omega t}\diff\omega.
    \end{aligned}
\end{equation}
As for \glspl{SWSH}, there are also two normalization conventions where either
\begin{equation}
\label{eq:SWSH_normalization_scheme1}
    \int_0^\pi \left|{_s}S_{\ell m}^{a\omega}(\theta)\right|^2\sin\theta\diff\theta=1,
\end{equation}
which we will refer to as the \gls{SWSH} normalization scheme 1, and
\begin{equation}
\label{eq:SWSH_normalization_scheme2}
    \int_0^\pi \left|{_s}S_{\ell m}^{a\omega}(\theta)\right|^2\sin\theta\diff\theta=\frac{1}{2\pi},
\end{equation}
which we will refer to as the \gls{SWSH} normalization scheme 2, respectively.

Table~\ref{tabscrA} shows the value of $\mathscr{A}$ with different choices of normalization conventions.
Although the exact choice does not have any impact on physics, care should be taken when comparing results from different papers since the value of $\mathscr{A}$ may vary literature to literature.
In this work, we use the nonunitary \gls{FT} convention and the \gls{SWSH} normalization scheme 2, and therefore $\mathscr{A}=-1$.
\begin{table}[h]
    \caption{The value of the constant $\mathscr{A}$ for different choices of \gls{FT} and \gls{SWSH} conventions. In this work, we have $\mathscr{A} = -1$.}
    \label{tabscrA}
    \centering
    \begin{ruledtabular}
    \begin{tabular}{ccc}
    $\mathscr{A}$ & \gls{SWSH} scheme 1 & \gls{SWSH} scheme 2 \\ 
    \hline
    Unitary \gls{FT} & $-1/\sqrt{2\pi}$ & $-\sqrt{2\pi}$ \\
    Nonunitary \gls{FT} & $-1/2\pi$ & $-1$\\
    \end{tabular}
    \end{ruledtabular}
\end{table}

\section{Asymptotic expansions for $Y^{\rm in, up}$}\label{Appendix_Ysolution}
In this Appendix, we derive the asymptotic expansions for $Y^{\rm in, up}$, respectively, for speeding up the numerical integration of Eq.~\eqref{Eq.ODEforY}, which is repeated here for reference as
\begin{equation}
\tag{\ref{Eq.ODEforY}}
    Y_{\ell m\omega}^{\rm in/up\ \prime\prime}(r)\equiv \frac{X_{\ell m\omega}^{\rm in/up}(r)}{r^2\sqrt{r^2+a^2}}\exp\left(-i\int^r\frac{K}{\Delta}\diff r\right).
\end{equation}

\subsection{$Y^{\rm in}$ solution}
\label{App:subsec:Y_in}
Recall that in Ref.~\cite{Lo:2023fvv}, we have shown that asymptotically as $r \to \infty$,
\begin{equation}
    X^{\rm in}(r\to\infty)=B_{\rm SN}^{\rm ref}e^{i\omega \rs}\sum_{w=0}^\infty\frac{\mathcal{C}^\infty_{+,w}}{r^w}+B_{\rm SN}^{\rm inc}e^{-i\omega r_*}\sum_{w=0}^\infty\frac{\mathcal{C}^\infty_{-,w}}{r^w}.
\end{equation}
The expressions of $\mathcal{C}^{\infty}_{\pm,1,2,3}$ can be found in the Appendix G of Ref.~\cite{Lo:2023fvv} (note that $\mathcal{C}^{\infty}_{\pm, 0} = 1$).
To write down an asymptotic expansion of $Y^{\rm in}(r \to \infty)$, we also need the series expansions of the two other terms, which are given by
\begin{subequations}
    \begin{align}
        \exp\left(-i\int^r\frac{K}{\Delta}\diff\tilde{r}\right) & =e^{-i\omega\rs}\sum_{j=0}^\infty\frac{a_j}{r^j},\\
        \frac{1}{r^2\sqrt{r^2+a^2}} & =\frac{1}{r^3}\sum_{j=0}^\infty\frac{b_j}{r^j},
    \end{align}
\end{subequations}
where
\begin{subequations}
    \begin{align}
        a_j & = \frac{1}{j!}B_j( P_1,\dots,P_j),\\
        P_j & = \frac{iam\left(\rp^j-\rmi^j\right)\Gamma(j)}{\rmi-\rp},\\
        b_j & = \frac{1+(-1)^j}{2} a^j \binom{-1/2}{j/2},
    \end{align}
\end{subequations}
and $B_j$ denotes the $j$th complete exponential Bell polynomial.
Therefore, the piece that is proportional to $B_{\rm SN}^{\rm ref}$, which we denote as ${Y^{\rm in}_+}''(r\to\infty)$, can be expressed as
\begin{equation}\label{Eq.Y''_inf_+}
    \begin{aligned}
        &{Y^{\rm in}_+}''(r\to\infty)\\
        =&\frac{X^\infty_+(r)}{r^2\sqrt{r^2+a^2}}\exp\left(-i\int^r\frac{K}{\Delta}\diff\tilde{r}\right)\\
        =&\frac{B_{\rm SN}^{\rm ref}}{r^3}\left(\sum_{j=0}^\infty\frac{a_j}{r^i}\right) \left(\sum_{v=0}^\infty\frac{b_v}{r^j}\right) \left(\sum_{w=0}^\infty\frac{\mathcal{C}^\infty_{+,w}}{r^k}\right)\\
        =&B_{\rm SN}^{\rm ref}\sum_{j=0}^\infty\frac{Y^{\infty,+}_{j}}{r^{j+3}},
    \end{aligned}
\end{equation}
where
\begin{equation} 
    Y^{\infty}_{+,j}=\sum_{v=0}^j\sum_{w=0}^{j-v}a_{v}  b_{w} \mathcal{C}^{\infty}_{+,j-v-w}.
\end{equation}
Notice that it is not oscillatory because $e^{-i \int^r\frac{K}{\Delta}\diff\tilde{r}} \sim e^{-i\omega \rs}$ [cf. Eq.~\eqref{eq:KoverDeltaIntegral}] cancels out the phase term $e^{i \omega \rs}$ coming from $X^{\rm in}$.

The other piece that is proportional to $B_{\rm SN}^{\rm inc}$, which we denote as ${Y^{\rm in}_-}''(r\to\infty)$, is more complicated because the phase terms do not cancel out each other.
Here, we need to expand also $r_{*}(r \to \infty)$, which is given by
\begin{equation}
    r_*=r+2\ln \frac{r}{2}-\sum_{v=1}^\infty\frac{2}{vr^v}\left(\sum_{j=0}^vr_+^jr_-^{v-j}\right).
\end{equation}
Therefore, we have
\begin{equation}
    e^{-2i\omega\rs}=e^{4i\omega\ln 2}\frac{e^{-2i\omega r}}{r^{4i\omega}}\sum_{j=0}^\infty\frac{d_j}{r^j},
\end{equation}
where
\begin{subequations}
    \begin{align}
        &d_j=\frac{1}{j!}B_j(Q_1,\dots,Q_j),\\
        &Q_j=4i\omega\Gamma(j)\left(\sum_{v=0}^j r_+^vr_-^{j-v}\right).
    \end{align}
\end{subequations}
Finally, we have
\begin{equation}\label{Eq.Y''_inf_-}
    \begin{aligned}
        &{Y_-^{\rm in}}''(r\to\infty)\\
        =&B_{\rm SN}^{\rm inc}\frac{e^{4i\omega\ln 2-2i\omega r}}{r^{3+4i\omega}}\left(\sum_{j=0}^\infty\frac{a_j}{r^j}\right) \left(\sum_{w=0}^\infty\frac{b_w}{r^w}\right)\\
        & \left(\sum_{v=0}^\infty\frac{\mathcal{C}^\infty_{-,v}}{r^v}\right) \left(\sum_{u=0}^\infty\frac{d_u}{r^u}\right)\\
        =&B_{\rm SN}^{\rm inc}\frac{e^{4i\omega\ln 2-2i\omega r}}{r^{4i\omega}}\sum_{j=0}^\infty\frac{Y_{-,j}^{\infty}}{r^{j+3}},
    \end{aligned}
\end{equation}
where
\begin{equation}
    Y_{-,j}^{\infty}=\sum_{v=0}^j\sum_{w=0}^{j-v}\sum_{u=0}^{j-v-w}a_v  b_w  \mathcal{C}^\infty_{-,u}  d_{j-v-w-u}.
\end{equation}
Combining Eq.~\eqref{Eq.Y''_inf_+} and Eq.~\eqref{Eq.Y''_inf_-}, we obtain Eq.~\eqref{Eq.Yin''AsymptoticExpansion}.
\vspace{3em} %

It is not difficult to show that $Y_{+,0}^{\infty}=Y_{-,0}^{\infty}=1$.
Here, we also give the explicit expressions of the next three coefficients, which are given by
\begin{widetext}
\begin{subequations}
    \begin{align}
        Y_{+,1}^{\infty}&=\mathcal{C}_{+,1}^\infty-iam,\\
        Y_{+,2}^{\infty}&=\mathcal{C}_{+,2}^\infty-iam\mathcal{C}_{+,1}^\infty-\frac{a}{2}\left(a+am^2+2im\right),\\
        Y_{+,3}^{\infty}&=\mathcal{C}_{+,3}^\infty-iam\mathcal{C}_{+,2}^\infty-\frac{a}{2}\left(a+am^2+2im\right)\mathcal{C}_{+,1}^\infty+\frac{iam}{6} \left[a^2\left(m^2+5\right)+6 ia m-8\right],\\
        Y_{-,1}^{\infty}&=\mathcal{C}_{-,1}^\infty-iam+8i\omega,\\
        Y_{-,2}^{\infty}&=\mathcal{C}_{-,2}^\infty-i\left(am-8\omega\right)\mathcal{C}_{-,1}^\infty-\frac{1}{2}a^2\left(m^2+4i\omega +1\right)+am(8\omega -i)+8\omega (-4\omega +i),\\
        Y_{-,3}^{\infty}&=\mathcal{C}_{-,3}^\infty-i\left(am-8\omega\right)\mathcal{C}_{-,2}^\infty+\frac{1}{2}\left[2am(8\omega -i)+16\omega (i-4\omega)-a^2\left(m^2+4i\omega +1\right)\right]\mathcal{C}_{-,1}^\infty\notag\\
        &+\frac{i}{6}\left\{a^3 m\left(m^2+12i\omega+5\right)-2 a^2\left[3 m^2(4\omega -i)+4\omega (7+12i\omega )\right]\right.\\
        &\left.+8 am\left(24\omega^2-12i\omega -1\right)+64\omega\left(1+6i\omega-8\omega ^2\right)\right\}\notag.
    \end{align}
\end{subequations}
\end{widetext}

The initial values of $Y^{\rm in}(r)$ and ${Y^{\rm in}}'(r)$ at large $r = r_{\rm out}$ can then be obtained by integrating Eq.~\eqref{Eq.Yin''AsymptoticExpansion}.
The ${Y_+^{\rm in}}$ piece is straightforward and is given by
\begin{subequations}
    \begin{align}
        &{Y_+^{\rm in}}'(r_{\rm out})=-B_{\rm SN}^{\rm ref}\sum_{j=0}^\infty\frac{Y_{+,j}^{\infty}}{j+2}\frac{1}{r_{\rm out}^{j+2}},\\
        &Y_+^{\rm in}(r_{\rm out})=B_{\rm SN}^{\rm ref}\sum_{j=0}^\infty\frac{Y_{+,j}^{\infty}}{(j+1)(j+2)}\frac{1}{r_{\rm out}^{j+1}}.
    \end{align}
\end{subequations}

While the $Y_-^{\rm in}$ piece is more complicated because the phase term is nonvanishing. We define
\begin{widetext}
    \begin{equation}
    \begin{aligned}
        y_{j}(r_{\rm out})& \equiv \int_{r_{\rm out}}^\infty\frac{e^{-2i\omega r}}{r^{j+4i\omega}}\\
        & = \frac{1}{r_{\rm out}^{j-1+4i\omega}}\left[\frac{_1F_2\left(\frac{1-j}{2}-2i\omega;\frac{1}{2},\frac{3-j}{2}-2i\omega;-\omega^2r_{\rm out}^2\right)}{j-1+4i\omega}\right.\\
        &\;\;\;\left.+\frac{2i\omega r_{\rm out}\times{}_1F_2\left(1-\frac{j}{2}-2i\omega;\frac{3}{2},2-\frac{j}{2}-2i\omega;-\omega^2r_{\rm out}^2\right)}{j-2+4i\omega}\right]\\
        &\;\;\;+\frac{\Gamma(1-j-4i\omega)|2\omega|^{j+4i\omega}}{2}\left[\frac{1}{|\omega|}\sin\frac{\pi(j+4i\omega)}{2}-\frac{i}{\omega}\cos\frac{\pi(j+4i\omega)}{2}\right],
    \end{aligned}
\end{equation}
\end{widetext}
where $_1F_2(a;b,c;x)$ is the hypergeometric function.
With these, we can write
\begin{subequations}
    \begin{align}
        {Y_-^{\rm in}}'(r_{\rm out})=&-e^{4i\omega \ln 2}B_{\rm SN}^{\rm inc}\sum_{j=0}^{\infty}Y_{-,j}^{\infty}y_{j+3}(r_{\rm out}),\\
        Y_-^{\rm in}(r_{\rm out})=&e^{4i\omega \ln 2}B_{\rm SN}^{\rm inc}\sum_{j=0}^{\infty}Y_{-,j}^{\infty}\left[y_{j+2}(r_{\rm out})\right.\notag\\
        &\left.-r_{\rm out}\cdot y_{j+3}(r_{\rm out})\right].
    \end{align}
\end{subequations}

Unfortunately, using the hypergeometric function implemented in \texttt{HypergeometricFunctions.jl} \cite{hypergeometricfunctionsjl} is too time-consuming for an acceptable precision when $\omega r_{\rm out}$ is a relatively large value. Therefore, we switch to an asymptotic expansion given by
\begin{multline}
        \frac{\Gamma(a_1)}{\Gamma(b_1)\Gamma(b_2)}{}_1F_2(a_1;b_1,b_2;-z)\\
        = {}_1H_2(z)+{}_1E_2(ze^{-\pi i})+{}_1E_2(ze^{\pi i}),
\end{multline}
where
\begin{subequations}
    \begin{align}
        {}_1H_2(z)=&\sum_{j=0}^\infty\frac{(-1)^j}{j!}\frac{\Gamma(a_1+j)}{\Gamma(b_1-a_1-j)\Gamma(b_2-a_1-j)}z^{-a_1-j},\\
        {}_1E_2(z)=&\frac{e^{2\sqrt{z}}}{\sqrt{\pi}}\sum_{j=0}^\infty c_k\frac{z^{(\nu-j)/2}}{2^{j+1}}\\
        =&\begin{cases}
            \frac{(-z)^{\nu/2}e^{-2i\sqrt{-z}}}{\sqrt{\pi}}\sum_{j=0}^\infty\frac{c_k(-z)^{-j/2}}{2^{j+1}}& z\to ze^{-\pi i}\\
            \frac{(-z)^{\nu/2}e^{2i\sqrt{-z}}}{\sqrt{\pi}}\sum_{j=0}^\infty\frac{c_k(-z)^{-j/2}}{2^{j+1}}& z\to ze^{\pi i}
        \end{cases},\notag
    \end{align}
\end{subequations}
with
\begin{subequations}
    \begin{align}
        \nu=&a_1-b_1-b_2+\frac{1}{2},\\
        c_0=&1,\\
        c_j=&-\frac{1}{4j}\sum_{w=0}^{j-1}c_w e_{j,w},\\
        e_{j,w}=&\frac{(1-\nu-2b_1+w)_{2+j-w}(a_1-b_1)}{(b_2-b_1)(1-b_1)}\notag\\
        &+\frac{(1-\nu-2b_2+w)_{2+j-w}(a_1-b_2)}{(b_1-b_2)(1-b_2)}\\
        &+\frac{(w-1-\nu)_{2+j-w}(a_1-1)}{(1-b_1)(1-b_2)}\notag.
    \end{align}
\end{subequations}
In our case, we have $b_2=a_1+1$. Therefore,
\begin{equation}
    \frac{\Gamma(a_1)}{\Gamma(b_1)\Gamma(b_2)}=\frac{1}{a_1\Gamma(b_1)}.
\end{equation}

Now we can construct the initial conditions for the \gls{ODE} in Eq.~\eqref{Eq.Y_rs_ODE} for $Y^{\rm in}$ and solve it inward to the horizon to get the values of $Y^{\rm in}$ and ${Y^{\rm in}}'$ across the entire domain of definition with
\begin{subequations}
    \begin{align}
        \left.Y\right|_{\rs=\rs^{\rm out}}& =Y^{\rm in}_+(r_{\rm out})+Y^{\rm in}_-(r_{\rm out}),\\
        \left.\frac{\diff Y}{\diff \rs}\right|_{\rs=\rs^{\rm out}} & =\frac{\Delta}{r_{\rm out}^2+a^2}\left[{Y^{\rm in}_+}'(r_{\rm out})+{Y^{\rm in}_-}'(r_{\rm out})\right].
    \end{align}
\end{subequations}

\begin{figure}[htpb!]
    \centering
    \includegraphics[width=\columnwidth]{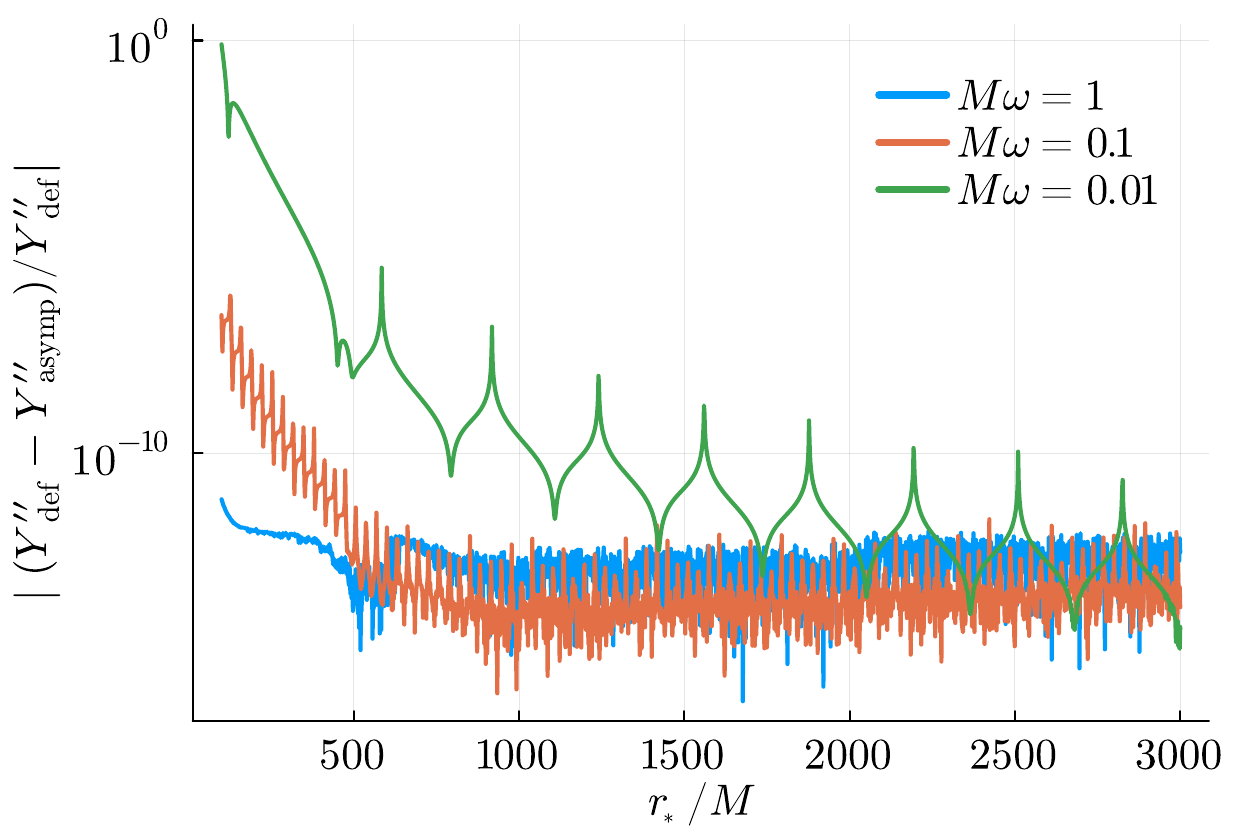}
    \caption{The residue of the asymptotic expansion in Eq.~\eqref{Eq.Yin''AsymptoticExpansion} compared with the definition in Eq.~\eqref{Eq.ODEforY} with $\ell=m=2$, $a/M=0.9$ and frequencies $M\omega=1,\  0.1,\ 0.01$. The asymptotic function is expanded to $\bigO(1/r^6)$ order.}
    \label{fig:relative_error}
\end{figure}

Figure~\ref{fig:relative_error} shows the relative error between the asymptotic expansion of ${Y^{\rm in}}''(r\to\infty)$ defined in Eq.~\eqref{Eq.Yin''AsymptoticExpansion} and its definition in Eq.~\eqref{Eq.ODEforY} by expanding up to $\sim\bigO(1/r^6)$ order. 
We can see that for larger values of $\omega$, the asymptotic expansion converges rapidly to the definition.
However, for smaller values of $\omega$, the convergence decreases, and we need to have a larger $\rs^{\rm out}$, or equivalently, increase the expansion order.
From our calculations, we find that generally setting $\rs^{\rm out}={\rm max}(1000, 10\pi/|\omega|)$ is sufficient to reach the $10^{-12}$ relative tolerance if we truncate the expansion at $\sim\bigO(1/r^6)$ order.

\subsection{$Y^{\rm up}$ solution}
\label{App:subsec:Y_up}
Recall that in Ref.~\cite{Lo:2023fvv}, we derived the asymptotic expansion of $X^{\rm up}$ for $r \to r_+$ as
\begin{multline}
        X^{\rm up}(r\to\rp) = C_{\rm SN}^{\rm inc}e^{i\kappa \rs}\sum_{w=0}^\infty\mathcal{C}^{\rm H}_{+,w}\left(r-\rp\right)^w\\
        +C_{\rm SN}^{\rm ref}e^{-i\kappa \rs}\sum_{w=0}^\infty\mathcal{C}^{\rm H}_{-,w}\left(r-\rp\right)^w.
\end{multline}
Following the same procedure in Appendix \ref{App:subsec:Y_in}, we obtain
\begin{subequations}\label{Eq.Y''_hor}
    \begin{align}
        {Y^{\rm up}_+}''(r\to\rp) & = C_{\rm SN}^{\rm inc}\sum_{j=0}^\infty Y^{\rm H}_{+,j}\left(r-\rp\right)^j,\label{Eq.Y''_hor_+}\\
        {Y^{\rm up}_-}''(r\to\rp) & = C_{\rm SN}^{\rm ref}\sum_{j=0}^\infty Y^{\rm H}_{-,j}\left(r-\rp\right)^{j+iq},\label{Eq.Y''_hor_-}
    \end{align}
\end{subequations}
where
\begin{equation}
    q=\frac{(a\rp m+2a^2\omega-4\rp\omega)}{\rp\sqrt{1-a^2}}.
\end{equation}
Unfortunately, the expressions of $Y^{\rm H}_{\pm,0,1,2}$ and $\mathcal{C}^{\rm H}_{\pm, 0,1,2}$ are too long to show directly here and are available in a Mathematica notebook \cite{10.5281/zenodo.17574059}. By combining Eq.~\eqref{Eq.Y''_hor_+} and Eq.~\eqref{Eq.Y''_hor_-}, we obtain Eq.~\eqref{Eq.Yup''AsymptoticExpansion}.

We then integrate Eqs.~\eqref{Eq.Y''_hor} to get the initial values as
\begin{subequations}
    \begin{align}
        &{Y_+^{\rm up}}'(r_{\rm in})=C_{\rm SN}^{\rm inc}\sum_{j=0}^\infty Y_{+,j}^{\rm H}\frac{\left(r_{\rm in}-\rp\right)^{j+1}}{j+1},\\
        &Y_+^{\rm up}(r_{\rm in})=C_{\rm SN}^{\rm inc}\sum_{j=0}^\infty Y_{+,j}^{\rm H}\frac{\left(r_{\rm in}-\rp\right)^{j+2}}{(j+1)(j+2)},\\
        &{Y_-^{\rm up}}'(r_{\rm in})=C_{\rm SN}^{\rm ref}\sum_{j=0}^\infty Y_{-,j}^{\rm H}\frac{\left(r_{\rm in}-\rp\right)^{j+1+iq}}{j+1+iq},\\
        &Y_-^{\rm up}(r_{\rm in})=C_{\rm SN}^{\rm ref}\sum_{j=0}^\infty Y_{-,j}^{\rm H}\frac{\left(r_{\rm in}-\rp\right)^{j+2+iq}}{(j+1+iq)(j+2+iq)}.
    \end{align}
\end{subequations}

Now we can also construct the initial conditions for the \gls{ODE} in Eq.~\eqref{Eq.Y_rs_ODE} for $Y^{\rm up}$ and solve it outward to infinity to get the initial values of $Y^{\rm up}(r)$ and ${Y^{\rm up}}'(r)$ across the entire domain of definition using
\begin{subequations}
    \begin{align}
        \left. Y^{\rm up} \right|_{\rs =\rs^{\rm in}} & =Y^{\rm up}_+(r_{\rm in})+Y^{\rm up}_-(r_{\rm in}),\\
        \left. \frac{\diff Y^{\rm up}}{\diff \rs}\right|_{\rs=\rs^{\rm in}} & =\frac{\Delta}{r_{\rm in}^2+a^2}\left[{Y^{\rm up}_+}'(r_{\rm in})+{Y^{\rm up}_-}'(r_{\rm in})\right].
    \end{align}
\end{subequations}

\section{Expressions in $\mathcal{W}$ integrals}
\label{Appendix_W_ingredients}
\begin{widetext}
\begin{subequations}\label{Eq.fgh_012}
    \begin{align}
        &f_0(r)=\frac{\mathscr{A}}{\omega^2}w_{nn}^{(0)}(r)\sim\bigO\left(u^r\right),\\
        &f_1(r)=\frac{\mathscr{A}}{\omega^2}\left[{w_{nn}^{(0)}}'(r)+i\xi(r)w_{nn}^{(0)}(r)+w_{nn}^{(1)}(r)\right]\sim\bigO\left(\frac{u^r}{r}\right),\\
        &f_2(r)=\frac{\mathscr{A}}{\omega^2}\left[{w_{nn}^{(1)}}'(r)+i\xi(r)w_{nn}^{(1)}(r)\right]\sim\bigO\left(\frac{u^r}{r^2}\right),\\
        &g_0(r)=-\frac{\mathscr{A}}{i\omega}w_{n\bar{m}}^{(0)}(r)\sim\bigO\left(1\right),\\
        &g_1(r)=-\frac{\mathscr{A}}{i\omega}\left[{w_{n\bar{m}}^{(0)}}'(r)+i\xi(r)w_{n\bar{m}}^{(0)}(r)-w_{n\bar{m}}^{(1)}(r)+w_{n\bar{m}}^{(2)}(r)\right]\sim\bigO\left(\frac{1}{r}\right),\\
        &g_2(r)=\frac{\mathscr{A}}{i\omega}\left[\left(w_{n\bar{m}}^{(1)}(r)-w_{n\bar{m}}^{(2)}(r)\right)'+i\xi(r)\left(w_{n\bar{m}}^{(1)}(r)-w_{n\bar{m}}^{(2)}(r)\right)\right]\sim\bigO\left(\frac{1}{r^2}\right),\\
        &h_0(r)=-\mathscr{A}\frac{Sr^2\bar{\rho}^4\bar{\mathcal{M}}^2}{2\rho^2 u^r}\sim\bigO\left(\frac{1}{u^r}\right),\\
        &h_1(r)=-\mathscr{A}\left[\left(\frac{r^2}{\rho}\right)'+\frac{\left(r^2\rho^3\right)'}{\rho^4}\right]\frac{S\bar{\rho}^4\bar{\mathcal{M}}^2}{2\rho u^r}\sim\bigO\left(\frac{1}{ru^r}\right),\\
        &h_2(r)=-\mathscr{A}\left[\frac{\left(r^2\rho^3\right)'}{\rho^4}\right]'\frac{S\bar{\rho}^4\bar{\mathcal{M}}^2}{2\rho u^r}\sim\bigO\left(\frac{1}{r^2u^r}\right),
    \end{align}
\end{subequations}
\end{widetext}
with
\begin{subequations}
    \begin{align}
        &w_{nn}^{(0)}(r)=\frac{1}{2}r^2\rho\bar{\rho}^2u^r \mathscr{L}_1^\dagger\left[\rho^{-4}\mathscr{L}_2^\dagger\left(\rho^3S\right)\right],\\
        &w_{nn}^{(1)}(r)=w_{nn}^{(0)}(r)\left(\frac{\mathcal{N}}{u^r}\right)'\frac{u^r}{\mathcal{N}}+{w_{nn}^{(0)}}'(r)+i\xi(r)w_{nn}^{(0)}(r),\\
        &w_{n\bar{m}}^{(0)}(r)=\frac{r^2\bar{\rho}^3}{\rho}\bar{\mathcal{M}}\left[\mathscr{L}_2^\dagger S+ia\left(\rho-\bar{\rho}\right)\sin\theta S\right],\\
        &w_{n\bar{m}}^{(1)}(r)=\frac{r^2\bar{\rho}\bar{\mathcal{M}}}{2}\mathscr{L}_2^\dagger\left[\rho^3S\left(\bar{\rho}^2\rho^{-4}\right)'\right],\\
        &w_{n\bar{m}}^{(2)}(r)=\bar{\rho}\bar{\mathcal{M}}\left\{\frac{r^2\bar{\rho}^2}{\rho}\left[\mathscr{L}_2^\dagger S+i a\left(\rho-\bar{\rho}\right)\sin\theta S\right]\right\}'.
    \end{align}
\end{subequations}

\section{Solving for geodesic motions for Kerr black holes}
\label{Appendix_geodesic_motions} %
The motion of a particle in a Kerr background is determined by the four constants of motion, namely, the mass $\mu$, energy $E$, angular momentum along the spin axis $L_z$, and Carter constant $Q$.
In test mass limit, i.e., $\mu\ll1$, the motion can be described by the following equations of motion in Kerr spacetime:
\begin{subequations}\label{Eq.4velocities}
    \begin{align}
        &\Sigma\frac{\diff t}{\diff\tau}=-a\left(a\mathcal{E}\sin^2\theta-\mathcal{L}_z\right)+\frac{r^2+a^2}{\Delta}P,\\
        &\Sigma\frac{\diff r}{\diff\tau}=\pm\sqrt{R},\\
        &\Sigma\frac{\diff\theta}{\diff\tau}=\pm\sqrt{\Theta},\\
        &\Sigma\frac{\diff\varphi}{\diff\tau}=-\left(a\mathcal{E}-\frac{\mathcal{L}_z}{\sin^2\theta}\right)+\frac{a}{\Delta}P,
    \end{align}
\end{subequations}
where
\begin{subequations}\label{Eq.4veloFunctions}
    \begin{align}
        &P=\mathcal{E}(r^2+a^2)-a\mathcal{L}_z,\\
        &R=P^2-\Delta\left[r^2+\left(\mathcal{L}_z-a\mathcal{E}\right)^2+\mathcal{Q}\right],\\
        &\Theta=\mathcal{Q}-\cos^2\theta\left[a^2\left(1-\mathcal{E}^2\right)+\frac{\mathcal{L}_z^2}{\sin^2\theta}\right].
    \end{align}
\end{subequations}
Here the constants are rescaled by $\mathcal{E} \equiv E/\mu$, $\mathcal{L}_z \equiv L_z/\left(M\mu\right)$, and $\mathcal{Q} \equiv Q/\left(M\mu\right)^2$. 

We follow the procedure in Ref.~\cite{Fujita:2009bp} to solve the equations. Here we outline the algorithm:
\begin{enumerate}
    \item For a given set of orbital parameters, namely, the spin parameter $a$, semi-latus rectum $p$, eccentricity $e$, and inclination parameter $x \equiv \cos\theta_{\rm inc}$, we follow Ref.~\cite{W_Schmidt_2002} to map them to the constants of motion $(\mathcal{E},\ \mathcal{L}_z,\ \mathcal{Q})$.
    \item Then we calculate the main frequencies of the motions by integrating the geodesic equations. By using the Mino time $\lambda$ where $\diff\lambda \equiv \diff\tau/\Sigma$, we can decouple the $r$- and $\theta$-direction motions \cite{Mino:2003yg} and obtain the Mino frequencies $\Upsilon_r$ and $\Upsilon_\theta$ using elliptic integrals. We calculate the $t$- and $\varphi$-direction Mino frequencies $\Gamma$ and $\Upsilon_\varphi$, respectively, based on the $r$ and $\theta$ motions.
    \item Finally, we solve the geodesic equations by integrating them over one period, analytically expressing them as elliptic integrals using the Mino time $\lambda$, and extending them to full domain from $-\infty$ to $\infty$.
\end{enumerate}
Following the above three steps, we obtain the solution of a generic timelike bound geodesic motion which can be written as Eqs.~\eqref{Eq.KerrGeoOrbit}.

\begin{figure}[htpb!]
    \centering
    \includegraphics[width=1.0\linewidth]{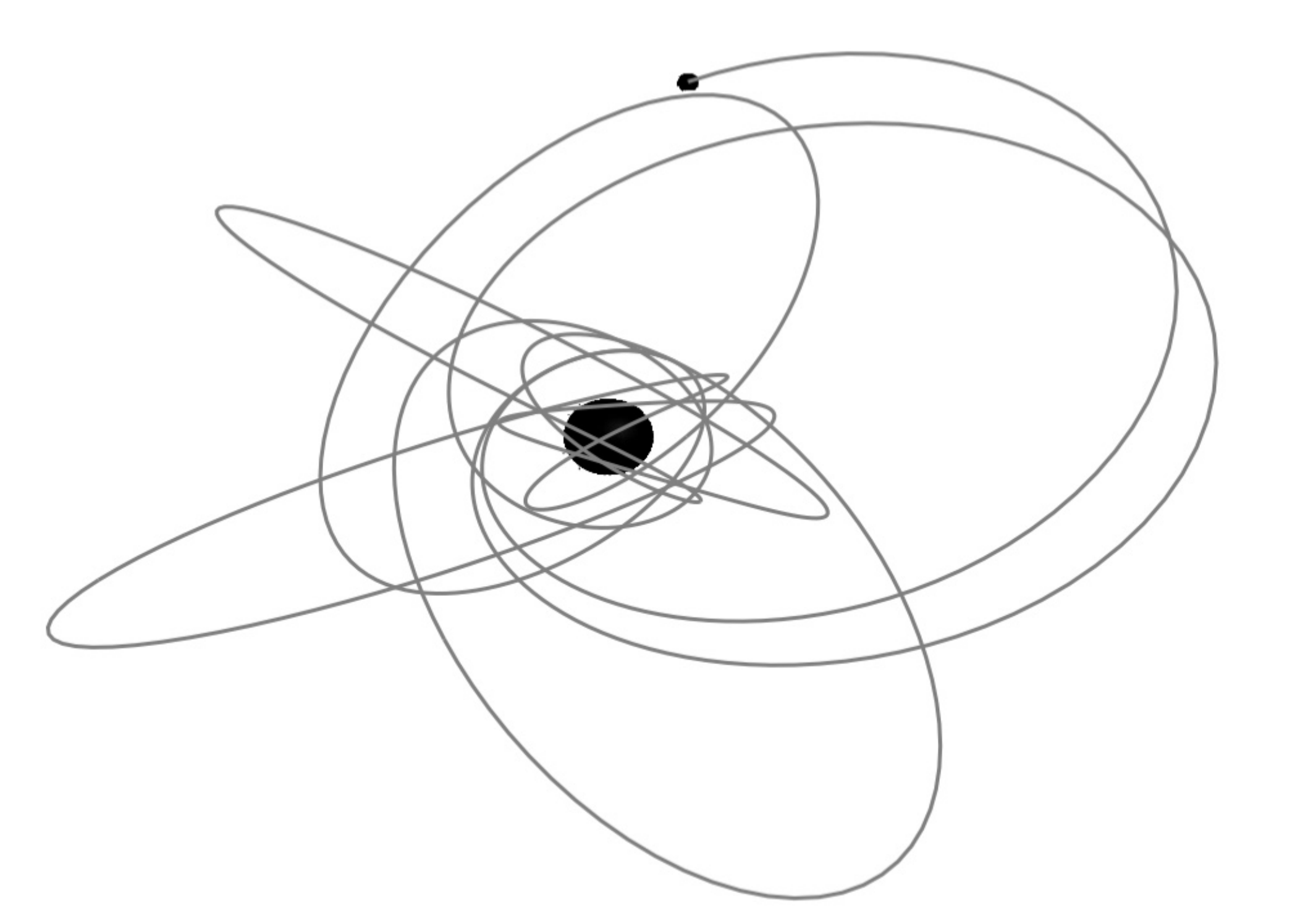}
    \caption{A timelike bound geodesic trajectory around a Kerr black hole.}
    \label{fig:generic_trjectory}
\end{figure}

We implemented \texttt{KerrGeodesics.jl}\footnote{\url{https://github.com/CuberYyc808/KerrGeodesics.jl}}, utilizing the high performance of \texttt{julia} to numerically compute the elliptic integrals (using the package \texttt{Elliptic.jl}).
One can obtain all the ingredients for calculating the solution in microseconds.
Figure~\ref{fig:generic_trjectory} shows the trajectory with $a=0.9M$, $p=6M$, $e=0.7$, $x=\cos\pi/4$, which corresponds to the waveform in Fig.~\ref{fig:wavform_bound}.

Currently, unbound geodesics are not available in the package.
We plan to include plunge orbits (see Ref.~\cite{Dyson:2023fws}) and scattering orbits (see Ref.~\cite{PhysRevD.96.084044}) in the future.

\section{Levin's method}\label{Appendix_Levin}
The evaluation of highly oscillatory integrals, such as Eq.~\eqref{Eq.J_double_integral} when $n$ and $k$ are large, is difficult and often encountered in a wide range of problems.
To tackle this issue, we employ Levin's method \cite{e14f9c91-61c1-39f6-9e5a-c0c05123dd15}, which converts the quadrature problem into an equivalent system of \glspl{ODE} that gives the antiderivative function of the integrand kernel. 
In the following, we briefly illustrate Levin's method for one-dimensional integrals.

For a one-dimensional integral of the form
\begin{equation}\label{eq:Levin_1D_integral}
    \mathbb{I}=\int_a^b f(r)e^{ig(r)}\diff r,
\end{equation}
where the phase function $g(r)$ varies rapidly while the kernel function $f(r)$ varies slowly, we want to find the solution of $p(r)$ that satisfies the following \gls{ODE}
\begin{equation}\label{eq:ODE_Levin}
    p'(r)+ig'(r)p(r)=f(r).
\end{equation}
With $p(r)$, $\mathbb{I}$ can be evaluated using simply
\begin{equation}
    \mathbb{I}=p(b)e^{ig(b)}-p(a)e^{ig(a)}.
\end{equation}
Following Ref.~\cite{10.1007/s00211-024-01443-6}, we solve Eq.~\eqref{eq:ODE_Levin} for $p(r)$ using a Chebyshev spectral method. The \gls{ODE} problem is further transformed into a problem of solving a system of linear equations given by
\begin{equation}
    \left[\overleftrightarrow{D}+i\overleftrightarrow{g'}\right] \; \vec{p}=\vec{f},
\end{equation}
where $\overleftrightarrow{D}$ is the differentiation matrix, $\overleftrightarrow{g'}$ and $\vec{f}$ are a diagonal matrix and a vector evaluated at the collocation points, respectively.
We refer readers to a detailed exposition of the algorithm for one-dimensional integrals and the two-dimensional generalization in Refs.~\cite{10.1007/s00211-024-01443-6} and~\cite{2025arXiv250602424C}, respectively.

To facilitate the calculations in this work, we implemented an optimized version of the adaptive Levin's algorithm following Refs.~\cite{10.1007/s00211-024-01443-6, 2025arXiv250602424C} in \texttt{julia}, which is publicly available as \texttt{AdaptiveLevin.jl}\footnote{\url{https://github.com/CuberYyc808/AdaptiveLevin.jl}}.

\bibliography{BHPT}%

\begin{thebibliography}{56}%
\makeatletter
\providecommand \@ifxundefined [1]{%
 \@ifx{#1\undefined}
}%
\providecommand \@ifnum [1]{%
 \ifnum #1\expandafter \@firstoftwo
 \else \expandafter \@secondoftwo
 \fi
}%
\providecommand \@ifx [1]{%
 \ifx #1\expandafter \@firstoftwo
 \else \expandafter \@secondoftwo
 \fi
}%
\providecommand \natexlab [1]{#1}%
\providecommand \enquote  [1]{``#1''}%
\providecommand \bibnamefont  [1]{#1}%
\providecommand \bibfnamefont [1]{#1}%
\providecommand \citenamefont [1]{#1}%
\providecommand \href@noop [0]{\@secondoftwo}%
\providecommand \href [0]{\begingroup \@sanitize@url \@href}%
\providecommand \@href[1]{\@@startlink{#1}\@@href}%
\providecommand \@@href[1]{\endgroup#1\@@endlink}%
\providecommand \@sanitize@url [0]{\catcode `\\12\catcode `\$12\catcode `\&12\catcode `\#12\catcode `\^12\catcode `\_12\catcode `\%12\relax}%
\providecommand \@@startlink[1]{}%
\providecommand \@@endlink[0]{}%
\providecommand \url  [0]{\begingroup\@sanitize@url \@url }%
\providecommand \@url [1]{\endgroup\@href {#1}{\urlprefix }}%
\providecommand \urlprefix  [0]{URL }%
\providecommand \Eprint [0]{\href }%
\providecommand \doibase [0]{https://doi.org/}%
\providecommand \selectlanguage [0]{\@gobble}%
\providecommand \bibinfo  [0]{\@secondoftwo}%
\providecommand \bibfield  [0]{\@secondoftwo}%
\providecommand \translation [1]{[#1]}%
\providecommand \BibitemOpen [0]{}%
\providecommand \bibitemStop [0]{}%
\providecommand \bibitemNoStop [0]{.\EOS\space}%
\providecommand \EOS [0]{\spacefactor3000\relax}%
\providecommand \BibitemShut  [1]{\csname bibitem#1\endcsname}%
\let\auto@bib@innerbib\@empty
\bibitem [{\citenamefont {Abbott}\ \emph {et~al.}(2016)\citenamefont {Abbott} \emph {et~al.}}]{LIGOScientific:2016aoc}%
  \BibitemOpen
  \bibfield  {author} {\bibinfo {author} {\bibfnamefont {B.~P.}\ \bibnamefont {Abbott}} \emph {et~al.} (\bibinfo {collaboration} {LIGO Scientific, Virgo}),\ }\bibfield  {title} {\bibinfo {title} {{Observation of Gravitational Waves from a Binary Black Hole Merger}},\ }\href {https://doi.org/10.1103/PhysRevLett.116.061102} {\bibfield  {journal} {\bibinfo  {journal} {Phys. Rev. Lett.}\ }\textbf {\bibinfo {volume} {116}},\ \bibinfo {pages} {061102} (\bibinfo {year} {2016})},\ \Eprint {https://arxiv.org/abs/1602.03837} {arXiv:1602.03837 [gr-qc]} \BibitemShut {NoStop}%
\bibitem [{\citenamefont {Abac}\ \emph {et~al.}(2025{\natexlab{a}})\citenamefont {Abac} \emph {et~al.}}]{LIGOScientific:2025pvj}%
  \BibitemOpen
  \bibfield  {author} {\bibinfo {author} {\bibfnamefont {A.~G.}\ \bibnamefont {Abac}} \emph {et~al.} (\bibinfo {collaboration} {LIGO Scientific, VIRGO, KAGRA}),\ }\href@noop {} {\bibinfo {title} {{GWTC-4.0: Population Properties of Merging Compact Binaries}}} (\bibinfo {year} {2025}{\natexlab{a}}),\ \Eprint {https://arxiv.org/abs/2508.18083} {arXiv:2508.18083 [astro-ph.HE]} \BibitemShut {NoStop}%
\bibitem [{\citenamefont {Abac}\ \emph {et~al.}(2025{\natexlab{b}})\citenamefont {Abac} \emph {et~al.}}]{LIGOScientific:2025rid}%
  \BibitemOpen
  \bibfield  {author} {\bibinfo {author} {\bibfnamefont {A.~G.}\ \bibnamefont {Abac}} \emph {et~al.} (\bibinfo {collaboration} {LIGO Scientific, Virgo, KAGRA}),\ }\bibfield  {title} {\bibinfo {title} {{GW250114: Testing Hawking{\textquoteright}s Area Law and the Kerr Nature of Black Holes}},\ }\href {https://doi.org/10.1103/kw5g-d732} {\bibfield  {journal} {\bibinfo  {journal} {Phys. Rev. Lett.}\ }\textbf {\bibinfo {volume} {135}},\ \bibinfo {pages} {111403} (\bibinfo {year} {2025}{\natexlab{b}})},\ \Eprint {https://arxiv.org/abs/2509.08054} {arXiv:2509.08054 [gr-qc]} \BibitemShut {NoStop}%
\bibitem [{\citenamefont {Aasi}\ \emph {et~al.}(2015)\citenamefont {Aasi} \emph {et~al.}}]{LIGOScientific:2014pky}%
  \BibitemOpen
  \bibfield  {author} {\bibinfo {author} {\bibfnamefont {J.}~\bibnamefont {Aasi}} \emph {et~al.} (\bibinfo {collaboration} {LIGO Scientific}),\ }\bibfield  {title} {\bibinfo {title} {{Advanced LIGO}},\ }\href {https://doi.org/10.1088/0264-9381/32/7/074001} {\bibfield  {journal} {\bibinfo  {journal} {Class. Quant. Grav.}\ }\textbf {\bibinfo {volume} {32}},\ \bibinfo {pages} {074001} (\bibinfo {year} {2015})},\ \Eprint {https://arxiv.org/abs/1411.4547} {arXiv:1411.4547 [gr-qc]} \BibitemShut {NoStop}%
\bibitem [{\citenamefont {Acernese}\ \emph {et~al.}(2015)\citenamefont {Acernese} \emph {et~al.}}]{VIRGO:2014yos}%
  \BibitemOpen
  \bibfield  {author} {\bibinfo {author} {\bibfnamefont {F.}~\bibnamefont {Acernese}} \emph {et~al.} (\bibinfo {collaboration} {VIRGO}),\ }\bibfield  {title} {\bibinfo {title} {{Advanced Virgo: a second-generation interferometric gravitational wave detector}},\ }\href {https://doi.org/10.1088/0264-9381/32/2/024001} {\bibfield  {journal} {\bibinfo  {journal} {Class. Quant. Grav.}\ }\textbf {\bibinfo {volume} {32}},\ \bibinfo {pages} {024001} (\bibinfo {year} {2015})},\ \Eprint {https://arxiv.org/abs/1408.3978} {arXiv:1408.3978 [gr-qc]} \BibitemShut {NoStop}%
\bibitem [{\citenamefont {Somiya}(2012)}]{Somiya:2011np}%
  \BibitemOpen
  \bibfield  {author} {\bibinfo {author} {\bibfnamefont {K.}~\bibnamefont {Somiya}} (\bibinfo {collaboration} {KAGRA}),\ }\bibfield  {title} {\bibinfo {title} {{Detector configuration of KAGRA: The Japanese cryogenic gravitational-wave detector}},\ }\href {https://doi.org/10.1088/0264-9381/29/12/124007} {\bibfield  {journal} {\bibinfo  {journal} {Class. Quant. Grav.}\ }\textbf {\bibinfo {volume} {29}},\ \bibinfo {pages} {124007} (\bibinfo {year} {2012})},\ \Eprint {https://arxiv.org/abs/1111.7185} {arXiv:1111.7185 [gr-qc]} \BibitemShut {NoStop}%
\bibitem [{\citenamefont {Colpi}\ \emph {et~al.}(2024)\citenamefont {Colpi} \emph {et~al.}}]{Colpi:2024xhw}%
  \BibitemOpen
  \bibfield  {author} {\bibinfo {author} {\bibfnamefont {M.}~\bibnamefont {Colpi}} \emph {et~al.},\ }\href@noop {} {\bibinfo {title} {{LISA Definition Study Report}}} (\bibinfo {year} {2024}),\ \Eprint {https://arxiv.org/abs/2402.07571} {arXiv:2402.07571 [astro-ph.CO]} \BibitemShut {NoStop}%
\bibitem [{\citenamefont {{Amaro-Seoane}}(2018)}]{2018LRR....21....4A}%
  \BibitemOpen
  \bibfield  {author} {\bibinfo {author} {\bibfnamefont {P.}~\bibnamefont {{Amaro-Seoane}}},\ }\bibfield  {title} {\bibinfo {title} {{Relativistic dynamics and extreme mass ratio inspirals}},\ }\href {https://doi.org/10.1007/s41114-018-0013-8} {\bibfield  {journal} {\bibinfo  {journal} {Living Reviews in Relativity}\ }\textbf {\bibinfo {volume} {21}},\ \bibinfo {eid} {4} (\bibinfo {year} {2018})},\ \Eprint {https://arxiv.org/abs/1205.5240} {arXiv:1205.5240 [astro-ph.CO]} \BibitemShut {NoStop}%
\bibitem [{\citenamefont {Babak}\ \emph {et~al.}(2017)\citenamefont {Babak}, \citenamefont {Gair}, \citenamefont {Sesana}, \citenamefont {Barausse}, \citenamefont {Sopuerta}, \citenamefont {Berry}, \citenamefont {Berti}, \citenamefont {Amaro-Seoane}, \citenamefont {Petiteau},\ and\ \citenamefont {Klein}}]{Babak:2017tow}%
  \BibitemOpen
  \bibfield  {author} {\bibinfo {author} {\bibfnamefont {S.}~\bibnamefont {Babak}}, \bibinfo {author} {\bibfnamefont {J.}~\bibnamefont {Gair}}, \bibinfo {author} {\bibfnamefont {A.}~\bibnamefont {Sesana}}, \bibinfo {author} {\bibfnamefont {E.}~\bibnamefont {Barausse}}, \bibinfo {author} {\bibfnamefont {C.~F.}\ \bibnamefont {Sopuerta}}, \bibinfo {author} {\bibfnamefont {C.~P.~L.}\ \bibnamefont {Berry}}, \bibinfo {author} {\bibfnamefont {E.}~\bibnamefont {Berti}}, \bibinfo {author} {\bibfnamefont {P.}~\bibnamefont {Amaro-Seoane}}, \bibinfo {author} {\bibfnamefont {A.}~\bibnamefont {Petiteau}},\ and\ \bibinfo {author} {\bibfnamefont {A.}~\bibnamefont {Klein}},\ }\bibfield  {title} {\bibinfo {title} {{Science with the space-based interferometer LISA. V: Extreme mass-ratio inspirals}},\ }\href {https://doi.org/10.1103/PhysRevD.95.103012} {\bibfield  {journal} {\bibinfo  {journal} {Phys. Rev. D}\ }\textbf {\bibinfo {volume} {95}},\ \bibinfo {pages} {103012} (\bibinfo {year} {2017})},\ \Eprint
  {https://arxiv.org/abs/1703.09722} {arXiv:1703.09722 [gr-qc]} \BibitemShut {NoStop}%
\bibitem [{\citenamefont {Berry}\ \emph {et~al.}(2019)\citenamefont {Berry}, \citenamefont {Hughes}, \citenamefont {Sopuerta}, \citenamefont {Chua}, \citenamefont {Heffernan}, \citenamefont {Holley-Bockelmann}, \citenamefont {Mihaylov}, \citenamefont {Miller},\ and\ \citenamefont {Sesana}}]{Berry:2019wgg}%
  \BibitemOpen
  \bibfield  {author} {\bibinfo {author} {\bibfnamefont {C.~P.~L.}\ \bibnamefont {Berry}}, \bibinfo {author} {\bibfnamefont {S.~A.}\ \bibnamefont {Hughes}}, \bibinfo {author} {\bibfnamefont {C.~F.}\ \bibnamefont {Sopuerta}}, \bibinfo {author} {\bibfnamefont {A.~J.~K.}\ \bibnamefont {Chua}}, \bibinfo {author} {\bibfnamefont {A.}~\bibnamefont {Heffernan}}, \bibinfo {author} {\bibfnamefont {K.}~\bibnamefont {Holley-Bockelmann}}, \bibinfo {author} {\bibfnamefont {D.~P.}\ \bibnamefont {Mihaylov}}, \bibinfo {author} {\bibfnamefont {M.~C.}\ \bibnamefont {Miller}},\ and\ \bibinfo {author} {\bibfnamefont {A.}~\bibnamefont {Sesana}},\ }\bibfield  {title} {\bibinfo {title} {{The unique potential of extreme mass-ratio inspirals for gravitational-wave astronomy}},\ }\href@noop {} {\bibfield  {journal} {\bibinfo  {journal} {Bull. Am. Astron. Soc.}\ }\textbf {\bibinfo {volume} {51}},\ \bibinfo {pages} {42} (\bibinfo {year} {2019})},\ \Eprint {https://arxiv.org/abs/1903.03686} {arXiv:1903.03686 [astro-ph.HE]} \BibitemShut
  {NoStop}%
\bibitem [{\citenamefont {Chandrasekhar}(1985)}]{Chandrasekhar:1985kt}%
  \BibitemOpen
  \bibfield  {author} {\bibinfo {author} {\bibfnamefont {S.}~\bibnamefont {Chandrasekhar}},\ }\href@noop {} {\emph {\bibinfo {title} {{The mathematical theory of black holes}}}}\ (\bibinfo {year} {1985})\BibitemShut {NoStop}%
\bibitem [{\citenamefont {Teukolsky}(1973)}]{Teukolsky:1973ha}%
  \BibitemOpen
  \bibfield  {author} {\bibinfo {author} {\bibfnamefont {S.~A.}\ \bibnamefont {Teukolsky}},\ }\bibfield  {title} {\bibinfo {title} {{Perturbations of a rotating black hole. 1. Fundamental equations for gravitational electromagnetic and neutrino field perturbations}},\ }\href {https://doi.org/10.1086/152444} {\bibfield  {journal} {\bibinfo  {journal} {Astrophys. J.}\ }\textbf {\bibinfo {volume} {185}},\ \bibinfo {pages} {635} (\bibinfo {year} {1973})}\BibitemShut {NoStop}%
\bibitem [{\citenamefont {Lo}(2024)}]{Lo:2023fvv}%
  \BibitemOpen
  \bibfield  {author} {\bibinfo {author} {\bibfnamefont {R.~K.~L.}\ \bibnamefont {Lo}},\ }\bibfield  {title} {\bibinfo {title} {{Recipes for computing radiation from a Kerr black hole using a generalized Sasaki-Nakamura formalism: Homogeneous solutions}},\ }\href {https://doi.org/10.1103/PhysRevD.110.124070} {\bibfield  {journal} {\bibinfo  {journal} {Phys. Rev. D}\ }\textbf {\bibinfo {volume} {110}},\ \bibinfo {pages} {124070} (\bibinfo {year} {2024})},\ \Eprint {https://arxiv.org/abs/2306.16469} {arXiv:2306.16469 [gr-qc]} \BibitemShut {NoStop}%
\bibitem [{\citenamefont {Mano}\ \emph {et~al.}(1996)\citenamefont {Mano}, \citenamefont {Suzuki},\ and\ \citenamefont {Takasugi}}]{Mano:1996vt}%
  \BibitemOpen
  \bibfield  {author} {\bibinfo {author} {\bibfnamefont {S.}~\bibnamefont {Mano}}, \bibinfo {author} {\bibfnamefont {H.}~\bibnamefont {Suzuki}},\ and\ \bibinfo {author} {\bibfnamefont {E.}~\bibnamefont {Takasugi}},\ }\bibfield  {title} {\bibinfo {title} {{Analytic solutions of the Teukolsky equation and their low frequency expansions}},\ }\href {https://doi.org/10.1143/PTP.95.1079} {\bibfield  {journal} {\bibinfo  {journal} {Prog. Theor. Phys.}\ }\textbf {\bibinfo {volume} {95}},\ \bibinfo {pages} {1079} (\bibinfo {year} {1996})},\ \Eprint {https://arxiv.org/abs/gr-qc/9603020} {arXiv:gr-qc/9603020} \BibitemShut {NoStop}%
\bibitem [{\citenamefont {Fujita}\ and\ \citenamefont {Tagoshi}(2004)}]{10.1143/PTP.112.415}%
  \BibitemOpen
  \bibfield  {author} {\bibinfo {author} {\bibfnamefont {R.}~\bibnamefont {Fujita}}\ and\ \bibinfo {author} {\bibfnamefont {H.}~\bibnamefont {Tagoshi}},\ }\bibfield  {title} {\bibinfo {title} {{New Numerical Methods to Evaluate Homogeneous Solutions of the Teukolsky Equation}},\ }\href {https://doi.org/10.1143/PTP.112.415} {\bibfield  {journal} {\bibinfo  {journal} {Progress of Theoretical Physics}\ }\textbf {\bibinfo {volume} {112}},\ \bibinfo {pages} {415} (\bibinfo {year} {2004})},\ \Eprint {https://arxiv.org/abs/https://academic.oup.com/ptp/article-pdf/112/3/415/5382220/112-3-415.pdf} {https://academic.oup.com/ptp/article-pdf/112/3/415/5382220/112-3-415.pdf} \BibitemShut {NoStop}%
\bibitem [{\citenamefont {Fujita}\ and\ \citenamefont {Tagoshi}(2005)}]{10.1143/PTP.113.1165}%
  \BibitemOpen
  \bibfield  {author} {\bibinfo {author} {\bibfnamefont {R.}~\bibnamefont {Fujita}}\ and\ \bibinfo {author} {\bibfnamefont {H.}~\bibnamefont {Tagoshi}},\ }\bibfield  {title} {\bibinfo {title} {{New Numerical Methods to Evaluate Homogeneous Solutions of the Teukolsky Equation. II: — Solutions of the Continued Fraction Equation —}},\ }\href {https://doi.org/10.1143/PTP.113.1165} {\bibfield  {journal} {\bibinfo  {journal} {Progress of Theoretical Physics}\ }\textbf {\bibinfo {volume} {113}},\ \bibinfo {pages} {1165} (\bibinfo {year} {2005})},\ \Eprint {https://arxiv.org/abs/https://academic.oup.com/ptp/article-pdf/113/6/1165/5285582/113-6-1165.pdf} {https://academic.oup.com/ptp/article-pdf/113/6/1165/5285582/113-6-1165.pdf} \BibitemShut {NoStop}%
\bibitem [{\citenamefont {Sasaki}\ and\ \citenamefont {Nakamura}(1982{\natexlab{a}})}]{SASAKI198268}%
  \BibitemOpen
  \bibfield  {author} {\bibinfo {author} {\bibfnamefont {M.}~\bibnamefont {Sasaki}}\ and\ \bibinfo {author} {\bibfnamefont {T.}~\bibnamefont {Nakamura}},\ }\bibfield  {title} {\bibinfo {title} {A class of new perturbation equations for the kerr geometry},\ }\href {https://doi.org/https://doi.org/10.1016/0375-9601(82)90507-2} {\bibfield  {journal} {\bibinfo  {journal} {Physics Letters A}\ }\textbf {\bibinfo {volume} {89}},\ \bibinfo {pages} {68} (\bibinfo {year} {1982}{\natexlab{a}})}\BibitemShut {NoStop}%
\bibitem [{\citenamefont {Sasaki}\ and\ \citenamefont {Nakamura}(1982{\natexlab{b}})}]{10.1143/PTP.67.1788}%
  \BibitemOpen
  \bibfield  {author} {\bibinfo {author} {\bibfnamefont {M.}~\bibnamefont {Sasaki}}\ and\ \bibinfo {author} {\bibfnamefont {T.}~\bibnamefont {Nakamura}},\ }\bibfield  {title} {\bibinfo {title} {{Gravitational Radiation from a Kerr Black Hole. I. Formulation and a Method for Numerical Analysis}},\ }\href {https://doi.org/10.1143/PTP.67.1788} {\bibfield  {journal} {\bibinfo  {journal} {Progress of Theoretical Physics}\ }\textbf {\bibinfo {volume} {67}},\ \bibinfo {pages} {1788} (\bibinfo {year} {1982}{\natexlab{b}})},\ \Eprint {https://arxiv.org/abs/https://academic.oup.com/ptp/article-pdf/67/6/1788/5332324/67-6-1788.pdf} {https://academic.oup.com/ptp/article-pdf/67/6/1788/5332324/67-6-1788.pdf} \BibitemShut {NoStop}%
\bibitem [{\citenamefont {Jiang}\ and\ \citenamefont {Han}(2025)}]{Jiang:2025mna}%
  \BibitemOpen
  \bibfield  {author} {\bibinfo {author} {\bibfnamefont {Y.}~\bibnamefont {Jiang}}\ and\ \bibinfo {author} {\bibfnamefont {W.-B.}\ \bibnamefont {Han}},\ }\href@noop {} {\bibinfo {title} {{A New High-Performing Method for Solving the Homogeneous Teukolsky Equation}}} (\bibinfo {year} {2025}),\ \Eprint {https://arxiv.org/abs/2507.15363} {arXiv:2507.15363 [gr-qc]} \BibitemShut {NoStop}%
\bibitem [{\citenamefont {Drasco}\ and\ \citenamefont {Hughes}(2006)}]{Drasco:2005kz}%
  \BibitemOpen
  \bibfield  {author} {\bibinfo {author} {\bibfnamefont {S.}~\bibnamefont {Drasco}}\ and\ \bibinfo {author} {\bibfnamefont {S.~A.}\ \bibnamefont {Hughes}},\ }\bibfield  {title} {\bibinfo {title} {{Gravitational wave snapshots of generic extreme mass ratio inspirals}},\ }\href {https://doi.org/10.1103/PhysRevD.73.024027} {\bibfield  {journal} {\bibinfo  {journal} {Phys. Rev. D}\ }\textbf {\bibinfo {volume} {73}},\ \bibinfo {pages} {024027} (\bibinfo {year} {2006})},\ \bibinfo {note} {[Erratum: Phys.Rev.D 88, 109905 (2013), Erratum: Phys.Rev.D 90, 109905 (2014)]},\ \Eprint {https://arxiv.org/abs/gr-qc/0509101} {arXiv:gr-qc/0509101} \BibitemShut {NoStop}%
\bibitem [{\citenamefont {Poisson}(1997)}]{Poisson:1996ya}%
  \BibitemOpen
  \bibfield  {author} {\bibinfo {author} {\bibfnamefont {E.}~\bibnamefont {Poisson}},\ }\bibfield  {title} {\bibinfo {title} {{Gravitational radiation from infall into a black hole: Regularization of the Teukolsky equation}},\ }\href {https://doi.org/10.1103/PhysRevD.55.639} {\bibfield  {journal} {\bibinfo  {journal} {Phys. Rev. D}\ }\textbf {\bibinfo {volume} {55}},\ \bibinfo {pages} {639} (\bibinfo {year} {1997})},\ \Eprint {https://arxiv.org/abs/gr-qc/9606078} {arXiv:gr-qc/9606078} \BibitemShut {NoStop}%
\bibitem [{\citenamefont {{Sasaki}}\ and\ \citenamefont {{Tagoshi}}(2003)}]{2003LRR.....6....6S}%
  \BibitemOpen
  \bibfield  {author} {\bibinfo {author} {\bibfnamefont {M.}~\bibnamefont {{Sasaki}}}\ and\ \bibinfo {author} {\bibfnamefont {H.}~\bibnamefont {{Tagoshi}}},\ }\bibfield  {title} {\bibinfo {title} {{Analytic Black Hole Perturbation Approach to Gravitational Radiation}},\ }\href {https://doi.org/10.12942/lrr-2003-6} {\bibfield  {journal} {\bibinfo  {journal} {Living Reviews in Relativity}\ }\textbf {\bibinfo {volume} {6}},\ \bibinfo {eid} {6} (\bibinfo {year} {2003})},\ \Eprint {https://arxiv.org/abs/gr-qc/0306120} {arXiv:gr-qc/0306120 [gr-qc]} \BibitemShut {NoStop}%
\bibitem [{\citenamefont {Watarai}(2024)}]{Watarai:2024huy}%
  \BibitemOpen
  \bibfield  {author} {\bibinfo {author} {\bibfnamefont {D.}~\bibnamefont {Watarai}},\ }\bibfield  {title} {\bibinfo {title} {{Ringdown of a postinnermost stable circular orbit of a rapidly spinning black hole: Mass ratio dependence of higher harmonic quasinormal mode excitation}},\ }\href {https://doi.org/10.1103/PhysRevD.110.124029} {\bibfield  {journal} {\bibinfo  {journal} {Phys. Rev. D}\ }\textbf {\bibinfo {volume} {110}},\ \bibinfo {pages} {124029} (\bibinfo {year} {2024})},\ \Eprint {https://arxiv.org/abs/2408.16747} {arXiv:2408.16747 [gr-qc]} \BibitemShut {NoStop}%
\bibitem [{\citenamefont {Nakamura}\ \emph {et~al.}(1987)\citenamefont {Nakamura}, \citenamefont {Oohara},\ and\ \citenamefont {Kojima}}]{10.1143/PTPS.90.1}%
  \BibitemOpen
  \bibfield  {author} {\bibinfo {author} {\bibfnamefont {T.}~\bibnamefont {Nakamura}}, \bibinfo {author} {\bibfnamefont {K.}~\bibnamefont {Oohara}},\ and\ \bibinfo {author} {\bibfnamefont {Y.}~\bibnamefont {Kojima}},\ }\bibfield  {title} {\bibinfo {title} {General relativistic collapse to black holes and gravitational waves from black holes},\ }\href {https://doi.org/10.1143/PTPS.90.1} {\bibfield  {journal} {\bibinfo  {journal} {Progress of Theoretical Physics Supplement}\ }\textbf {\bibinfo {volume} {90}},\ \bibinfo {pages} {1} (\bibinfo {year} {1987})},\ \Eprint {https://arxiv.org/abs/https://academic.oup.com/ptps/article-pdf/doi/10.1143/PTPS.90.1/5201911/90-1.pdf} {https://academic.oup.com/ptps/article-pdf/doi/10.1143/PTPS.90.1/5201911/90-1.pdf} \BibitemShut {NoStop}%
\bibitem [{\citenamefont {Shibata}(1993{\natexlab{a}})}]{10.1143/ptp/90.3.595}%
  \BibitemOpen
  \bibfield  {author} {\bibinfo {author} {\bibfnamefont {M.}~\bibnamefont {Shibata}},\ }\bibfield  {title} {\bibinfo {title} {Gravitational waves induced by a particle orbiting around a rotating black hole: Effect of orbital precession},\ }\href {https://doi.org/10.1143/ptp/90.3.595} {\bibfield  {journal} {\bibinfo  {journal} {Progress of Theoretical Physics}\ }\textbf {\bibinfo {volume} {90}},\ \bibinfo {pages} {595} (\bibinfo {year} {1993}{\natexlab{a}})},\ \Eprint {https://arxiv.org/abs/https://academic.oup.com/ptp/article-pdf/90/3/595/5161685/90-3-595.pdf} {https://academic.oup.com/ptp/article-pdf/90/3/595/5161685/90-3-595.pdf} \BibitemShut {NoStop}%
\bibitem [{\citenamefont {Sago}\ and\ \citenamefont {Tanaka}(2020)}]{10.1093/ptep/ptaa149}%
  \BibitemOpen
  \bibfield  {author} {\bibinfo {author} {\bibfnamefont {N.}~\bibnamefont {Sago}}\ and\ \bibinfo {author} {\bibfnamefont {T.}~\bibnamefont {Tanaka}},\ }\bibfield  {title} {\bibinfo {title} {Gravitational wave echoes induced by a point mass plunging into a black hole},\ }\href {https://doi.org/10.1093/ptep/ptaa149} {\bibfield  {journal} {\bibinfo  {journal} {Progress of Theoretical and Experimental Physics}\ }\textbf {\bibinfo {volume} {2020}},\ \bibinfo {pages} {123E01} (\bibinfo {year} {2020})}\BibitemShut {NoStop}%
\bibitem [{\citenamefont {Glampedakis}\ and\ \citenamefont {Kennefick}(2002)}]{PhysRevD.66.044002}%
  \BibitemOpen
  \bibfield  {author} {\bibinfo {author} {\bibfnamefont {K.}~\bibnamefont {Glampedakis}}\ and\ \bibinfo {author} {\bibfnamefont {D.}~\bibnamefont {Kennefick}},\ }\bibfield  {title} {\bibinfo {title} {Zoom and whirl: Eccentric equatorial orbits around spinning black holes and their evolution under gravitational radiation reaction},\ }\href {https://doi.org/10.1103/PhysRevD.66.044002} {\bibfield  {journal} {\bibinfo  {journal} {Phys. Rev. D}\ }\textbf {\bibinfo {volume} {66}},\ \bibinfo {pages} {044002} (\bibinfo {year} {2002})}\BibitemShut {NoStop}%
\bibitem [{\citenamefont {Hughes}(2000)}]{PhysRevD.61.084004}%
  \BibitemOpen
  \bibfield  {author} {\bibinfo {author} {\bibfnamefont {S.~A.}\ \bibnamefont {Hughes}},\ }\bibfield  {title} {\bibinfo {title} {Evolution of circular, nonequatorial orbits of kerr black holes due to gravitational-wave emission},\ }\href {https://doi.org/10.1103/PhysRevD.61.084004} {\bibfield  {journal} {\bibinfo  {journal} {Phys. Rev. D}\ }\textbf {\bibinfo {volume} {61}},\ \bibinfo {pages} {084004} (\bibinfo {year} {2000})}\BibitemShut {NoStop}%
\bibitem [{\citenamefont {Shibata}(1993{\natexlab{b}})}]{PhysRevD.48.663}%
  \BibitemOpen
  \bibfield  {author} {\bibinfo {author} {\bibfnamefont {M.}~\bibnamefont {Shibata}},\ }\bibfield  {title} {\bibinfo {title} {Gravitational waves induced by a particle orbiting around a rotating black hole: Spin-orbit interaction effect},\ }\href {https://doi.org/10.1103/PhysRevD.48.663} {\bibfield  {journal} {\bibinfo  {journal} {Phys. Rev. D}\ }\textbf {\bibinfo {volume} {48}},\ \bibinfo {pages} {663} (\bibinfo {year} {1993}{\natexlab{b}})}\BibitemShut {NoStop}%
\bibitem [{\citenamefont {Shibata}(1994)}]{PhysRevD.50.6297}%
  \BibitemOpen
  \bibfield  {author} {\bibinfo {author} {\bibfnamefont {M.}~\bibnamefont {Shibata}},\ }\bibfield  {title} {\bibinfo {title} {Gravitational waves by compact star orbiting around rotating supermassive black holes},\ }\href {https://doi.org/10.1103/PhysRevD.50.6297} {\bibfield  {journal} {\bibinfo  {journal} {Phys. Rev. D}\ }\textbf {\bibinfo {volume} {50}},\ \bibinfo {pages} {6297} (\bibinfo {year} {1994})}\BibitemShut {NoStop}%
\bibitem [{BHP()}]{BHPToolkit}%
  \BibitemOpen
  \href@noop {} {\bibinfo {title} {{Black Hole Perturbation Toolkit}}},\ \bibinfo {howpublished} {(\href{http://bhptoolkit.org/}{bhptoolkit.org})}\BibitemShut {NoStop}%
\bibitem [{\citenamefont {Nasipak}(2022)}]{PhysRevD.106.064042}%
  \BibitemOpen
  \bibfield  {author} {\bibinfo {author} {\bibfnamefont {Z.}~\bibnamefont {Nasipak}},\ }\bibfield  {title} {\bibinfo {title} {Adiabatic evolution due to the conservative scalar self-force during orbital resonances},\ }\href {https://doi.org/10.1103/PhysRevD.106.064042} {\bibfield  {journal} {\bibinfo  {journal} {Phys. Rev. D}\ }\textbf {\bibinfo {volume} {106}},\ \bibinfo {pages} {064042} (\bibinfo {year} {2022})}\BibitemShut {NoStop}%
\bibitem [{\citenamefont {Nasipak}(2024)}]{PhysRevD.109.044020}%
  \BibitemOpen
  \bibfield  {author} {\bibinfo {author} {\bibfnamefont {Z.}~\bibnamefont {Nasipak}},\ }\bibfield  {title} {\bibinfo {title} {Adiabatic gravitational waveform model for compact objects undergoing quasicircular inspirals into rotating massive black holes},\ }\href {https://doi.org/10.1103/PhysRevD.109.044020} {\bibfield  {journal} {\bibinfo  {journal} {Phys. Rev. D}\ }\textbf {\bibinfo {volume} {109}},\ \bibinfo {pages} {044020} (\bibinfo {year} {2024})}\BibitemShut {NoStop}%
\bibitem [{\citenamefont {Mitsou}(2011)}]{PhysRevD.83.044039}%
  \BibitemOpen
  \bibfield  {author} {\bibinfo {author} {\bibfnamefont {E.}~\bibnamefont {Mitsou}},\ }\bibfield  {title} {\bibinfo {title} {Gravitational radiation from radial infall of a particle into a schwarzschild black hole: A numerical study of the spectra, quasinormal modes, and power-law tails},\ }\href {https://doi.org/10.1103/PhysRevD.83.044039} {\bibfield  {journal} {\bibinfo  {journal} {Phys. Rev. D}\ }\textbf {\bibinfo {volume} {83}},\ \bibinfo {pages} {044039} (\bibinfo {year} {2011})}\BibitemShut {NoStop}%
\bibitem [{\citenamefont {Smarr}(1977)}]{PhysRevD.15.2069}%
  \BibitemOpen
  \bibfield  {author} {\bibinfo {author} {\bibfnamefont {L.}~\bibnamefont {Smarr}},\ }\bibfield  {title} {\bibinfo {title} {Gravitational radiation from distant encounters and from head-on collisions of black holes: The zero-frequency limit},\ }\href {https://doi.org/10.1103/PhysRevD.15.2069} {\bibfield  {journal} {\bibinfo  {journal} {Phys. Rev. D}\ }\textbf {\bibinfo {volume} {15}},\ \bibinfo {pages} {2069} (\bibinfo {year} {1977})}\BibitemShut {NoStop}%
\bibitem [{\citenamefont {Berti}\ \emph {et~al.}(2010)\citenamefont {Berti}, \citenamefont {Cardoso}, \citenamefont {Hinderer}, \citenamefont {Lemos}, \citenamefont {Pretorius}, \citenamefont {Sperhake},\ and\ \citenamefont {Yunes}}]{PhysRevD.81.104048}%
  \BibitemOpen
  \bibfield  {author} {\bibinfo {author} {\bibfnamefont {E.}~\bibnamefont {Berti}}, \bibinfo {author} {\bibfnamefont {V.}~\bibnamefont {Cardoso}}, \bibinfo {author} {\bibfnamefont {T.}~\bibnamefont {Hinderer}}, \bibinfo {author} {\bibfnamefont {M.}~\bibnamefont {Lemos}}, \bibinfo {author} {\bibfnamefont {F.}~\bibnamefont {Pretorius}}, \bibinfo {author} {\bibfnamefont {U.}~\bibnamefont {Sperhake}},\ and\ \bibinfo {author} {\bibfnamefont {N.}~\bibnamefont {Yunes}},\ }\bibfield  {title} {\bibinfo {title} {Semianalytical estimates of scattering thresholds and gravitational radiation in ultrarelativistic black hole encounters},\ }\href {https://doi.org/10.1103/PhysRevD.81.104048} {\bibfield  {journal} {\bibinfo  {journal} {Phys. Rev. D}\ }\textbf {\bibinfo {volume} {81}},\ \bibinfo {pages} {104048} (\bibinfo {year} {2010})}\BibitemShut {NoStop}%
\bibitem [{\citenamefont {Cardoso}\ and\ \citenamefont {Lemos}(2003)}]{Cardoso:2002jr}%
  \BibitemOpen
  \bibfield  {author} {\bibinfo {author} {\bibfnamefont {V.}~\bibnamefont {Cardoso}}\ and\ \bibinfo {author} {\bibfnamefont {J.~P.~S.}\ \bibnamefont {Lemos}},\ }\bibfield  {title} {\bibinfo {title} {{Gravitational radiation from the radial infall of highly relativistic point particles into Kerr black holes}},\ }\href {https://doi.org/10.1103/PhysRevD.67.084005} {\bibfield  {journal} {\bibinfo  {journal} {Phys. Rev. D}\ }\textbf {\bibinfo {volume} {67}},\ \bibinfo {pages} {084005} (\bibinfo {year} {2003})},\ \Eprint {https://arxiv.org/abs/gr-qc/0211094} {arXiv:gr-qc/0211094} \BibitemShut {NoStop}%
\bibitem [{\citenamefont {Chua}\ \emph {et~al.}(2021)\citenamefont {Chua}, \citenamefont {Katz}, \citenamefont {Warburton},\ and\ \citenamefont {Hughes}}]{Chua:2020stf}%
  \BibitemOpen
  \bibfield  {author} {\bibinfo {author} {\bibfnamefont {A.~J.~K.}\ \bibnamefont {Chua}}, \bibinfo {author} {\bibfnamefont {M.~L.}\ \bibnamefont {Katz}}, \bibinfo {author} {\bibfnamefont {N.}~\bibnamefont {Warburton}},\ and\ \bibinfo {author} {\bibfnamefont {S.~A.}\ \bibnamefont {Hughes}},\ }\bibfield  {title} {\bibinfo {title} {{Rapid generation of fully relativistic extreme-mass-ratio-inspiral waveform templates for LISA data analysis}},\ }\href {https://doi.org/10.1103/PhysRevLett.126.051102} {\bibfield  {journal} {\bibinfo  {journal} {Phys. Rev. Lett.}\ }\textbf {\bibinfo {volume} {126}},\ \bibinfo {pages} {051102} (\bibinfo {year} {2021})},\ \Eprint {https://arxiv.org/abs/2008.06071} {arXiv:2008.06071 [gr-qc]} \BibitemShut {NoStop}%
\bibitem [{\citenamefont {Katz}\ \emph {et~al.}(2021)\citenamefont {Katz}, \citenamefont {Chua}, \citenamefont {Speri}, \citenamefont {Warburton},\ and\ \citenamefont {Hughes}}]{Katz:2021yft}%
  \BibitemOpen
  \bibfield  {author} {\bibinfo {author} {\bibfnamefont {M.~L.}\ \bibnamefont {Katz}}, \bibinfo {author} {\bibfnamefont {A.~J.~K.}\ \bibnamefont {Chua}}, \bibinfo {author} {\bibfnamefont {L.}~\bibnamefont {Speri}}, \bibinfo {author} {\bibfnamefont {N.}~\bibnamefont {Warburton}},\ and\ \bibinfo {author} {\bibfnamefont {S.~A.}\ \bibnamefont {Hughes}},\ }\bibfield  {title} {\bibinfo {title} {{Fast extreme-mass-ratio-inspiral waveforms: New tools for millihertz gravitational-wave data analysis}},\ }\href {https://doi.org/10.1103/PhysRevD.104.064047} {\bibfield  {journal} {\bibinfo  {journal} {Phys. Rev. D}\ }\textbf {\bibinfo {volume} {104}},\ \bibinfo {pages} {064047} (\bibinfo {year} {2021})},\ \Eprint {https://arxiv.org/abs/2104.04582} {arXiv:2104.04582 [gr-qc]} \BibitemShut {NoStop}%
\bibitem [{\citenamefont {Speri}\ \emph {et~al.}(2024)\citenamefont {Speri}, \citenamefont {Katz}, \citenamefont {Chua}, \citenamefont {Hughes}, \citenamefont {Warburton}, \citenamefont {Thompson}, \citenamefont {Chapman-Bird},\ and\ \citenamefont {Gair}}]{Speri:2023jte}%
  \BibitemOpen
  \bibfield  {author} {\bibinfo {author} {\bibfnamefont {L.}~\bibnamefont {Speri}}, \bibinfo {author} {\bibfnamefont {M.~L.}\ \bibnamefont {Katz}}, \bibinfo {author} {\bibfnamefont {A.~J.~K.}\ \bibnamefont {Chua}}, \bibinfo {author} {\bibfnamefont {S.~A.}\ \bibnamefont {Hughes}}, \bibinfo {author} {\bibfnamefont {N.}~\bibnamefont {Warburton}}, \bibinfo {author} {\bibfnamefont {J.~E.}\ \bibnamefont {Thompson}}, \bibinfo {author} {\bibfnamefont {C.~E.~A.}\ \bibnamefont {Chapman-Bird}},\ and\ \bibinfo {author} {\bibfnamefont {J.~R.}\ \bibnamefont {Gair}},\ }\bibfield  {title} {\bibinfo {title} {{Fast and Fourier: Extreme Mass Ratio Inspiral Waveforms in the Frequency Domain}},\ }\bibfield  {journal} {\bibinfo  {journal} {Front. Appl. Math. Stat.}\ }\textbf {\bibinfo {volume} {9}},\ \href {https://doi.org/10.3389/fams.2023.1266739} {10.3389/fams.2023.1266739} (\bibinfo {year} {2024}),\ \Eprint {https://arxiv.org/abs/2307.12585} {arXiv:2307.12585 [gr-qc]} \BibitemShut {NoStop}%
\bibitem [{\citenamefont {Chapman-Bird}\ \emph {et~al.}(2025)\citenamefont {Chapman-Bird} \emph {et~al.}}]{Chapman-Bird:2025xtd}%
  \BibitemOpen
  \bibfield  {author} {\bibinfo {author} {\bibfnamefont {C.~E.~A.}\ \bibnamefont {Chapman-Bird}} \emph {et~al.},\ }\href@noop {} {\bibinfo {title} {{The Fast and the Frame-Dragging: Efficient waveforms for asymmetric-mass eccentric equatorial inspirals into rapidly-spinning black holes}}} (\bibinfo {year} {2025}),\ \Eprint {https://arxiv.org/abs/2506.09470} {arXiv:2506.09470 [gr-qc]} \BibitemShut {NoStop}%
\bibitem [{\citenamefont {Lo}\ and\ \citenamefont {Yin}(2026)}]{Lo:2025lpo}%
  \BibitemOpen
  \bibfield  {author} {\bibinfo {author} {\bibfnamefont {R.~K.~L.}\ \bibnamefont {Lo}}\ and\ \bibinfo {author} {\bibfnamefont {Y.}~\bibnamefont {Yin}},\ }\bibfield  {title} {\bibinfo {title} {{Near-horizon gravitational perturbations of rotating black holes}},\ }\href {https://doi.org/10.1103/bljh-l413} {\bibfield  {journal} {\bibinfo  {journal} {Phys. Rev. D}\ }\textbf {\bibinfo {volume} {113}},\ \bibinfo {pages} {L061505} (\bibinfo {year} {2026})},\ \Eprint {https://arxiv.org/abs/2512.07937} {arXiv:2512.07937 [gr-qc]} \BibitemShut {NoStop}%
\bibitem [{\citenamefont {Zhang}\ \emph {et~al.}(2013)\citenamefont {Zhang}, \citenamefont {Berti},\ and\ \citenamefont {Cardoso}}]{Zhang:2013ksa}%
  \BibitemOpen
  \bibfield  {author} {\bibinfo {author} {\bibfnamefont {Z.}~\bibnamefont {Zhang}}, \bibinfo {author} {\bibfnamefont {E.}~\bibnamefont {Berti}},\ and\ \bibinfo {author} {\bibfnamefont {V.}~\bibnamefont {Cardoso}},\ }\bibfield  {title} {\bibinfo {title} {{Quasinormal ringing of Kerr black holes. II. Excitation by particles falling radially with arbitrary energy}},\ }\href {https://doi.org/10.1103/PhysRevD.88.044018} {\bibfield  {journal} {\bibinfo  {journal} {Phys. Rev. D}\ }\textbf {\bibinfo {volume} {88}},\ \bibinfo {pages} {044018} (\bibinfo {year} {2013})},\ \Eprint {https://arxiv.org/abs/1305.4306} {arXiv:1305.4306 [gr-qc]} \BibitemShut {NoStop}%
\bibitem [{\citenamefont {Lo}\ \emph {et~al.}(2025)\citenamefont {Lo}, \citenamefont {Sabani},\ and\ \citenamefont {Cardoso}}]{Lo:2025njp}%
  \BibitemOpen
  \bibfield  {author} {\bibinfo {author} {\bibfnamefont {R.~K.~L.}\ \bibnamefont {Lo}}, \bibinfo {author} {\bibfnamefont {L.}~\bibnamefont {Sabani}},\ and\ \bibinfo {author} {\bibfnamefont {V.}~\bibnamefont {Cardoso}},\ }\bibfield  {title} {\bibinfo {title} {{Quasinormal modes and excitation factors of Kerr black holes}},\ }\href {https://doi.org/10.1103/PhysRevD.111.124002} {\bibfield  {journal} {\bibinfo  {journal} {Phys. Rev. D}\ }\textbf {\bibinfo {volume} {111}},\ \bibinfo {pages} {124002} (\bibinfo {year} {2025})},\ \Eprint {https://arxiv.org/abs/2504.00084} {arXiv:2504.00084 [gr-qc]} \BibitemShut {NoStop}%
\bibitem [{\citenamefont {Jakobsen}\ \emph {et~al.}(2023)\citenamefont {Jakobsen}, \citenamefont {Mogull}, \citenamefont {Plefka},\ and\ \citenamefont {Sauer}}]{Jakobsen:2023hig}%
  \BibitemOpen
  \bibfield  {author} {\bibinfo {author} {\bibfnamefont {G.~U.}\ \bibnamefont {Jakobsen}}, \bibinfo {author} {\bibfnamefont {G.}~\bibnamefont {Mogull}}, \bibinfo {author} {\bibfnamefont {J.}~\bibnamefont {Plefka}},\ and\ \bibinfo {author} {\bibfnamefont {B.}~\bibnamefont {Sauer}},\ }\bibfield  {title} {\bibinfo {title} {{Dissipative Scattering of Spinning Black Holes at Fourth Post-Minkowskian Order}},\ }\href {https://doi.org/10.1103/PhysRevLett.131.241402} {\bibfield  {journal} {\bibinfo  {journal} {Phys. Rev. Lett.}\ }\textbf {\bibinfo {volume} {131}},\ \bibinfo {pages} {241402} (\bibinfo {year} {2023})},\ \Eprint {https://arxiv.org/abs/2308.11514} {arXiv:2308.11514 [hep-th]} \BibitemShut {NoStop}%
\bibitem [{\citenamefont {Warburton}(2025)}]{Warburton:2025ymy}%
  \BibitemOpen
  \bibfield  {author} {\bibinfo {author} {\bibfnamefont {N.}~\bibnamefont {Warburton}},\ }\bibfield  {title} {\bibinfo {title} {{Gravitational radiation from hyperbolic orbits: comparison between self-force, post-Minkowskian, post-Newtonian, and numerical relativity results}},\ }\href@noop {} {\  (\bibinfo {year} {2025})},\ \Eprint {https://arxiv.org/abs/2512.02274} {arXiv:2512.02274 [gr-qc]} \BibitemShut {NoStop}%
\bibitem [{\citenamefont {Yin}\ \emph {et~al.}()\citenamefont {Yin}, \citenamefont {Lo},\ and\ \citenamefont {Chen}}]{10.5281/zenodo.17574059}%
  \BibitemOpen
  \bibfield  {author} {\bibinfo {author} {\bibfnamefont {Y.}~\bibnamefont {Yin}}, \bibinfo {author} {\bibfnamefont {R.~K.~L.}\ \bibnamefont {Lo}},\ and\ \bibinfo {author} {\bibfnamefont {X.}~\bibnamefont {Chen}},\ }\href {https://doi.org/10.5281/zenodo.17574059} {\bibinfo {title} {10.5281/zenodo.17574059}}\BibitemShut {NoStop}%
\bibitem [{\citenamefont {Slevinsky}\ and\ \citenamefont {other contributors}(2025)}]{hypergeometricfunctionsjl}%
  \BibitemOpen
  \bibfield  {author} {\bibinfo {author} {\bibfnamefont {R.~M.}\ \bibnamefont {Slevinsky}}\ and\ \bibinfo {author} {\bibnamefont {other contributors}},\ }\href@noop {} {\bibinfo {title} {\texttt{HypergeometricFunctions.jl}}},\ \bibinfo {howpublished} {\url{https://github.com/JuliaMath/HypergeometricFunctions.jl}} (\bibinfo {year} {2018-2025})\BibitemShut {NoStop}%
\bibitem [{\citenamefont {Fujita}\ and\ \citenamefont {Hikida}(2009)}]{Fujita:2009bp}%
  \BibitemOpen
  \bibfield  {author} {\bibinfo {author} {\bibfnamefont {R.}~\bibnamefont {Fujita}}\ and\ \bibinfo {author} {\bibfnamefont {W.}~\bibnamefont {Hikida}},\ }\bibfield  {title} {\bibinfo {title} {{Analytical solutions of bound timelike geodesic orbits in Kerr spacetime}},\ }\href {https://doi.org/10.1088/0264-9381/26/13/135002} {\bibfield  {journal} {\bibinfo  {journal} {Class. Quant. Grav.}\ }\textbf {\bibinfo {volume} {26}},\ \bibinfo {pages} {135002} (\bibinfo {year} {2009})},\ \Eprint {https://arxiv.org/abs/0906.1420} {arXiv:0906.1420 [gr-qc]} \BibitemShut {NoStop}%
\bibitem [{\citenamefont {Schmidt}(2002)}]{W_Schmidt_2002}%
  \BibitemOpen
  \bibfield  {author} {\bibinfo {author} {\bibfnamefont {W.}~\bibnamefont {Schmidt}},\ }\bibfield  {title} {\bibinfo {title} {Celestial mechanics in kerr spacetime},\ }\href {https://doi.org/10.1088/0264-9381/19/10/314} {\bibfield  {journal} {\bibinfo  {journal} {Classical and Quantum Gravity}\ }\textbf {\bibinfo {volume} {19}},\ \bibinfo {pages} {2743} (\bibinfo {year} {2002})}\BibitemShut {NoStop}%
\bibitem [{\citenamefont {Mino}(2003)}]{Mino:2003yg}%
  \BibitemOpen
  \bibfield  {author} {\bibinfo {author} {\bibfnamefont {Y.}~\bibnamefont {Mino}},\ }\bibfield  {title} {\bibinfo {title} {{Perturbative approach to an orbital evolution around a supermassive black hole}},\ }\href {https://doi.org/10.1103/PhysRevD.67.084027} {\bibfield  {journal} {\bibinfo  {journal} {Phys. Rev. D}\ }\textbf {\bibinfo {volume} {67}},\ \bibinfo {pages} {084027} (\bibinfo {year} {2003})},\ \Eprint {https://arxiv.org/abs/gr-qc/0302075} {arXiv:gr-qc/0302075} \BibitemShut {NoStop}%
\bibitem [{\citenamefont {Dyson}\ and\ \citenamefont {van~de Meent}(2023)}]{Dyson:2023fws}%
  \BibitemOpen
  \bibfield  {author} {\bibinfo {author} {\bibfnamefont {C.}~\bibnamefont {Dyson}}\ and\ \bibinfo {author} {\bibfnamefont {M.}~\bibnamefont {van~de Meent}},\ }\bibfield  {title} {\bibinfo {title} {{Kerr-fully diving into the abyss: analytic solutions to plunging geodesics in Kerr}},\ }\href {https://doi.org/10.1088/1361-6382/acf552} {\bibfield  {journal} {\bibinfo  {journal} {Class. Quant. Grav.}\ }\textbf {\bibinfo {volume} {40}},\ \bibinfo {pages} {195026} (\bibinfo {year} {2023})},\ \Eprint {https://arxiv.org/abs/2302.03704} {arXiv:2302.03704 [gr-qc]} \BibitemShut {NoStop}%
\bibitem [{\citenamefont {Bini}\ \emph {et~al.}(2017)\citenamefont {Bini}, \citenamefont {Geralico},\ and\ \citenamefont {Vines}}]{PhysRevD.96.084044}%
  \BibitemOpen
  \bibfield  {author} {\bibinfo {author} {\bibfnamefont {D.}~\bibnamefont {Bini}}, \bibinfo {author} {\bibfnamefont {A.}~\bibnamefont {Geralico}},\ and\ \bibinfo {author} {\bibfnamefont {J.}~\bibnamefont {Vines}},\ }\bibfield  {title} {\bibinfo {title} {Hyperbolic scattering of spinning particles by a kerr black hole},\ }\href {https://doi.org/10.1103/PhysRevD.96.084044} {\bibfield  {journal} {\bibinfo  {journal} {Phys. Rev. D}\ }\textbf {\bibinfo {volume} {96}},\ \bibinfo {pages} {084044} (\bibinfo {year} {2017})}\BibitemShut {NoStop}%
\bibitem [{\citenamefont {Levin}(1982)}]{e14f9c91-61c1-39f6-9e5a-c0c05123dd15}%
  \BibitemOpen
  \bibfield  {author} {\bibinfo {author} {\bibfnamefont {D.}~\bibnamefont {Levin}},\ }\bibfield  {title} {\bibinfo {title} {Procedures for computing one- and two-dimensional integrals of functions with rapid irregular oscillations},\ }\href {http://www.jstor.org/stable/2007287} {\bibfield  {journal} {\bibinfo  {journal} {Mathematics of Computation}\ }\textbf {\bibinfo {volume} {38}},\ \bibinfo {pages} {531} (\bibinfo {year} {1982})}\BibitemShut {NoStop}%
\bibitem [{\citenamefont {Chen}\ \emph {et~al.}(2024)\citenamefont {Chen}, \citenamefont {Serkh},\ and\ \citenamefont {Bremer}}]{10.1007/s00211-024-01443-6}%
  \BibitemOpen
  \bibfield  {author} {\bibinfo {author} {\bibfnamefont {S.}~\bibnamefont {Chen}}, \bibinfo {author} {\bibfnamefont {K.}~\bibnamefont {Serkh}},\ and\ \bibinfo {author} {\bibfnamefont {J.}~\bibnamefont {Bremer}},\ }\bibfield  {title} {\bibinfo {title} {On the adaptive levin method},\ }\href {https://doi.org/10.1007/s00211-024-01443-6} {\bibfield  {journal} {\bibinfo  {journal} {Numer. Math.}\ }\textbf {\bibinfo {volume} {156}},\ \bibinfo {pages} {1927–1985} (\bibinfo {year} {2024})}\BibitemShut {NoStop}%
\bibitem [{\citenamefont {{Chen}}\ \emph {et~al.}(2025)\citenamefont {{Chen}}, \citenamefont {{Serkh}}, \citenamefont {{Bremer}},\ and\ \citenamefont {{Aubry}}}]{2025arXiv250602424C}%
  \BibitemOpen
  \bibfield  {author} {\bibinfo {author} {\bibfnamefont {S.}~\bibnamefont {{Chen}}}, \bibinfo {author} {\bibfnamefont {K.}~\bibnamefont {{Serkh}}}, \bibinfo {author} {\bibfnamefont {J.}~\bibnamefont {{Bremer}}},\ and\ \bibinfo {author} {\bibfnamefont {M.}~\bibnamefont {{Aubry}}},\ }\href {https://doi.org/10.48550/arXiv.2506.02424} {\bibinfo {title} {{An adaptive delaminating Levin method in two dimensions}}} (\bibinfo {year} {2025}),\ \Eprint {https://arxiv.org/abs/2506.02424} {arXiv:2506.02424 [cs.NA]} \BibitemShut {NoStop}%
\end{thebibliography}%

\end{document}